%% file: main.tex
\documentclass[acmsmall,screen]{acmart}
\usepackage{ifthen}
\usepackage{tikz}
\usepackage{algorithmic}
\usepackage{graphicx}
\usepackage{textcomp}
\usepackage{xspace}
\usepackage{xcolor}
\usepackage{makecell}
\usepackage{paralist}
\usepackage{mdframed}
\newcommand{\ie}{\emph{i.e.,}\xspace}

\newcommand{\etal}{\emph{et~al.}\xspace}

\usepackage{float}     
\usepackage{tabularx}
\usepackage{array}
\usepackage{ragged2e}
\usepackage{makecell}
\usepackage{placeins}
\usepackage{subcaption}
\usepackage{listings}
\usepackage{longtable}
\usepackage{booktabs}
\usepackage{xltabular}
\usepackage[most]{tcolorbox}
\usepackage{ifthen} 
\usepackage{pdflscape}

\AtBeginDocument{%
  \providecommand\BibTeX{{%
    Bib\TeX}}}

\newboolean{showcomments}
\setboolean{showcomments}{true}

\ifthenelse{\boolean{showcomments}}
  {\newcommand{\nb}[2]{
    \fbox{\bfseries\sffamily\scriptsize#1}
    {\sf\small$\blacktriangleright$\textit{#2}$\blacktriangleleft$}
   }
  }
  {\newcommand{\nb}[2]{}
  }

\newtcolorbox{promptbox}{colback=white, arc=0.5mm, top=1mm, bottom=1mm, left=1mm, right=1mm, title=System prompt used for generation}

\newtcolorbox{optimizationbox}{colback=white, arc=0.5mm, top=1mm, bottom=1mm, left=1mm, right=1mm, title=Optimization Prompt}

\newtcolorbox{finalbox}{colback=white, arc=0.5mm, top=1mm, bottom=1mm, left=1mm, right=1mm, title=Prompt Generation}

\def\BibTeX{{\rm B\kern-.05em{\sc i\kern-.025em b}\kern-.08em
    T\kern-.1667em\lower.7ex\hbox{E}\kern-.125emX}}

\newmdenv[
  backgroundcolor=gray!10,
  linecolor=gray!45,
  linewidth=0.6pt,
  roundcorner=2pt,
  innertopmargin=4pt,
  innerbottommargin=4pt,
  innerleftmargin=6pt,
  innerrightmargin=6pt,
  skipabove=6pt,
  skipbelow=10pt
]{keyfindingbox}

\newcommand{\keyfinding}[3]{%
\begin{keyfindingbox}
\textbf{Key finding:} #1\\
\textbf{Occurrence:} #2\\
\textbf{Domains:} #3
\end{keyfindingbox}
}

\newcommand{\sourcefinding}[3]{%
\begin{keyfindingbox}
\textbf{Key finding:} #1\\
\textbf{Occurrence:} #2\\
\textbf{Representative sources:} #3
\end{keyfindingbox}
}
\begin{document}

\title{Uncertainty in Cyber-Physical Systems: A  Literature Survey}

\author{Amna Khan}
\orcid{0009-0008-5030-8974}
\email{akhan@mondragon.edu}
\affiliation{%
	\institution{Mondragon University}
	\country{Spain}
}

\author{Beena}
\orcid{0009-0002-2989-1399}
\email{b.rai@studenti.unisannio.it}
\affiliation{%
	\institution{University of Sannio}
	\country{Italy}
}

\author{Aitor Arrieta}
\orcid{0000-0001-7507-5080}
\email{aarrieta@mondragon.edu}
\affiliation{%
	\institution{Mondragon University}
	\country{Spain}
}

\author{Massimiliano Di Penta}
\orcid{0000-0002-0340-9747}
\email{dipenta@unisannio.it}
\affiliation{%
\institution{University of Sannio}
	\country{Italy}
}

\begin{abstract}
Cyber-Physical Systems (CPSs) operate in dynamic, dependable environments where sensing, communication, decision-making, control, and assurance are affected by uncertainty. 
This paper characterizes CPS uncertainty through a literature survey comprising 186 primary studies. We (i) characterize uncertainty types, (ii) identify uncertainty sources, (iii) relate uncertainty to life-cycle phases in which they occur, and (iv) overview the possible uncertain mitigation strategies.
The findings show that uncertainty is most often conceptualized as an assurance issue. Safety and specification uncertainty is the most prevalent category, followed by environmental and model/epistemic uncertainty. Domain mapping reveals broad cross-domain engagement, while safety-critical sectors such as autonomous systems, energy systems, and robotics CPSs display domain-specific uncertainty patterns. The life-cycle analysis indicates that uncertainty is not limited to the development or testing phases but also arises during runtime operation, underscoring the need for continuous assurance approaches. Mitigation strategies primarily include runtime monitoring, robust and adaptive control, and probabilistic reasoning. In summary, this survey contributes to a thorough characterization of CPS uncertainty and provides a discussion of articles on this topic.
\end{abstract}

\ccsdesc[500]{Software and its engineering~Extra-functional properties}
\ccsdesc[500]{Software and its engineering~Software functional properties}
\ccsdesc[500]{Software and its engineering~Software testing and debugging}

\keywords{}

\maketitle

\input{sections/introduction}
\FloatBarrier
\input{sections/background}

\FloatBarrier
\input{sections/design}

\FloatBarrier
\input{sections/results}

\FloatBarrier
\input{sections/discussion}
\FloatBarrier
\input{sections/threats}
\FloatBarrier
\input{sections/related}

\FloatBarrier
\input{sections/conclusion}
\FloatBarrier

\bibliographystyle{ACM-Reference-Format}
\bibliography{Biblio.bib}
\end{document}

%% file: sections/introduction.tex
\section{Introduction}
\label{sec:intro}


Cyber Physical Systems (CPSs) integrate digital cyber computations with physical processes~\cite{derler2011modeling}. They continuously sense the environment, process information, make decisions, and actuate the physical world through feedback loops that connect cyber and physical components~\cite{rajkumar2010cps,4519604}. This tight coupling has made CPSs central to many safety- and mission-critical domains, including autonomous vehicles, robotics, unmanned aerial vehicles, , healthcare systems, and industrial automation. In such domains, software does not merely process information: it directly influences physical behavior. As a result, uncertainty in CPSs can affect not only functional correctness and system performance, but also safety, reliability, resilience, and trustworthiness.

Uncertainty is an inherent property of CPSs. These systems operate in open, dynamic, and only partially observable environments, where sensor readings may be noisy, models may be incomplete, communication may be delayed, and physical conditions may change unexpectedly. For instance, a robot may face uncertain terrain, a UAV may encounter wind disturbances, an autonomous vehicle may observe unfamiliar traffic scenarios, and an industrial control system may experience changing process dynamics. In addition, CPSs are increasingly incorporating data-driven and learning-enabled components, which introduce further sources of uncertainty related to data quality, model generalization, distribution shift, and prediction confidence. Consequently, uncertainty can emerge at different layers of a CPS, from perception and communication to computation, control, and actuation.

The literature traditionally distinguishes between aleatoric and epistemic uncertainty~\cite{derkiureghian2009aleatory,hullermeier2021aleatoric}. Aleatoric uncertainty refers to the intrinsic variability of the physical world, such as sensor noise, environmental disturbances, or stochastic process behavior. Epistemic uncertainty, in contrast, stems from incomplete knowledge, such as unknown parameters, simplified models, incomplete requirements, or insufficient training data. While this distinction is useful, it does not fully capture how uncertainty is discussed and managed across the CPS literature. In practice, uncertainty is described using many overlapping terms, including environmental uncertainty, sensor uncertainty, model uncertainty, runtime context uncertainty, communication uncertainty, scenario uncertainty, human-interaction uncertainty, and safety or specification uncertainty. These terms are often used differently across research communities, making it difficult to compare findings and understand how uncertainty affects CPSs as a whole.

This fragmentation is particularly visible across disciplinary perspectives. In control and robotics, uncertainty is commonly associated with sensor noise, model mismatch, state-estimation errors, disturbances, and robust control. In software engineering, uncertainty often appears in requirements, architecture, design trade-offs, testing, verification, runtime adaptation, and assurance. In AI and machine learning, uncertainty is often associated with training data, prediction confidence, distribution shift, model generalization, and the behavior of learning-enabled components. Each perspective provides valuable insights, but the resulting body of work remains scattered. Existing studies often focus on a specific domain, layer, technique, or type of uncertainty, rather than providing an integrated view of how uncertainty arises, propagates, and is mitigated throughout the CPS lifecycle.

This lack of integration creates several challenges. First, researchers and practitioners lack a consolidated taxonomy that connects different uncertainty types across CPS domains. Second, it is unclear which sources of uncertainty are most frequently reported in the literature and how they relate to CPS layers, domains, and lifecycle stages. Third, uncertainty may arise during development, testing, deployment, or runtime operation, but these phases are often studied separately. Finally, mitigation strategies are diverse, ranging from probabilistic modeling and robust control to simulation, testing, runtime monitoring, digital twins, and learning-based adaptation. However, there is limited synthesis of how these strategies are used and combined across CPS research.

To address these gaps, this paper presents a literature survey about uncertainty in CPSs. The goal is to provide a structured and reproducible synthesis of how uncertainty is conceptualized, where it originates, when it manifests during the CPS lifecycle, and how it is mitigated. 

 The survey follows established literature review guidelines and combines database search with citation chasing. We focus on primary studies published between 2015 and 2026 in peer-reviewed venues relevant to software engineering, CPSs, dependable systems, testing and verification, model-driven engineering, reliability, and AI-enabled software-intensive systems.
Note that this survey is not strictly a systematic literature review, especially because, as said, we focused our attention on venues belonging to specific areas of research, unless other venues were identified through snowballing.

The study is guided by four research questions. First, we investigate which types of uncertainty exist in CPSs and how they are described across different domains. Second, we analyze the main sources that cause uncertainty in CPSs. Third, we examine where uncertainty arises or manifests across the CPS lifecycle. Fourth, we identify the mitigation strategies reported in the literature to manage or reduce uncertainty in CPSs. Specifically, this paper makes the following contributions:

\begin{itemize}
    \item A structured characterization of uncertainty types in CPSs. We identify and normalize uncertainty categories reported across the literature, enabling comparison between studies that use different terminology.
    \item A cross-domain analysis of uncertainty sources. We relate uncertainty categories to the underlying causes discussed in the literature, including sensing limitations, model mismatch, environmental variation, communication issues, cyber threats, human interaction, data limitations, and runtime context changes.
    \item A lifecycle-oriented view of CPS uncertainty. We analyze whether uncertainty is addressed during development, testing, deployment, operation, or across multiple phases, highlighting the need to consider uncertainty as a cross-cutting CPS concern.
    \item A characterization of uncertainty mitigation strategies. We identify and organize the approaches used to manage uncertainty, including probabilistic modeling, robust and adaptive control, simulation and digital twins, scenario-based testing, runtime monitoring, formal verification, fault tolerance, uncertainty quantification, and learning-based adaptation.
\end{itemize}

The rest of the paper is structured as follows: Section~\ref{sec:back} provides background on the concepts of systems and software uncertainty. Section~\ref{sec:design} describes the study design and planning. Then, results are reported in Section~\ref{sec:results}, while their implications are discussed in Section~\ref{sec:discussion}, and their threats to validity in Section~\ref{sec:threats}. After a discussion of related literature studies on CPS (Section~\ref{sec:related}), Section~\ref{sec:conclusion} concludes the paper and outlines directions for future work.

%% file: sections/background.tex
\section{Background: Cyber Physical Systems and their Uncertainty}
\label{sec:back}
This section provides the conceptual foundation for the review. It introduces CPS as the target system class and then clarifies the role of uncertainty in system behavior, software assurance, and CPS operation.

\subsection{Cyber-Physical Systems: Conceptual Foundations}
CPSs are physical systems that use computers and communication networks to monitor, coordinate, and control their operations~\cite{rajkumar2010cps}. More broadly, CPSs integrate these functional operations with physical processes through continuous, feedback-driven interaction between the cyber and physical layers. This interaction enables the monitoring of physical states, real-time assessment, and the influence of physical behavior across various temporal and spatial scales~\cite{4519604}.

\begin{figure}[ht]
\centering
\includegraphics[width=0.85\textwidth]{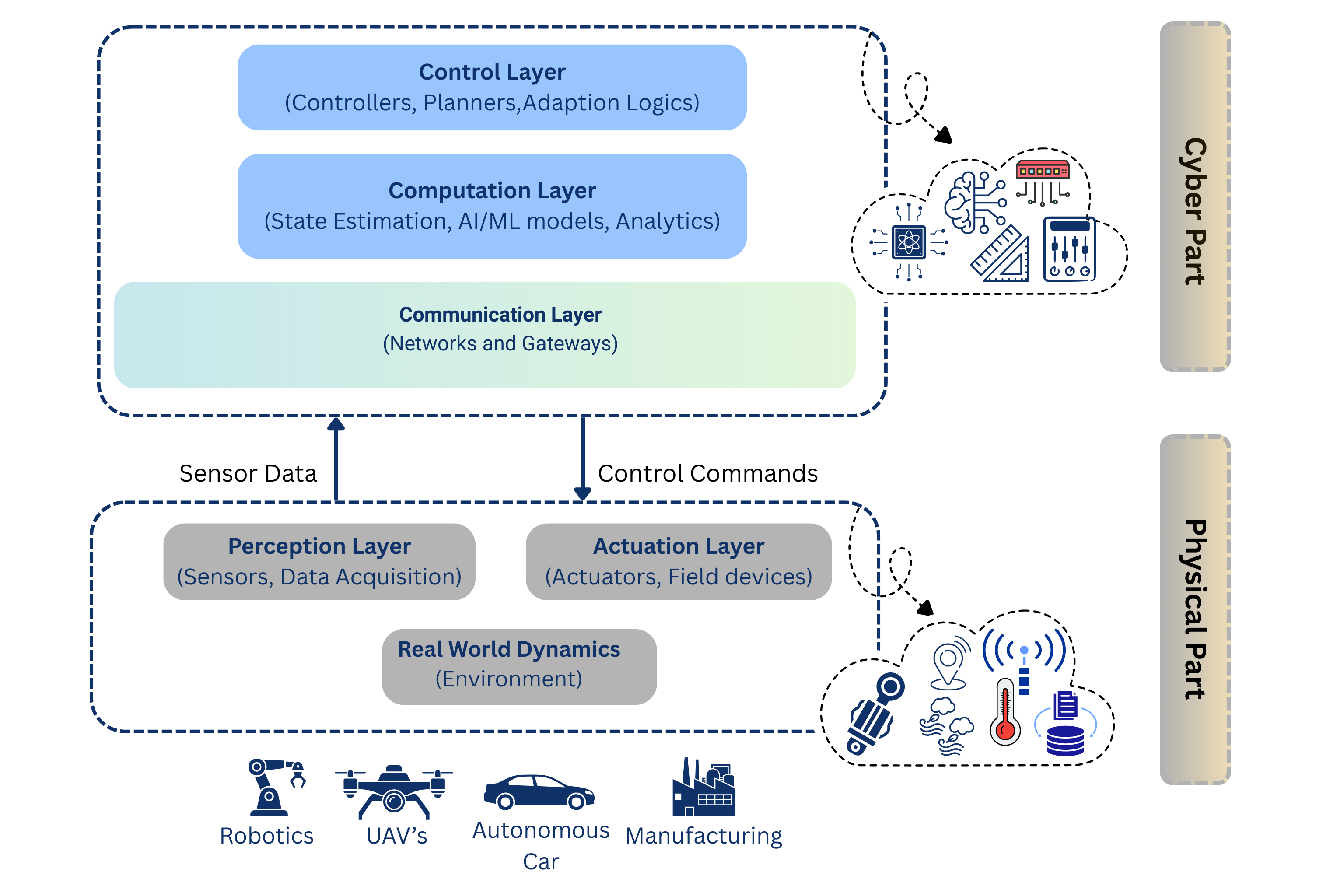}
\caption{Layered Architecture of a CPS}
\label{fig:Architecture_3}
\end{figure}

To understand the provenance of CPS uncertainty, we represent a CPS (independently of its actual software architecture) through a layered functional structure, as illustrated in Fig.~\ref{fig:Architecture_3}. Within this framework, CPS functions are divided into five conceptual layers: (1) perception, (2) communication, (3) computation, (4) control, and (5) actuation. The perception layer acquires data from the physical environment using sensors. The communication layer transfers all the information between distributed components and gateways. The computation layer processes data and performs analytics or state estimation. The control layer determines appropriate system responses. The actuation layer implements control commands within the physical system~\cite{rajkumar2010cps,baheti2011cps,lee2017introduction}. Hence, the figure below, which we used as a reference frame, helps clarify the flow of information and control within the system and provides a clear way to think about system behavior across the digital and physical domains.

CPSs are being adopted across different areas, domains, and industries. For example, in robotics, CPSs enable real-time sensing and actuation, which directly support robots in navigation and manipulation of the physical world, and other human-robot interactions. For example, Unmanned Aerial Vehicles (UAVs) have a CPS architecture that combines sensing, estimated states via filters, and flight control algorithms for stable, robust UAV flight and operation. Autonomous vehicles exemplify CPSs due to their integration of sensing, computation, communication, control, and actuation, enabling environmental perception, decision-making, and safe path following. These are of utmost importance in dynamic environments such as highly unpredictable environments encountered on traffic roads. Smart manufacturing and Cyber-physical production systems are other domains where CPSs play an active role in maximizing system efficiency. The use of CPSs allows for Industry 4.0 features, including adaptive production, monitoring, and optimization. The feedback cycle of CPS enforces a continuous flow of information and control between the Cyber and physical parts of the system, while maintaining a clear separation between sensing, decision-making, and actuation~\cite{inbook}.

\subsection{Uncertainties and their Types in CPS}
Uncertainty in CPSs arises when the current state of the system, current environmental conditions, or the mathematical model's representation cannot be known with full confidence. As we know, CPSs operate in open and dynamic environments; they use imperfect sensing and, most of the time, leverage data-driven components that behave in ways that cannot be understood with utmost certainty beforehand.

In recent years, uncertainty has evolved from a known background assumption catered to simulation levels by introducing parameters to a major issue in software engineering, adaptive systems, and CPSs. This concern has been amplified since the inclusion of data-driven and machine-learning components in systems that interact with the real world. Researchers emphasized the need to deal with uncertainty explicitly throughout the software development life-cycle~\cite{camara2022_interaction_problem_self,al2022guide}. 
Uncertainty is commonly categorized as aleatoric uncertainty, which arises from inherent variability, and epistemic uncertainty, which results from incomplete knowledge or imperfect models~\cite{derkiureghian2009aleatory,hullermeier2021aleatoric}.
In CPS, this distinction remains relevant; however uncertainty is typically manifested at the system level through sources such as the environment, sensing, modeling, communication, runtime context, and faults. For example, environmental uncertainty is considered aleatoric when it reflects variable weather or traffic conditions, but it is epistemic when these conditions are inadequately modeled or insufficiently tested.

\textbf{\textit{Aleatoric Uncertainty:}} Refers to the natural variability in measurements and physical processes that comes from randomness within a CPS. In practice, this often shows up as sensor noise, environmental disturbances, or random changes in how a system behaves. For example, mobile robots may experience measurement noise in their range sensors, Inertial Measurement unit (IMUs) may pick up noise, and UAVs may face unpredictable disturbances from weather, such as wind or gusts. Similarly, variability in material properties, which becomes apparent in manufacturing processes, is also a type of uncertainty. As this variability is inherent to the physical world, aleatoric uncertainty is usually seen as irreducible, although its effects can be described statistically. 

\textbf{\textit{Epistemic uncertainty:}}   Epistemic uncertainty occurs because of the partial and imperfect knowledge about the system and its environment. More specifically, it is linked to data bias, assumptions about the system's modeling, and unknown parameters. This happens when the system is mathematically modeled, but the model cannot fully capture its accurate dynamics of interaction with its environment. For example, measuring parameters such as servo-control gains depends entirely on the modeled dynamics, which introduces inherent uncertainty if the model does not fully represent the system and its environment within its mathematical framework. Friction coefficient is an example of a parameter that is often not modeled accurately and within a range of certainty, since it is known approximately rather than accurately. Apart from this, epistemic uncertainty also arises when learning-based systems are trained on limited or inaccurate data. In contrast to aleatoric uncertainty, epistemic uncertainty can, in theory, be minimized through refined system identification methodologies. Similarly, by using enhanced learning techniques, learning-based systems can be rid of these uncertainties or at least they can be minimized drastically. 

Both of these uncertainties lead to some common issues for CPSs. These issues include parametric uncertainty, measurement errors, and process errors. These uncertainties leave a major impact on numerous stages of the system operations. State estimation is particularly affected by these uncertainties, which naturally propagate into future state predictions thereby directly impacting computation and decision-making, and ultimately actuation. Consequently, uncertainty that develops at one stage propagates to subsequent stages and affects the system's overall behavior. 

There are multiple perspectives presented to clarify why uncertainty in CPS cannot be captured by a single definition. Different research communities focuses on various sources of uncertainty, including incomplete requirements, model mismatch, sensor noise, runtime disturbances, and assurance evidence. This diversity underscores the need to normalize uncertainty terminology across the reviewed CPS literature.

\subsection{Disciplinary Perspectives on Uncertainty in CPSs}
Uncertainty in CPSs has been investigated from various disciplinary perspectives, such as control and robotics, software engineering, testing and verification, and AI/ML-based approaches~\cite{bandyszak2020_orthogonal_modeling_engineering,zhang2018specifying,an2020_modeling_verification_vehicles,catak2022_prediction_validator_deep,zhang2019_wise_test_case}. In particular, control and robotics perspectives deal with the uncertainty that arises from sensing and the environmental parameters. Instead, the software engineering perspective (which will be studied in this article) focuses on the uncertainty that arises during the software life-cycle. 

\begin{figure}[ht]
\centering
\includegraphics[width=0.85\textwidth]{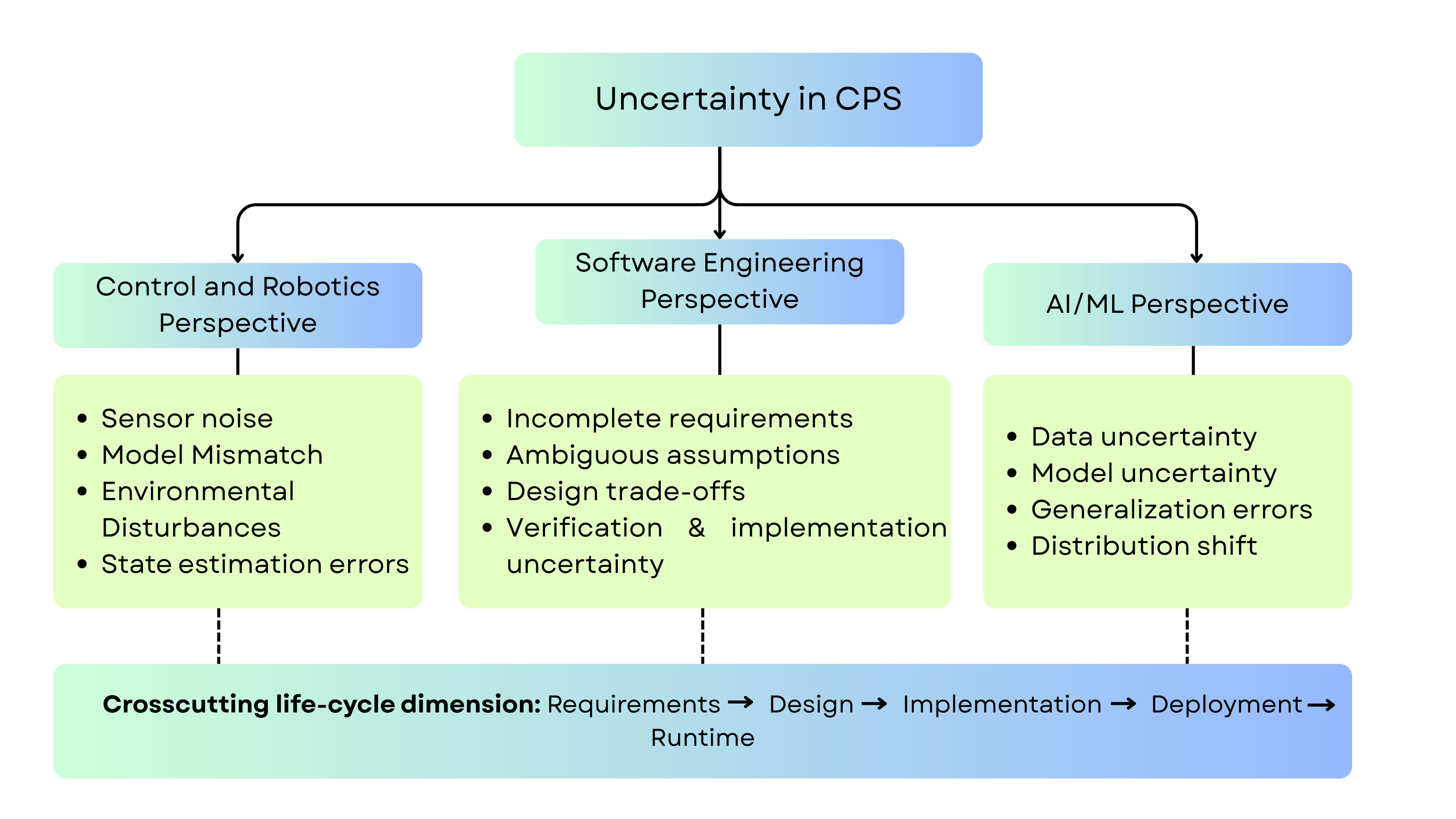}
\caption{Disciplinary perspectives}
\label{fig:disciplinary_perspective_1}
\end{figure}

Fig.~\ref{fig:disciplinary_perspective_1} presents interpretations of uncertainty in CPSs from various disciplinary perspectives, and highlights the integrative role of the life-cycle dimension.

The uncertainty analysis across the CPS life cycle allows one to focus on specific types of uncertainty occurring during requirements specification, design, implementation, integration, deployment, and runtime operation. The life-cycle perspective is not considered a separate or standalone viewpoint. Instead, it serves as a cross-cutting dimension that intersects all disciplinary perspectives. This paper specifically addresses uncertainties that emerge during development and runtime operation, or both. 

Development-time uncertainty occurs before the system's deployment. It usually occurs during requirement solicitation and elicitation, during the design stage, and during development time. At this point, software engineers often lack complete information about the system's goals, i.e., how it will operate or its physical behavior. 
As a result, they rely on simplified models and assumptions to support early design, analysis, and verification.
For example, developers of robots often use approximate kinematic and dynamic models. UAV designers might assume standard day conditions, and autonomous vehicle developers may use idealized friction values to model road conditions. These uncertainties are mostly epistemic, showing the limits of knowledge early in development. 
We discuss these perspectives in detail below.

\subsubsection{Control and Robotics Perspective:}

From a control and robotics perspective, uncertainty is an inherent characteristic of real-world systems. No matter how carefully a system is modeled or designed, a real CPS will never behave exactly like the equations that govern its simulations, and similarly, the measurements that the CPS relies on will never be perfectly accurate or precise. Therefore, the gap between simulation models and real-world behavior can be considered as a source of model, measurement, and disturbance uncertainty. Control strategies must therefore be designed to remain stable and robust under such deviations.
Similarly, estimation error and sensor noise are among the most common sources of uncertainty. 
Specifically, sensor information may yield measurements affected by noise, bias, and delays. Combining all these issues causes a lack of trust/confidence in the system's state, for example, its position, velocity, or orientation. This is a primary reason why filters such as the Kalman Filter use probability as a weight to determine which information to trust. Controllers and computers base their computation on these estimated states, which may have significant errors. External disturbances represent another important source of uncertainty. Robots and UAVs operate in highly dynamic and unpredictable environments. Uneven terrain can affect and disturb ground robots and unmanned ground vehicles. These disturbances introduce unmodeled phenomena that were not explicitly included in the original model used for the control design. However, these must be handled by the controller to ensure stability and meet performance metrics. The mathematical models used in controller design are simplified representations of reality. Variables such as mass, inertia, drag coefficients, motor constants, and delays are often estimated this way, which may introduce inaccuracies and, in general, fail to perfectly represent the system.

In addition, nonlinearities or secondary dynamics may be ignored to keep the model tractable. For example, in the aerospace industry, linear models are used as a basis for control design, and these are not only linear but also time-invariant i.e., the passage of time does not affect these linear representations of a highly nonlinear setup. Over time, parameters may also vary due to payload changes, wear, or battery depletion, increasing the mismatch between the model and the actual system. For these reasons, these systems rely heavily on feedback control and state estimation. Feedback control continuously compares the desired output with the current estimate of the output and applies a control action to steer the current estimate towards the desired output. State estimation techniques filter noisy sensor data to produce more reliable estimates of the system's true state. 

For example, let us consider a quadcopter attempting to maintain a constant altitude of 10 meters using a PID controller. The altitude estimation, which in turn be based on the barometer, would yield measurements that fluctuate due to sensor noise. Similarly, a sudden gust may push the drone upward or downward, a phenomenon often called disturbance. The controller uses the estimated thrust model to compare desired and estimated altitude, then adjusts the throttle to minimize error. If the filters on the sensor measurement and the controller tuning are inadequate, the drone may oscillate around the desired height rather than settling smoothly. This would also result from control gains that do not accurately model the physics of vertical flight. If the control gains are too aggressive or too sluggish, the same phenomenon might be observed. Thus, in practice, uncertainty cannot be eliminated; it can, however, be managed. Effective control design emphasizes robustness and margins, ensuring that systems continue to function reliably even when measurements are noisy, models are imperfect, and margins are severely reduced. 

\subsubsection{Software Engineering Perspective:}
From a software engineering perspective, uncertainty often originates from the user requirements. Most of the time, these requirements are incomplete, ambiguous, or still evolving. Similarly, architectural-level uncertainty arises when numerous design options are available, but it is not obvious which are best suited during the design phase. For instance, the designers of autonomous systems might not know how to balance safety, functionality, and efficiency, or which communication middleware would best meet real-time requirements. 

In general, from a software engineering perspective, uncertainty is mostly viewed as a form of ambiguity that becomes apparent during system design and can affect later stages, such as operation. Methods such as model-driven engineering, architectural patterns, and uncertainty analysis at the design phase help make these uncertainties explicit and easier to manage. 
From a software engineering perspective, 
uncertainty does not always stem from physical processes: Human decisions, including requirement specifications or design decisions, may be possible uncertainty causes.

\subsubsection{AI/ML Perspective:}
In the AI/ML Perspective, uncertainty mainly arises from the data-driven techniques employed, the interfaces used, and the predictions made. As we discussed earlier, limited data for the training phase in the learning-based systems will most certainly introduce uncertainty into the system. Similarly, biases in the data and noisy labels. and distribution shift may also be the reason for this uncertainty. For example, if we train a vision-based model on clear-weather or sunny-day images, it might behave unexpectedly in cloudy or foggy conditions or at other times of day.  From a different perspective, if we train a classifier on very limited data samples, it may develop outliers and make incorrect predictions. As we know, aleatoric uncertainty arises from data noise, whereas epistemic uncertainty reflects the lack of knowledge about the model. These approaches facilitate risk-aware decision-making by generating confidence estimates in conjunction with predictions, thereby quantifying the reliability of obstacle-detection outputs in autonomous driving scenarios.

%% file: sections/design.tex
\section{Study Design}
\label{sec:design}
This section outlines the design of the literature survey. It provides details on the research questions, search strategy, selection process, quality assessment, and data extraction, ensuring transparency and reproducibility.

\subsection{Research Questions}
To characterize uncertainty in CPSs, the study aims to investigate the following research questions (RQs):
\textit{\begin{itemize}
  \item \textbf{RQ1:} What type of uncertainties exist in  CPSs, and how are these uncertainties described in different domains?
  \item \textbf{RQ2:} What are the main sources of uncertainty in CPSs mentioned in the current CPS literature?
  \item \textbf{RQ3:} Where does uncertainty arise or manifest in the life cycle of a CPS? 
  \item \textbf{RQ4:} What mitigation strategies are reported in the literature to deal with uncertainties in CPS?  
\end{itemize}}

\subsection{Review Protocol}
While, as explained in the introduction, this is not strictly a systematic literature review, we structured our analysis by following Kitchenham’s systematic literature review protocol
\cite{KitchenhamBBTBL09}
and Webster's literature review method~\cite{art}.

This study used a detailed review protocol to clarify and improve the reproducibility of identifying and selecting relevant literature on CPS and uncertainty. The protocol outlines the review scope, search sources, eligibility criteria, and a step-by-step screening process to be conducted before searches are initiated. The study selection process was recorded using the PRISMA 2020 framework~\cite{Haddaway2022PRISMA2020}. It offers a standard way to report how records move through different phases, such as identification, screening, eligibility checks, and inclusion criteria.

The study uses two ways to find evidence. 
The database-search stage used Scopus as the primary bibliographic source because of its broad cross-publisher coverage. Searches in the ACM Digital Library and IEEE Xplore were conducted as complementary, publisher-specific searches to identify potentially relevant records that might not be indexed or retrieved equivalently through Scopus. The records obtained from all three sources were subsequently merged and subjected to the same deduplication, venue filtering, eligibility criteria, and screening procedure.
Secondly, citation chasing was carried out, which checks both references and citations, also known as forward and backward snowballing~\cite{Wohlin14}. Conducting snowballing helped us in finding articles that database searches might have missed. We used the CitationChaser tool (Lens-based)~\cite{Haddaway2022Citationchaser} to perform snowballing in both directions. We provided an initial list of seed papers with their DOIs. We exported (i) the list of references of the seed papers for backward snowballing, (ii) the papers that cite the seed papers for the forward snowballing. After the first iteration, we reviewed the papers using inclusion/exclusion criteria, and those that met the criteria were selected as the seed papers for the second iteration. We repeated the process until the final snowballing iteration yielded no further eligible articles.

All records were combined, duplicates were removed, and the remaining papers underwent screening in accordance with the PRISMA process. Predefined exclusion criteria were consistently applied during the study selection process. Reasons for exclusion were recorded at the full-text review stage. Papers were excluded if they did not primarily address CPS or uncertainty, were non-primary literature (e.g., books, posters, surveys, or systematic reviews), or were duplicate entries in the bibliographic files. Fig.~\ref{fig:prisma} illustrates the complete selection process and details the number of records and reports at each stage.

\begin{table}[ht]
\centering
\caption{Inclusion and exclusion criteria used for study selection.}
\label{tab:rules}
\renewcommand{\arraystretch}{1.15}
\setlength{\tabcolsep}{8pt}

\begin{tabular}{p{0.22\textwidth} p{0.73\textwidth}}
\hline
\textbf{Category} & \textbf{Description} \\
\hline

\raisebox{0pt}[0pt][0pt]{\parbox[t]{0.22\textwidth}{\textbf{Inclusion}\\\textbf{Criteria}}} &
\parbox[t]{0.73\textwidth}{
\textbullet\ Primary research articles that specifically address \textbf{CPS} and \textbf{uncertainty}.\\
\textbullet\ Publications within the time window of \textbf{2015--2026}.\\
\textbullet\ Full text available to enable eligibility assessment and data extraction.\\
\textbullet\ Peer-reviewed journal or conference papers written in English.
}
\\
\hline

\raisebox{0pt}[0pt][0pt]{\parbox[t]{0.22\textwidth}{\textbf{Exclusion}\\\textbf{Criteria}}} &
\parbox[t]{0.73\textwidth}{
\textbullet\ Articles that are only focused on CPS but not uncertainty, or papers with uncertainty but without a CPS context.\\
\textbullet\ Secondary papers (surveys, SLRs/mapping studies, books/chapters, posters, editorials).\\
\textbullet\ Publications outside the specified time range (before 2015 or after 2026).\\
\textbullet\ Duplicate records (within or across sources).\\
\textbullet\ Unavailability of full text, or insufficient detail about methodology to support extraction.\\
\textbullet\ Non-English publications.
}
\\
\hline
\end{tabular}
\end{table}

\subsection{Inclusion and Exclusion Criteria}

For this study, inclusion and exclusion criteria were used to ensure we included only papers highly relevant to our work. The included papers identified uncertainties in CPSs and clearly addressed them in the system or its components. We excluded articles that did not cover both CPS and uncertainty to keep the review focused on its goals. Table~\ref{tab:rules} gives the details of the rules we set for the inclusion and exclusion criteria, whereas the keywords used for searching purposes are mentioned in Table~\ref{tab:keywords}.

\FloatBarrier

\subsection{Search Strings and Search Strategy}
The Scopus query constituted the main database search and was designed to retrieve a broad candidate set. The ACM Digital Library and IEEE Xplore queries were used as complementary searches rather than as exact replicas of the Scopus query. Their formulation was adapted to the syntax, indexing mechanisms, and retrieval behaviour of each platform, while retaining the two central concepts of the review: CPSs and uncertainty. Database-specific terms were included in the third query block to target relevant CPS contexts supported by each search interface. All retrieved records were merged and evaluated using an identical selection procedure.

Search String used for Scopus:
\begin{tcolorbox}[
  colback=gray!15,
  colframe=gray!15,
  boxrule=0pt,
  arc=0mm,
  left=6pt,right=6pt,top=4pt,bottom=4pt,
  width=\linewidth
]
\begin{minipage}{\linewidth}
\ttfamily\small\raggedright
("cyber-physical system*" OR "CPS" OR "cyber physical system*")\\
AND\\
( "uncertainty*" OR "Uncertainties*" OR "Unknowns*" )\\
AND\\
("layer*" OR "layers*" OR "perception layer*" OR "control layer*" OR "communication layer*" OR "actuation layer*" )\\
\end{minipage}
\end{tcolorbox}

Search String used for ACM:
\begin{tcolorbox}[
  colback=gray!15,
  colframe=gray!15,
  boxrule=0pt,
  arc=0mm,
  left=6pt,right=6pt,top=4pt,bottom=4pt,
  width=\linewidth
]
\begin{minipage}{\linewidth}
\ttfamily\small\raggedright
("cyber-physical system*" OR "cyber physical system*")\\
AND\\
("uncertainty*" OR "aleatoric" OR "epistemic")\\
AND\\
("perception layer" OR "control layer" OR "communication layer" OR "actuation layer"\\
 OR "system architecture")
\end{minipage}
\end{tcolorbox}

Search String used for IEEE:
\begin{tcolorbox}[
  colback=gray!15,
  colframe=gray!15,
  boxrule=0pt,
  arc=0mm,
  left=6pt,right=6pt,top=4pt,bottom=4pt,
  width=\linewidth
]
\begin{minipage}{\linewidth}
\ttfamily\small\raggedright
("cyber-physical system*" OR "CPS")\\
AND\\
("uncertainty" OR "uncertainty quantification" OR "epistemic" OR "aleatoric")\\
AND\\
("runtime monitoring" OR "runtime assurance" OR "online monitoring" OR "safety monitor")\\
\end{minipage}
\end{tcolorbox}

After the retrieval phase, we removed duplicate records and applied an automated metadata filter to exclude records that were not in the predefined year range and venue list prior to title and abstract screening. This process is documented in the PRISMA box labeled ``Records removed before screening'' in Fig.~\ref{fig:prisma}. Similarly, we screened the remaining records by title and abstract, then conducted a full-text assessment for eligibility.

\begin{table}[H]
\centering
\caption{Research Query Table}
\label{tab:keywords}
\renewcommand{\arraystretch}{1.15}
\setlength{\tabcolsep}{8pt}

\begin{tabular}{p{0.22\textwidth} p{0.73\textwidth}}
\hline
\textbf{Concept} & \textbf{Searched Keywords} \\
\hline

\raisebox{0pt}[0pt][0pt]{\parbox[t]{0.22\textwidth}{\textbf{Uncertainty in CPS}}} &
\parbox[t]{0.73\textwidth}{
"Cyber-physical system*", CPS, "Cyber physical system*", "uncertainty*","Uncertainties*", "Unknowns*","layer*", "perception layer*", "control layer*", "communication layer*", "actuation layer*"
}
\\
\hline
\end{tabular}
\end{table}

During pre-screening, we applied script-based metadata filtering, using a predefined venue list and a year range, after retrieving records from multiple databases. We discussed all related details below in the study selection process for more clarity. The keywords used to maximize the precision in the searched results are mentioned in Table~\ref{tab:keywords}.

\FloatBarrier

\subsection{Selected Venues}
A venue-based filtering step was implemented following the initial search and screening stages to enhance the quality and topical relevance of the final study set. Instead of including papers from all publication outlets, the study intentionally focused on established peer-reviewed venues within software engineering and CPS-related fields. This approach aligns with the paper's disciplinary scope, which investigates uncertainty in CPS from the perspectives of software engineering, testing and verification, model-driven engineering, reliability, and AI/ML-enabled software systems.

The venues we selected for this research were deemed suitable for three primary reasons. First, they are established peer-reviewed journals/conferences, which strengthen the reliability and scholarly quality of the reviewed evidence. Second, their scopes correspond to key engineering activities where uncertainty is explicitly investigated, such as requirements specification, architecture and design, testing, verification, runtime assurance, software reliability, and adaptive behavior. Third, several of these venues connect software engineering with CPS, dependable systems, industrial informatics, automation, and data-driven methods, thereby increasing their relevance for a survey of uncertainty in software-intensive CPSs.

This filtering strategy also reduced noise from venues that were less relevant to the study's goals. Given the greater focus on uncertainty from disciplinary perspectives, particularly those associated with software-intensive CPSs, it was necessary to prioritize venues that address uncertainty across requirements, modeling, design trade-offs, testing, verification, runtime monitoring, reliability, and data-driven system behavior.

\begingroup
\small
\setlength{\LTleft}{0pt}
\setlength{\LTright}{0pt}

\begin{longtable}{>{\raggedright\arraybackslash}p{0.24\textwidth} >{\raggedright\arraybackslash}p{0.72\textwidth}}
\caption{Publication venues included in the venue-based filtering step}
\label{tab:selected_venues}\\
\toprule
\textbf{Category} & \textbf{Included venues} \\
\midrule
\endfirsthead

\toprule
\textbf{Category} & \textbf{Included venues} \\
\midrule
\endhead

\bottomrule
\endfoot

Core software engineering journals &
IEEE Transactions on Software Engineering (TSE); ACM Transactions on Software Engineering and Methodology (TOSEM); Journal of Systems and Software (JSS); Information and Software Technology (IST); Software Quality Journal (SQJ); Software Testing, Verification and Reliability (STVR); Empirical Software Engineering (EMSE); Software and Systems Modeling (SoSyM); Science of Computer Programming (SCP). \\

Core software engineering conferences and symposia &
International Conference on Software Engineering (ICSE); ACM SIGSOFT International Symposium on the Foundations of Software Engineering (FSE); International Conference on Automated Software Engineering (ASE); International Symposium on Software Testing and Analysis (ISSTA); IEEE International Conference on Software Testing, Verification and Validation (ICST); International Symposium on Software Engineering for Adaptive and Self-Managing Systems (SEAMS); International Symposium on Software Reliability Engineering (ISSRE); International Conference on Model-Driven Engineering Languages and Systems (MODELS). \\

Requirements, testing, and automated engineering venues &
IEEE International Requirements Engineering Conference (RE); Requirements Engineering; Journal of Automated Software Engineering (JASE). \\

Related software and systems venues &
International Systems and Software Product Line Conference (SPLC); International Conference on Performance Engineering (ICPE); International Conference on Distributed Event-Based Systems (DEBS); Engineering Digital Twins Conference (EDTConf); System Analysis and Modeling Conference (SAM); International Conference on Software Reuse (ICSR); International Conference on Software and Systems Reuse (ISEC). \\

AI/software and emerging adjacent venues &
AiWare; Internetware; International Conference on AI Foundation Models and Software Engineering (FORGE); ACM Proceedings on Software Engineering (PACMSE). \\

Programming languages and formal/software foundations venues &
Object-Oriented Programming, Systems, Languages, and Applications (OOPSLA); Programming Language Design and Implementation (PLDI); Symposium on Principles of Programming Languages (POPL); Proceedings of the ACM on Programming Languages (PACMPL); ACM Transactions on Programming Languages and Systems (TOPLAS). \\

CPS, dependable systems, and adjacent engineering venues &
ACM Transactions on Cyber-Physical Systems (TCPS); IEEE Transactions on Dependable and Secure Computing (TDSC); IEEE Transactions on Industrial Informatics (TII); IEEE Transactions on Automation Science and Engineering (T-ASE). \\

\end{longtable}
\endgroup

Table~\ref{tab:selected_venues} provides a summary of the publication venues included in this filtering step. Descriptive statistics on publication counts across venues are presented in the Results section.

\subsection{Article Selection Process}

\begin{figure}[ht]
\centering
\includegraphics[width=\textwidth]{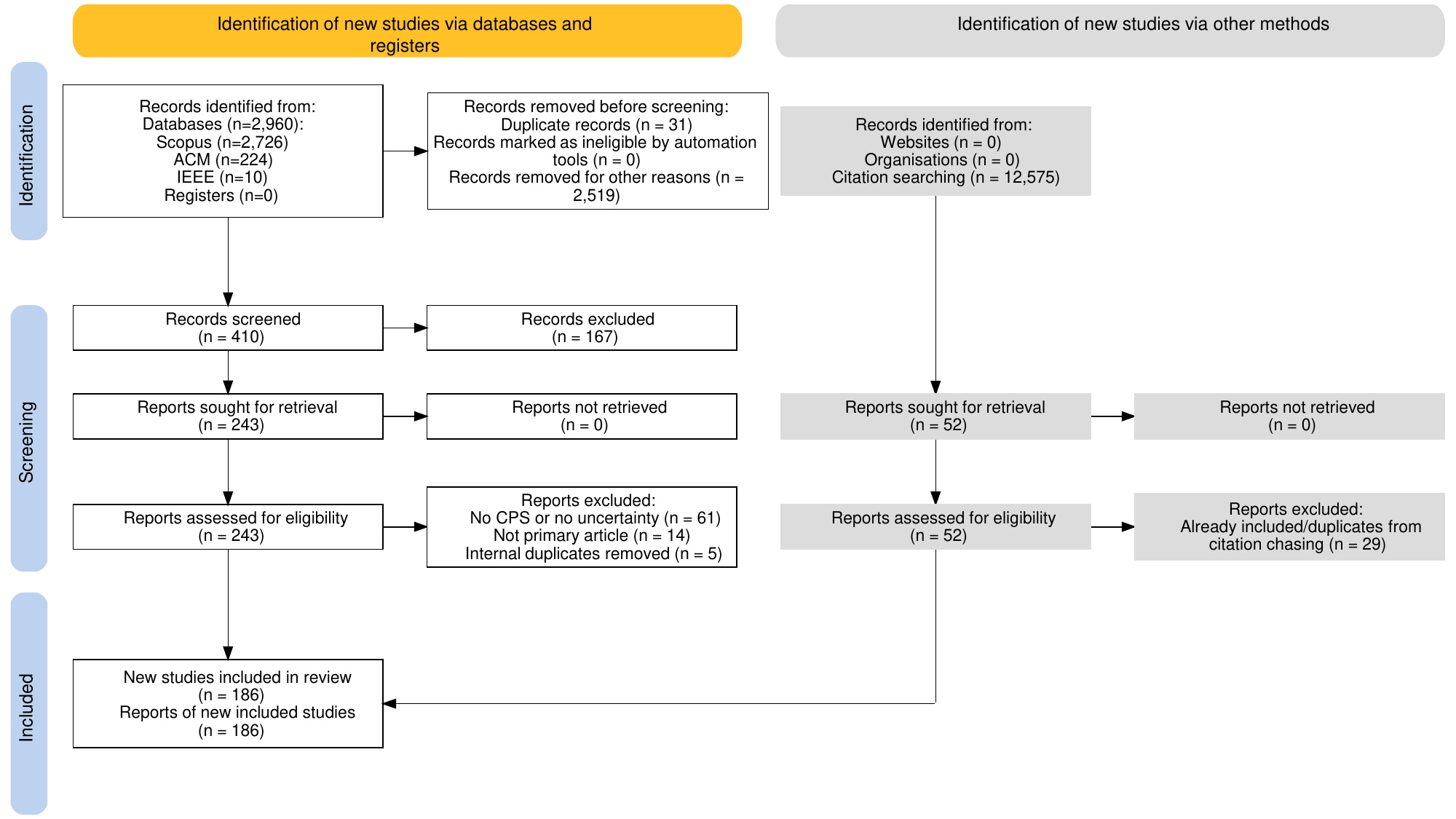}
\caption{Prisma Flow Diagram of the papers' selection process}
\label{fig:prisma}
\end{figure}

The Prisma flow diagram depicted in Fig. \ref{fig:prisma} documents the article selection process. Three bibliographic databases (Scopus, ACM Digital Library, and IEEE Xplore) were used, and 2,960 research papers in total were obtained with the following breakdown: 

\begin{itemize}
    \item Scopus: 2,726 papers
    \item ACM: 224 papers
    \item IEEE: 10 papers
\end{itemize}

Before screening, 31 duplicate papers were removed: 2 from Scopus and 29 from ACM. 

Additionally, 2,519 records were excluded for other reasons (e.g., automated filtering based on metadata, such as a selected list of venues and a year range). Scopus filtering removed 2,517 records, and 2 IEEE records were out of the target year range. As a result, 410 records were left for the screening stage. 

From that, we conducted a manual screening of titles and abstracts to remove irrelevant papers. To this end, two of the authors analyzed the articles, dividing them into two pools. For each pool, one author determined the relevance (using a Boolean tag), and another one performed a double-check.
During the title and abstract (quick relevance check) screening stage, 167 records were excluded, leaving 243 articles for retrieval. Full texts were successfully obtained for the detailed relevance check. For all targeted reports, 0 remain unretrieved, whereas all 243 were further assessed for eligibility. At the eligibility stage, 78 articles were excluded since 61 did not address CPS and/or did not discuss uncertainty. After full-text assessment, 14 articles were not primary (e.g., books, surveys, posters, or systematic literature reviews), and 5 were removed as duplicates across bibliographic files. 

After the exclusions described above, 163 articles were included from database searching. In parallel, citation chasing was carried out as an additional identification method. Citation searching yielded 12,575 records, of which 52 reports were selected and assessed for eligibility. Among these, 29 articles were excluded due to duplication of articles that were already included, and 23 additional papers were added. Overall, the selection process resulted in 186 included articles. 

\subsection{Data Extraction and Synthesis}
\label{sec:synthesis}
After selecting the primary papers, we extracted and standardized the information needed to address each research question. The following subsections describe how the data were coded, standardized, and synthesized for each RQ. In this case, unlike the selection phase, the article coding was performed by a single author (with different authors working on different research questions). However, because RQ2-RQ4 depend on the output of RQ1 (which characterizes uncertainty types), the analysis for such RQs was conducted after completing RQ1. Last but not least, the set of categories elicited for each RQ was refined and finally double-checked with contributions from all paper authors, both offline and in online meetings.

\textbf{RQ1: Uncertainty Categories and Domain.}
To answer RQ1, we combined a paper-level frequency analysis with a structured normalization and coding workflow. The main aim was to make results comparable across primary papers that use different uncertainty terminology and often discuss multiple uncertainty sources simultaneously.

We considered each primary study as a single observation. At first, we extracted and gathered uncertainty-related statements from the abstract. We examined the full text only if the abstract was not clear. These statements were then mapped to a fixed set of 16 normalized uncertainty categories using a functional-role interpretation, which focuses on what the uncertainty affects in the CPS life-cycle, rather than keyword matching. Articles were coded using a multi-label rule, allowing assignment of multiple uncertainty categories to a single study when it explicitly addressed several uncertainty sources, such as environmental uncertainty and runtime context uncertainty. Minor label variations, including differences in capitalization or near-synonym phrasing, were combined into canonical category names to make sure that the counting remains consistent.
In parallel, each study was assigned one or more normalized CPS domains (domain coding was also multi-label, where applicable). This enables reporting RQ1 both as (i) overall category prevalence and (ii) category–domain intersections.

For instance, suppose $N$ represents the total number of primary articles. In this case, $N = 186$. For each study $s$, uncertainty category $c$, and domain $d$, we define binary indicators (\ref{eq:indicators}), $I_{i,c}$ and $D_{i,d}$ as mentioned in Eq.~(\ref{eq:indicators})

\begin{equation}
\left\{
\begin{aligned}
I_{i,c} &=
\begin{cases}
1, & \text{if study } i \text{ addresses uncertainty category } c \\
0, & \text{otherwise}
\end{cases} \\
D_{i,d} &=
\begin{cases}
1, & \text{if study } i \text{ is mapped to domain } d \\
0, & \text{otherwise}
\end{cases}
\end{aligned}
\right.
\label{eq:indicators}
\end{equation}

Eq.(\ref{eq:count_category}) calculates the category frequency $n_{c}$
by adding up $I_{i,c}$ for all N primary articles. This gives the total number of papers in category c.

\begin{equation}
n_c=\sum_{i=1}^{N} I_{i,c}
\label{eq:count_category}
\end{equation}

Similarly, Eq. (\ref{eq:percent_category}) computes category percentage $p_c$ as the count of the papers that discuss uncertainty category $c$ out of the total $N$ papers, which are 186 in this case. 

\begin{equation}
p_c=\frac{n_c}{N}\times 100
\label{eq:percent_category}
\end{equation}

Finally, Eq.(\ref{eq:count_category_domain}) calculates the category $x$ domain intersection count $n_c,_d$. Simply speaking, the number of articles that are mapped to the uncertainty category $c$ and the domain $d$ at the same time. We used these values for the bubble chart of the domain $x$ category.

\begin{equation}
n_{c,d}=\sum_{i=1}^{N} I_{i,c}\,D_{i,d}
\label{eq:count_category_domain}
\end{equation}

\begin{figure}[H]
\centering
\includegraphics[width=0.90\textwidth]{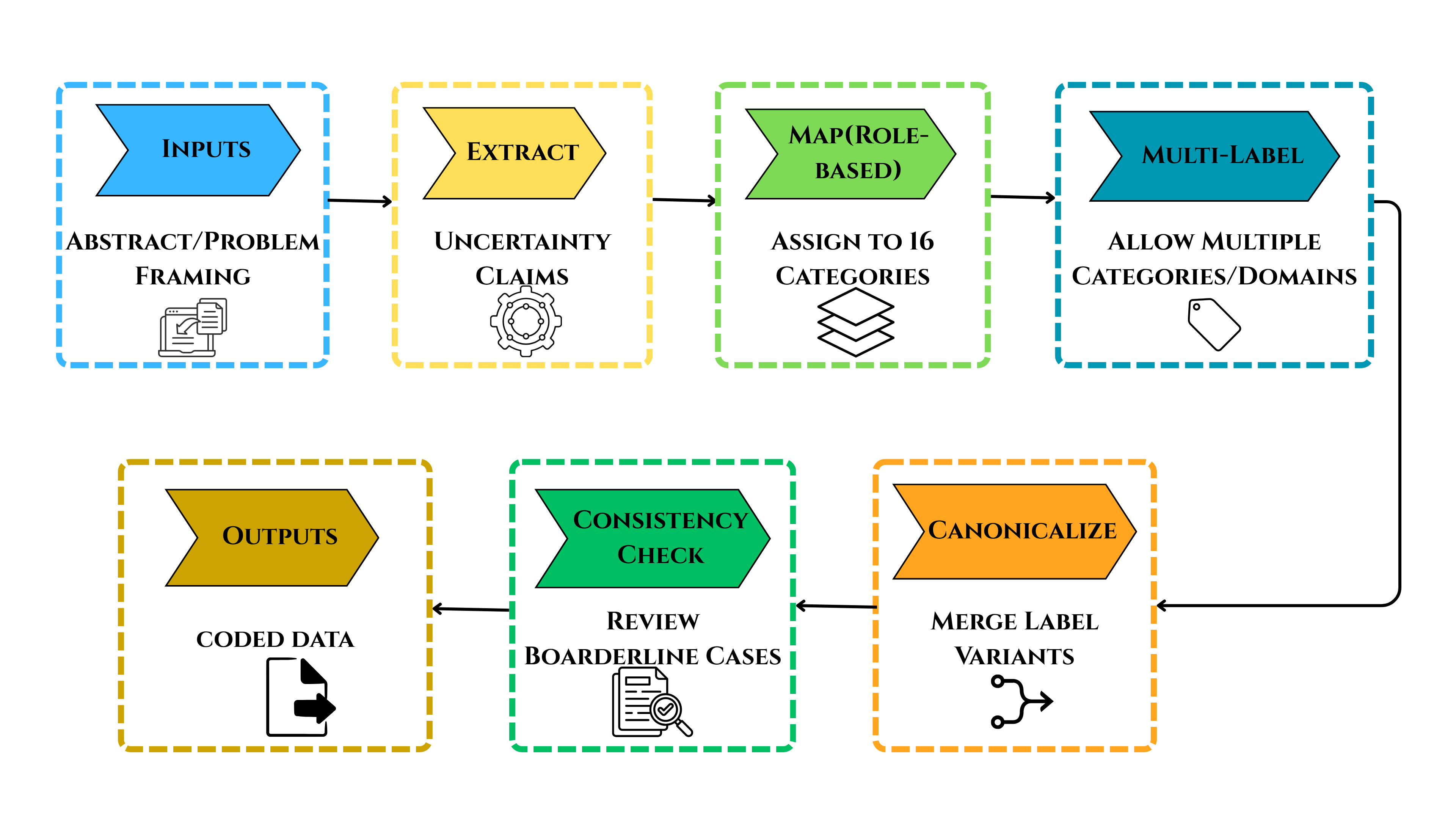}
\caption{RQ1 Mapping}
\label{fig:Mapping}
\end{figure}

We used multiple labels for category and domain coding. Therefore, a single paper may be classified into multiple categories and domains, leading to multiple counts for the same paper.
For affected-domain percentages, the denominator is the category-specific total \(n_c\), not the full corpus size \(N=186\). Thus, \(n\) varies across category subsections because each uncertainty category appears in a different number of papers. Since domain coding is multi-label, within-category domain percentages may not sum to 100\%.The steps are shown in Fig.~\ref{fig:Mapping}

\textbf{RQ2: Sources that Cause Uncertainty in CPSs.}
To address RQ2, we applied the normalized uncertainty categories established in RQ1 and conducted a further analysis of the information extracted from each primary study to identify the underlying sources of these uncertainties. Coding was conducted by examining each article for evidence of what caused or introduced the uncertainty. These extracted sources were then mapped to the 16 normalized uncertainty categories.

First, we examined the uncertainty descriptions reported in each paper and grouped similar causes using open coding. For example, weather, traffic variability, road conditions, environmental disturbances, and unknown operating contexts were categorized as \textit{dynamic physical environment}. Similarly, noisy sensor readings, incomplete observations, perception errors, and estimation errors were classified as \textit{sensing and perception limitations}. In this step, we prioritized preserving the meaning of the original study descriptions rather than relying solely on exact keywords.

Second, we consolidated overlapping codes into broader source categories. For example, model drift, unknown parameters, simplified assumptions, and model obsolescence were grouped as \textit{incomplete knowledge and model limitations}. Similarly, DoS attacks, false data injection, adversarial machine learning inputs, and unpredictable attacker behavior were categorized as \textit{cyber threats and adversarial actions}. This normalization enabled reducing paper-specific descriptions into a consistent set of recurring uncertainty sources.

Third, each source category was mapped to the uncertainty categories identified in RQ1. This step was necessary because a single source may span multiple uncertainty categories. For example, environmental variation can result in environmental, process, and dynamic uncertainty, as well as scenario uncertainty. Similarly, sensing limitations can cause sensor or measurement uncertainty, data uncertainty, and downstream decision uncertainty.

Finally, we refined the coding iteratively by comparing the source categories with examples from the primary articles. For example, UAV safety papers emphasized uncertainty arising from runtime environmental interactions and control behavior. Autonomous driving papers highlighted incompleteness in the scenario space and perception uncertainty. Smart-grid articles focused on communication, cyber-attack, and state-estimation uncertainty. Human-CPS articles identified uncertainty caused by human attention, preferences, and decision-making. This iterative process ensured that the final source categories were grounded in the extracted data and remained consistent with the normalized categories reported in RQ1.

\textbf{RQ3: Mapping of uncertainties onto CPS Lifecycle.}
To answer RQ3, we reviewed each study to indicate whether uncertainty was discussed as arising during runtime/operation, during development/testing. Life-cycle placement was categorized into two broad phases: development/testing and runtime/operation. This approach was adopted because the reviewed articles did not consistently distinguish clearly about the life-cycle stages such as design, implementation, verification, deployment, and maintenance. Employing these broader phases avoided artificial precision and enabled consistent comparison across studies. Articles addressing uncertainty in both phases were coded as covering both development/testing and runtime/operation.

The important information was then combined into a consistent and non-redundant set of lifecycle-stage categories.

For RQ3, the unit of analysis was determined at the paper level. Each primary study was then treated as a single unit and assigned a final lifecycle-stage category based on how the paper characterized the presence of uncertainty. Multiple mentions of uncertainty within the same paper were not counted separately. Rather, the complete details in each study were interpreted altogether and mapped to a single normalized category.
The free-text entries were standardized into three labels. 
\begin{itemize}
    \item The runtime/operation label was applied when a paper described uncertainty arising during live execution, deployment, real-time sensing, online decision-making, environmental interaction, cyber-attacks, disturbances, failures, or operational adaptation.
    
    \item The development/testing label was used when uncertainty was identified primarily in design, modeling, simulation, verification and validation, controller synthesis, scenario construction, testing, or other pre-deployment activities.
    
    \item Both development and runtime labels were assigned when the description explicitly indicated uncertainty in both phases or clearly linked development-time assumptions, abstractions, or models to their runtime manifestations.
\end{itemize}

Apart from this, one paper was excluded from the list as it wasn’t discussing uncertainty, like whether it occurred at runtime, development time, or both. Unlike the other RQs, RQ3 was not treated as a multi-label coding task after normalization. The lifecycle-stage distribution was intentionally set to be mutually exclusive, ensuring that each paper contributed to only one category. If a study addressed both phases, it was coded under the combined category, both development and runtime, rather than being counted separately under development and runtime. This approach was necessary to produce a clear lifecycle-stage distribution and to prevent double-counting or repetition. The percentage for each lifecycle category was calculated using \ref{eq:rq3_percentage}:

\begin{equation}
\text{Percentage of category} =
\frac{\text{Number of papers in the category}}
{\text{Total number of papers included for RQ3}} \times 100
\label{eq:rq3_percentage}
\end{equation}

\textbf{RQ4: Mitigation Strategies.}
To address RQ4, every included study was systematically examined to determine whether it proposed an explicit strategy, mechanism, framework, or approach for mitigating uncertainty in CPSs. 
The coding process focused on claims related to mitigation strategies. We assessed initially in the abstract and, when necessary, in the problem framing, method description, and reported solution. Instead of treating the terminology of each paper as a distinct category, semantically relevant mitigation approaches were gathered into broader categories to facilitate cross-study synthesis.

Conceptually related mitigation strategies were grouped under categories such as optimization and resource management, simulation and digital-twin-based mitigation, details are mentioned in subsection \ref{subsec:RQ4}.
Articles that define the uncertainty but do not propose a concrete mitigation approach were coded as ``reporting no explicit mitigation strategy''.

\begin{equation}
M_{i,m} =
\begin{cases}
1, & \text{if study } i \text{ proposes mitigation category } m \\
0, & \text{otherwise}
\end{cases}
\label{eq:rq4_binary}
\end{equation}

In accordance with the coding scheme that is binary and is presented in Eq. \ref{eq:indicators} and formalized for mitigation strategies as $M_i,_m$ in Eq. \ref{eq:rq4_binary}. RQ4 was coded at the paper level using a presence or absence indicator for each normalized mitigation category. Each study receives a value of 1 if it explicitly proposes mitigation category $m$ and 0 otherwise. Each paper was counted only once within a given mitigation category, regardless of how many times the same strategy was mentioned. This approach makes sure that the coding reflects the presence of a mitigation strategy in a study rather than its frequency of discussion.

Multi-label coding was applied to RQ4 because several articles proposed multiple mitigation strategies. As a result, a single paper could be assigned to several categories if it explicitly combined different mitigation mechanisms. For example, a study might apply probabilistic reasoning, runtime monitoring, and formal verification, and would therefore be counted once in each relevant category. This approach was necessary because mitigation in CPSs is frequently layered rather than limited to a single method. Restricting each paper to a single category would have oversimplified the literature and obscured the integrated nature of many proposed solutions.The percentage was calculated as for RQ1. 
In other words, the percentage indicates the share of reviewed articles that mentioned a particular mitigation strategy.

%% file: sections/results.tex
\section{Study Results}
\label{sec:results}
This section presents findings from the selected primary papers. We begin with a descriptive overview of the corpus, followed by results for each research question, using the normalized coding and synthesis procedure outlined in the study design.

\subsection{Descriptive overview of the selected articles}

\begin{figure}[ht]
    \centering
    
    \begin{subfigure}[b]{0.48\textwidth}
        \centering
        \includegraphics[width=\textwidth]{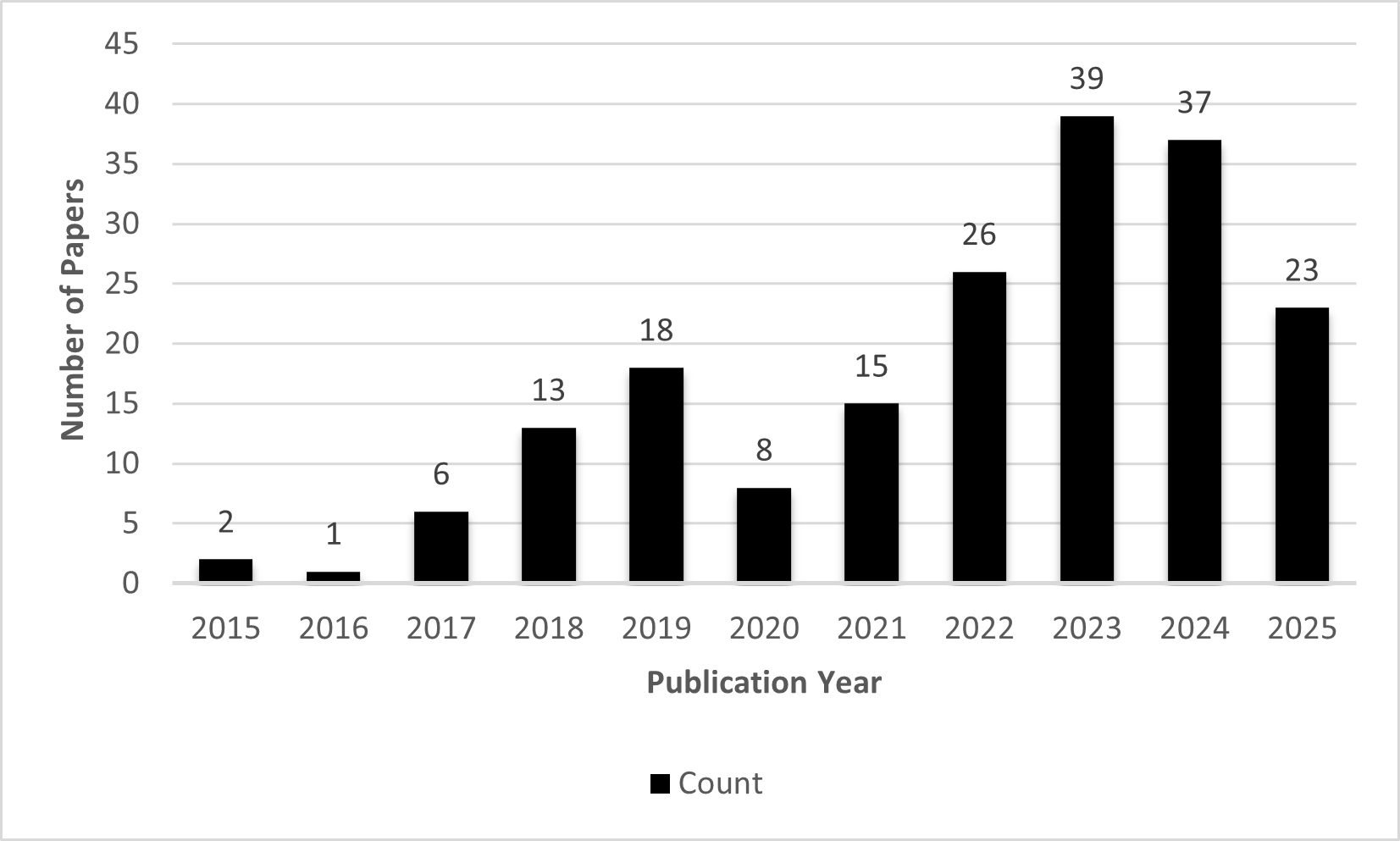}
        \caption{Years}
        \label{fig:years}
    \end{subfigure}
    \hfill
    \begin{subfigure}[b]{0.48\textwidth}
        \centering
        \includegraphics[width=\textwidth]{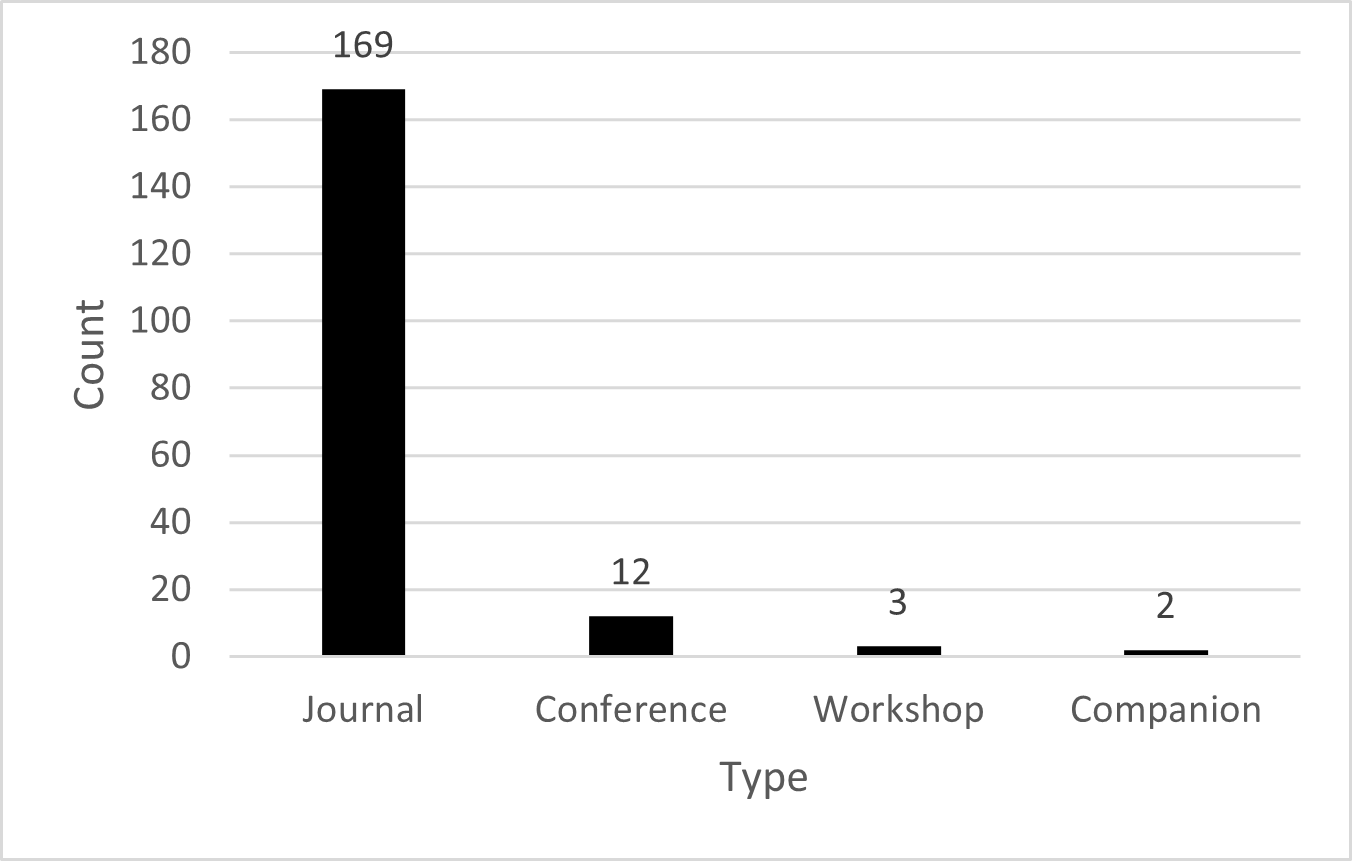}
        \caption{Type}
        \label{fig:type}
    \end{subfigure}
    
    \caption{Distribution of selected articles by publication year and type.}
    \label{fig:year_type_distribution}
\end{figure}

Fig.~\ref{fig:year_type_distribution} summarizes the number of papers published each year with the type e.g., conference papers, journals, workshops, and companions. These graphs in the figure clearly show that the volume of reviewed literature has increased over time and is mainly concentrated in journal publications.

\begin{figure}[ht]
\centering
\includegraphics[width=0.60\textwidth]{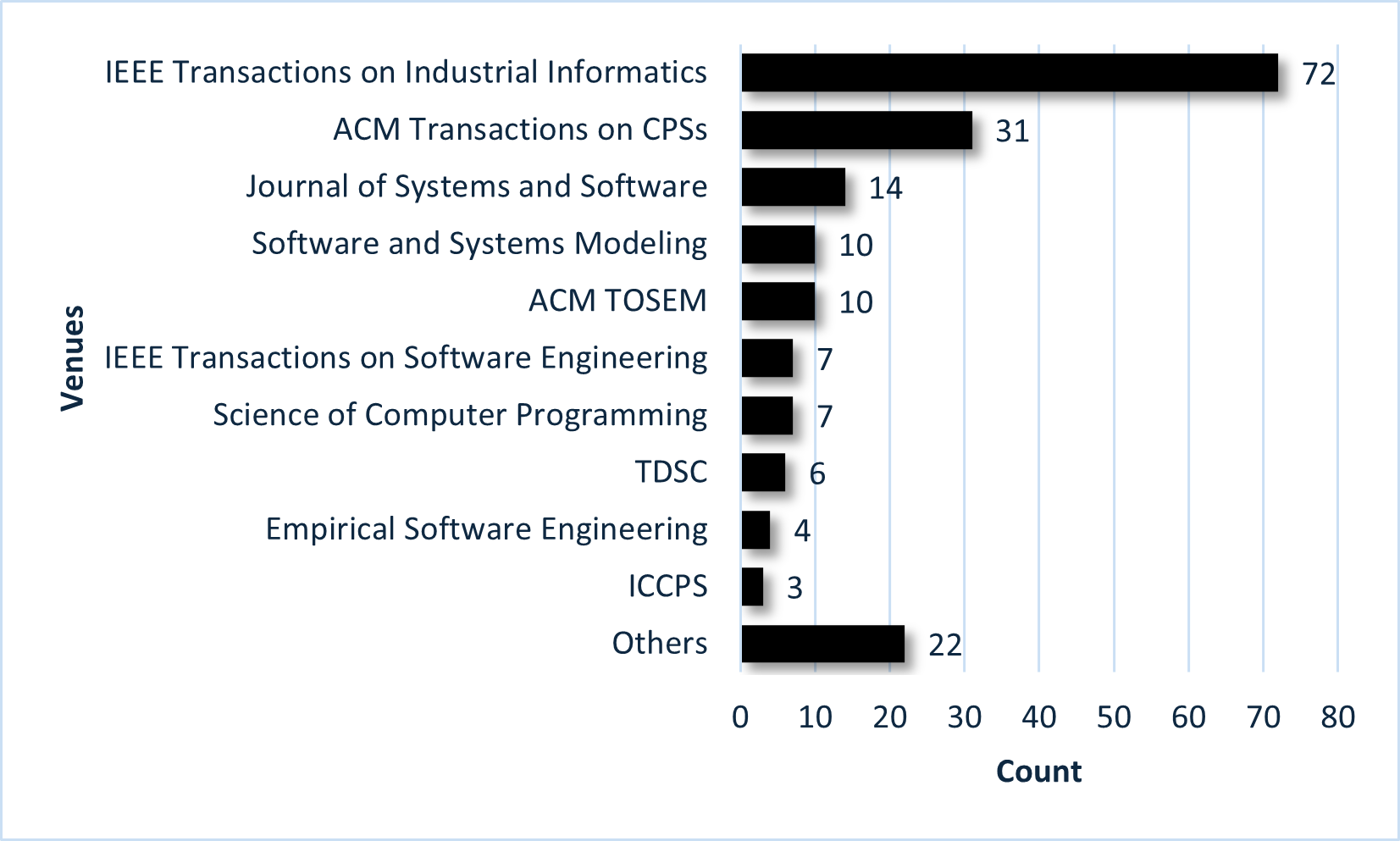}
\caption{Distribution of included articles across publication venues}
\label{fig:venue}
\end{figure}

Looking at Fig.~\ref{fig:venue}, we can notice that most papers were published in
IEEE Transactions on Industrial Informatics, with 72 articles (38.7\%), followed by ACM Transactions on CPSs  with 31 articles (16.7\%). Collectively, these two journals accounted for 103 papers, representing 55.4\% of the total. Other prominent journals included Journal of Systems and Software (14 articles, 7.5\%), Software and Systems Modeling (10, 5.4\%), and ACM Transactions on Software Engineering and Methodology ``ACM-TOSEM" (10, 5.4\%). Additional contributions appeared in IEEE Transactions on Software Engineering (7, 3.8\%), Science of Computer Programming (7, 3.8\%), IEEE Transactions on Dependable and Secure Computing ``TDSC" (6, 3.2\%), Empirical Software Engineering (4, 2.2\%), and ACM/IEEE International Conference on CPSs ``ICCPS" 3, 1.6\%). The remaining publications were categorized as 'Others,' comprising 22 articles (11.8\%). These findings specify that most of the research is denser in a small number of leading journals, with the remainder distributed across various other conferences and workshops. 

This section discusses how the findings from our SLR address the RQs formulated in Section \ref{sec:design}.

\subsection{RQ1: Uncertainty Types across Multiple CPS Domains}

Our analysis identified different types of uncertainties in CPSs, which vary across the domains listed in Table \ref{tab:rq1_mapping_clickable}. For example, model or parametric-based uncertainties arise in the industrial or manufacturing sector, in autonomous cars, in power grids, and in robotics.

\begin{table}[H]
\centering
\scriptsize
\setlength{\tabcolsep}{3pt}
\renewcommand{\arraystretch}{1.05}
\caption{Mapping of normalized uncertainty categories to primary articles.}
\label{tab:rq1_mapping_clickable}
\begin{tabularx}{\textwidth}{@{}
>{\vspace{0pt}\raggedright\arraybackslash}p{0.28\textwidth}
>{\vspace{0pt}\raggedright\arraybackslash}X@{}}
\toprule
\textbf{Uncertainty category} & \textbf{Used by} \\
\bottomrule
\end{tabularx}
\end{table}

 In the following, we discuss each uncertainty type and its occurrence across different CPS domains, as shown in detail in Fig. \ref{fig:rq1_bubble} below. For each uncertainty type, we provide a description, discuss its prevalence in the analyzed papers, the affected domains, and examples from those papers.

\begin{figure}[H]
  \centering
  \includegraphics[width=\textwidth]{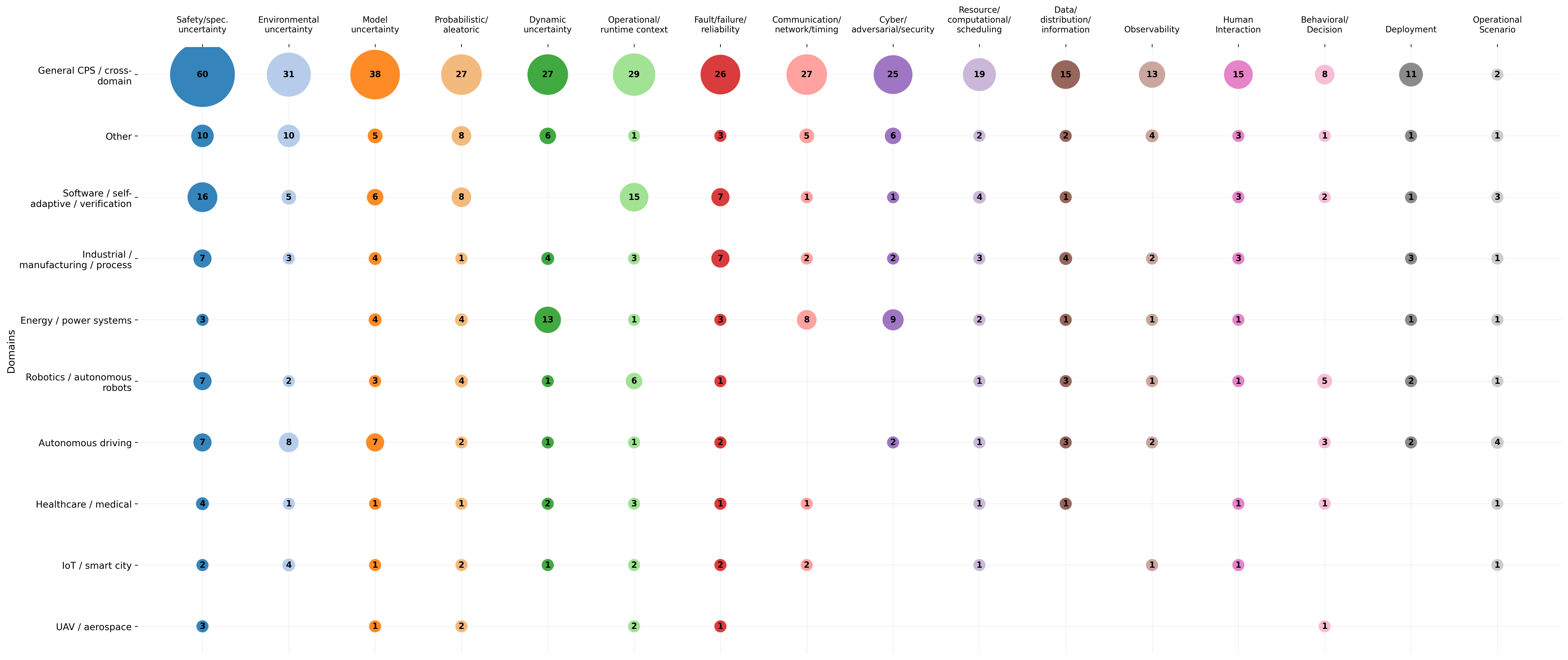}
  \caption{Domain $\times$ uncertainty-category bubble chart for RQ1. Bubble area is proportional to the number of primary articles mapped to each domain–category pair; numeric labels indicate the exact counts.
  }
  \label{fig:rq1_bubble}
\end{figure}
\textit{Note:} The \emph{Other} domains include mixed, or infrequent CPS application contexts that do not fit the main normalized domain categories. The replication package includes all domain-wise bubble-chart data for the distribution analysis.

\subsubsection{Category 1 -- Safety/Specification Uncertainty:}\hfill\break
\noindent
\textbf{Description and Occurrences:} 
Uncertainty in safety or specification reduces confidence in determining whether a CPS meets its safety goals and requirements, particularly when evidence is incomplete, ambiguous, or influenced by uncertain operating conditions. Most articles address this by using verification, validation, and testing. This category is observed in both traditional CPS and Learning-Enabled CPS (LE-CPS) and spans the design-time, test-time, and runtime phases.

The category includes behavioral uncertainty during runtime operations, environmental factors, assumptions, and the methods used for testing scenarios or monitoring schemes, ensuring that the uncertainty is relevant to safety, specification compliance, or confidence in testing and verification. More specifically this category is identified in nearly half of the articles we reviewed (89 out of 186,  47.8\%). 



Because domain normalization is multi-label, these percentages are based on the total (n = 89) and do not sum to 100\%. This distribution shows that safety, specification, and testing uncertainties exist across multiple domains, including safety-critical applications.

\textbf{Examples of work discussing this uncertainty:}
The primary concern here is implementing runtime safety assurance under uncertainty, specifically in domains where unsafe behavior is associated with randomness. As an example, in the aerospace applications, Khatiri et al.~\cite{khatiri2025uav_uncertainty} established a direct link between behavioral uncertainty in the control action of UAVs and unsafe operation during flights, making use of considerable PX4-Autopilot simulations (over 5,000 flights) and a detector for runtime uncertainties to oversee the safety-critical behavior. From a wide-ranging, cross-domain, and runtime assurance perspective, Camilli et al.~\cite{camilli2023_enforcing_resilience_equilibrium} propose a framework that checks whether the CPS remains within a feasibility zone despite incomplete knowledge, and illustrate it using robotics and UAV case studies. These are included in this category because operational uncertainty is crucial, especially when it includes safety guarantees and demands assurance procedures.

Some works in the literature, focus on testing safety/specification uncertainty.
Crespo-Rodriguez et al.~\cite{cresporodriguez2025_instance_space_vehicles} handle autonomous vehicle testing by making use of simulated scenarios, which deal with the challenge of testing autonomous vehicles where the test environment and scenario feature space both have uncertainties associated with them. Identifying the key scenario features associated with the fault-revealing tests is the primary contribution of this study. In parallel, Shin et al.~\cite{shin2018_hitecs_uml_profile} discuss uncertainties addressed using a hardware-in-the-loop (HIL) setup, where test behavior is influenced by environmental uncertainty, and HIL runs ensure the hardware’s real-time performance is not degraded beyond an acceptable criterion. Ali et al.~\cite{ali2017_evaluation_mutation_crossover} additionally supports the perspective of testing the system by investigating multi-objective test-case rationalization under the effects of uncertainty. Uncertainties are incorporated explicitly into the framework. Overall, these works indicate that, within this category, testing under uncertainty basically ensures test adequacy, efficiency, and confidence in test evidence in unexpected scenarios.

This is more prevalent in work about autonomous driving, where uncertainty is frequently re-evaluated as a decision-making and monitoring challenge. Zhao et al.~\cite{zhao2025_behavioral_decision_making} introduce Bayesian-network-based hazard probability inference and a decision-making model for autonomous driving under unsafe or extreme scenarios, including harsh or unpredictable weather, with decision safety verification. Ayerdi et al.~\cite{ayerdi2025_marmot_metamorphic_monitoring} focus specifically on estimating runtime uncertainty for autonomous driving systems (ADSs) by using metamorphic relations (MarMot), with the aim of detecting unforeseen situations that can lead to system misbehavior, such as driving off the road. In these articles, uncertainty is explicitly recognized as a crucial factor that must be examined, understood, and validated to ensure the system's behavior is close to what we expect. It does not simply assume uncertainty as variance in prediction.

In autonomous robotic systems, this category is evident in both assurance- and learning-based CPS safety-monitoring approaches. Zhou et al.~\cite{zhou2024_building_nvidia_isaac} investigate uncertainty in the exploitation of AI-enabled robotics and propose a standard framework within NVIDIA Omniverse Isaac Sim to analyze safety-related failures in simulation. Lin et al.~\cite{lin2025_monitoring_learning_enabled} examine LE-CPS out-of-distribution (OOD) inputs regarding safety, shifting the focus from just detecting OOD data to direct observation of the violations of safety specifications using predicted trajectories, conformal prediction, and gradual learning. This approach shows the logic or rationale for combining safety, specification, and testing within the category label: uncertainty is addressed rather than relying solely on generic uncertainty estimation, through specification-oriented empirical evaluation.

In industrial settings, Kaur and Chanak~\cite{kaur2023_intelligent_fault_tolerant} frame Safety in Industry 4.0 environments as encompassing unpredictable/unforeseen hazardous events and sensor/network faults, proposing an intelligent scheme that is fault-tolerant, more reliable, and provides timely feedback. 

At the specification level, one example is the work presented by Li et al.~\cite{li2017_stochastic_contracts_probabilistic}, which proposes a contract framework for CPS design that uses Stochastic Signal Temporal Logic (StSTL). This CPS is designed to meet probabilistic requirements within domains such as aerospace and power networks. This is the best example of specification uncertainty arising from probabilistic requirements, handled through formal reasoning and controller design. Similarly, Zhang et al.~\cite{zhang2019_wise_test_modeling} propose a framework for modeling related uncertainties being handled during testing in CPS. It addresses the restrictions that most ready-to-test models fail to express.

\keyfinding{Safety/specification uncertainty is mainly viewed as assurance uncertainty: Where the quality of the requirements and supporting evidence is the main aspect on which the safety and specification compliance relies.}{89/186 (47.8\%)}{General CPS/cross-domain (60/89, 67.4\%); Software/self-adaptive/verification (16/89, 18.0\%); Other (10/89, 11.2\%)}

\subsubsection{Category 2 -- Environmental Uncertainty:}\hfill\break
\noindent
\textbf{Description:}
This category refers to the uncertainty that arises from the environment where a CPS operates. This highlights any condition that is not under the system's control and that impacts observation, decision-making, testing, and the control of the system's performance. There are multiple sources of uncertainty in the environment, for example, harsh or unpredictable weather, road obstacles, traffic, and other external factors that disturb the environment, as well as the use of simulations or digital twins of systems that imperfectly represent environmental parameters.
This category is best described as extrinsic-condition uncertainty, rather than as general runtime uncertainty. The primary difference between other uncertainties and this type is that this uncertainty, surrounding the CPS, is assigned to the physical or operational environment rather than to internal software state transitions, scheduler behavior, or system mode changes. In the upcoming categories, these are being classified more appropriately as operational or runtime uncertainty. Maintaining this distinction ensures that category definitions remain analytically distinct throughout the analysis of RQ1.

\textbf{Occurrences:} Environmental uncertainty 
appeared in 53 out of 186 papers we studied (28.5\%). 
In the reviewed articles, environmental uncertainty is frequently examined alongside other types of uncertainty, particularly when environmental factors directly affect assurance and validation outcomes. Among the 53 papers classified in this category, environmental uncertainty most often co-occurs with safety, specification, or testing uncertainty (28/53, 52.8\%), followed by probabilistic or aleatoric uncertainty ((18/53, 34.0\%), fault, failure, or reliability uncertainty (13/53, 24.5\%), model, parametric, or epistemic uncertainty (12/53, 22.6\%), and communication, network, or timing uncertainty (12/53, 22.6\%). This pattern suggests that environmental uncertainty is rarely addressed as an isolated modeling concern; rather, it is typically investigated because uncertain environments can compromise system safety, reduce test confidence, or interact with stochastic behavior and communication reliability.

A topic commonly discussed in this category is treating environmental uncertainty as a challenge for testing and validation. Most articles in the literature identified not only uncertainties arising from the real environment but also inadequate testing procedures that fail to account for these uncertainties.

\textbf{Examples of work discussing this uncertainty:}
The concept of environmental uncertainty is a challenge in ensuring that test cases accurately replicate external environmental parameters. In this regard, Sartaj et al.~\cite{sartaj2025envdt} note the basic uncertainty related to the environment, not just in digital twin applications, and state that, if we remove environmental conditions in simulations and testing, the system's functionalities may remain untested. Hence, uncertainty arises from differences between the real world and the simulated environmental parameters, leading to a loss of trust and confidence in the system.

Runtime monitoring and safety-awareness challenge in autonomous driving is another aspect of this uncertainty. Most uncertainties coexist, as one may cause others. Ayerdi et al.~\cite{ayerdi2025_marmot_metamorphic_monitoring} 
observe estimation of runtime uncertainty for autonomous driving systems (ADSs) that operate under unpredictable conditions. The existence of uncertainty during operation introduces variations in the external driving scenarios. This requires continuous online monitoring to identify anomalous situations, rather than ignoring them, which can lead to other unsafe situations. Therefore, this type of environmental uncertainty plays a major role in affecting the system's behavior and its actions.
In robotics, environmental uncertainty affects the effectiveness of planning and the success of the task. For example, Chen et al.~\cite{chen2023_fast_online_planning} address motion planning in unknown environments through the use of uncertainty-aware planning logic, such as belief-based or Partially Observable Markov Decision Process (POMDP) reasoning.

\keyfinding{Uncertainty that manifests from external operating conditions (e.g., environment/scenarios) that passes into sensing, planning, and assurance confidence.}{53/186 (29.3\%)}{General CPS/cross-domain (31/53, 58.5\%); Other (10/53, 18.9\%); Autonomous driving (8/53, 15.1\%)}

\subsubsection{Category 3 -- Model Uncertainty:}\hfill\break
\noindent
\textbf{Description and Occurrences:}
Incomplete or faulty information about the system is the main reason behind this type of uncertainty. It comprises uncertainty related to different model structures, parameter values, and other assumptions used to test and control the system. The core concept is that whenever a system is not fully understood, it also leads to modeling issues that further spread the model/parametric uncertainty, even if the process behind it is not random.
In simpler terms, it is best understood as limited knowledge-based uncertainty. It is not similar to aleatoric uncertainty, which is caused by external factors that naturally exist in different forms, but rather a type of uncertainty that is fully controllable with precise and sufficient knowledge and understanding of the system. It depends entirely on whether the CPS uses valid or precise models or parameters for operations that require dependability. In the reviewed literature, this category is identified in 53 of 186 papers (28.5\%). 


\textbf{Examples of work discussing this uncertainty:}
This uncertainty arises from problems with model updating and validation, particularly in digital twin or data-driven CPS systems. These are the main systems (digital twins, learning-based systems/data-driven systems) that are mostly discussed as being more prone to such issues. One example of this uncertainty is highlighted in an industrial process. Wang et al.~\cite{wang2024_mutual_active_learning} assess this uncertainty as a model-mismatch issue because, over time, changing environments require updated data to be added to the system. If the model data becomes outdated, its accuracy decreases. If the model mismatch occurs because of insufficient or outdated information, it is due to epistemic/model/parametric uncertainty. It is therefore necessary to ensure the reliability of such systems, as they are also involved in decision-making.
In the context of self-adaptive systems, Zhang et al.~\cite{zhang2019_wise_test_modeling} 
looks at uncertainty-aware model-based testing and addresses the constraint that traditional models often ignore or fail to detect the system's unexpected behavior under different circumstances. The main focus of this category is not on testing the system itself (as it connects to category 1), but rather on the model behavior and its specifications, which are uncertain or incomplete.

In energy or power control CPS, model/parametric uncertainty is often treated as an estimation problem or, under uncertain parameters and the nonlinear behavior of the system, as a control problem as well. Rahimilarki et al.~\cite{rahimilarki2019_neural_network_fault} exemplify this pattern by addressing robust fault estimation for nonlinear systems under uncertainty and validating the approach in an energy-related system (wind turbine context). It is crucial that the treatment of uncertain parameters/model behavior be limited to or estimated for reliable control and monitoring.

\keyfinding{Epistemic/model uncertainty comes from incomplete knowledge and other inconsistencies in the model/parameters, which restricts prediction accuracy, verification, and controller confidence.}{53/186 (28.5\%)}{General CPS/cross-domain (38/53, 71.7\%); Autonomous driving (7/53, 13.2\%); Software/verification (6/53, 11.3\%)}

\subsubsection{Category 4 --
Probabilistic/Aleatoric Uncertainty:}
\textbf{Description and Occurrences:}
This uncertainty relates to randomness, unpredictable changes within systems that cannot be precisely predicted, or probabilistic behavior, including noise, random events, unexpected transitions, and the evolution of a probabilistic system. 
The primary concern of this category is not the unknown aspects of the model, but rather the probabilistic portrayal of uncertainty for purposes such as analysis, decision-making, control, or verification. This uncertainty occurs in 48  of the 186 studied papers (25.8\%). 



\textbf{Examples of work discussing this uncertainty:}
Probabilistic or aleatoric uncertainty most commonly exists in decision-making and control under stochastic conditions. Zhu et al.~\cite{zhu2025_observer_free_vehicles}
exemplifies this approach by addressing uncertain changes/variations and noise through a probabilistic, chance-constrained control formulation for safe motion, rather than relying solely on deterministic control design. This category is also observable in software and verification research, more specifically through stochastic formal modeling and probabilistic verification. Du et al.~\cite{du2018_pcssl_stochastic_extension} 
is one of the examples here, as it models CPS timing and behavioral alternatives using probabilistic semantics and enables statistical or probabilistic verification of system properties.

In robotics, autonomous systems, and learning-enabled CPS (LE-CPS) safety contexts, probabilistic uncertainty informs uncertainty-aware prediction and confidence-bounded monitoring. Lin et al.~\cite{lin2025_monitoring_learning_enabled} employed uncertainty-aware prediction, including probabilistic and confidence-based mechanisms, to estimate safety violations under uncertain inputs. Although this paper may also be relevant to safety or specification categories, its significance for this category of uncertainty lies in its probabilistic approach to prediction and monitoring, rather than in the broader safety category.

\keyfinding{Probabilistic/Aleatoric uncertainty highlights fundamental stochastic variability or noise that is usually addressed through probabilistic risk or uncertainty quantification.}{48/186 (25.8\%)}
{General CPS/cross-domain (27/48, 56.3\%); Other (8/48, 16.7\%); Software/verification (8/48, 16.7\%)}

\subsubsection{Category 5 -- Dynamics Uncertainty:}\hfill\break
\noindent
\textbf{Description and Occurrences:}
This category concerns the physical process itself, such as dynamics, nonlinear effects, disturbances, actuator constraints and saturation, changing operating rules, and variability in plant response over time. The main characteristic is that this uncertainty comes from the evolving physical behavior of the controlled process and its interaction with the cyber layer.
This category is most precisely expressed as physical behavior uncertainty. It differs from Environmental Uncertainty, which focuses on uncertainty in external factors such as weather, terrain, or traffic. In contrast, physical-behavior uncertainty focuses on the response of process and system dynamics during CPS operation.
This uncertainty category is recognized in 48 of 186 papers (25.8\%). 


\textbf{Examples of work discussing this uncertainty:}
Process and dynamics uncertainty is usually addressed as a robust control problem involving nonlinearities and disturbances. Wu et al.~\cite{wu2024_mixed_textbraceleft_textbraceright} 
exemplifies this approach by leveraging its robust model predictive control (MPC) framework to handle nonlinear CPS behavior under unexpected disturbances and actuator or input constraints. 

In energy and power systems, this category is often operationalized as updating estimation and reliability problems within physically evolving systems. Rahimilarki et al.~\cite{rahimilarki2019_neural_network_fault} (Energy/power systems) provides a strong example of robust fault and state estimation for nonlinear dynamic systems, specifically using the wind-turbine benchmark. The primary challenge is uncertainty about the system’s variable behavior and responses, which are influenced by faults. In this regard, the focus is on the evolution of the process itself rather than solely on the epistemic quality of the model.

Furthermore, uncertainty in process dynamics within CPSs frequently exhibits as a cyber-physical coordination challenge when plant behavior is affected by communication or network conditions. Ma et al.~\cite{ma2019_holistic_management_dependable} is another example suitable for this type of uncertainty. It highlights that the uncertain communication conditions are important to consider because they degrade control over the physical process, thereby altering the plant's effective dynamic behavior and control quality.

Another example of this category is Wang et al.~\cite{wang2024_data_guaranteed_trajectory}, as it examines uncertainty in the evolving behavior of the controlled system during operation. A key issue concerns how different changes in the system state influence stability, performance, or reliability over time. This accounts for the uncertainty inherent in process evolution, rather than limiting it to aspects such as sensing or communication.

\keyfinding{Dynamics/disturbance uncertainty arises because of uncertain physical processes and other disturbances, motivating robust or adaptive estimation and control.}{48/186 (25.8\%)}{General CPS/cross-domain (27/48, 56.2\%); Energy/power systems (13/48, 27.1\%); other (6/48, 12.5\%)
}

\subsubsection{Category 6 -- Operational/Runtime Context Uncertainty:}\hfill\break
\noindent
\textbf{Description and Occurrences:}
This category is expressed more precisely as execution-context uncertainty.
It relates to the system’s current operating situation during execution. This comprises changes in runtime conditions, context shifts, evolving operational modes, and uncertainty about whether the system continues to operate within the conditions expected at design time. The basic challenge is not only environmental changes but also the system's ability to interpret current conditions to ensure a safe and correct operation.

This category is identified in 45 of 186 papers (24.2\%). 
It is distinct from Category 2, as this type of uncertainty concerns issues in the system’s interpretation of real-time operations and adaptation, rather than external conditions alone. Furthermore, it differs from Category 3, since the primary concern is not model incompleteness but rather runtime context alignment, such as the continued validity of assumptions, monitors, or adaptation strategies during operation.


\textbf{Examples of work discussing this uncertainty:}
Gerostathopoulos et al.~\cite{gerostathopoulos2016_self_adaptation_intensive} addressed operational and runtime context uncertainty by establishing connections among runtime situations, system requirements, and architectural configurations to enable self-adaptation. In this context, uncertainty arises in interpreting the current operating situation and selecting an appropriate adaptation. This uncertainty extends beyond environmental changes to include challenges in the system’s real-time interpretation. 

Camilli et al.~\cite{camilli2023_enforcing_resilience_equilibrium} evaluated at runtime whether the CPS remains within an acceptable operating or viability region. The basic challenge addressed in this study is the uncertainty related to the current operational state and the need for corrective action. Consequently, the paper emphasizes runtime assessment, supervision, and enforcement in dynamic environments.

\keyfinding{Uncertainty in runtime context reflects the risk of assumption drift and other varying operational conditions, necessitating continuous monitoring and adaptation to maintain system assurance.}{45/186 (24.2\%)}{General CPS/cross-domain (29/45, 64.4\%); Software/self-adaptive/verification (15/45, 33.3\%); Robotics/autonomous robots (6/45, 13.3\%)
}

\subsubsection{Category 7 --
Fault/Failure/Reliability Uncertainty:}\hfill\break
\noindent
\textbf{Description and Occurrences:}
This category is most accurately interpreted as dependability uncertainty arising from faults or conditions induced by degradation. It relates to the uncertainty in fault occurrence, how faults propagate, the reliability of the methods used for detection and estimation, and the system's ability to maintain dependable operation. Fault, failure, and reliability uncertainty is the seventh normalized uncertainty category, occurring in 44 of 186 papers (23.7\%). 



\textbf{Examples of work discussing this uncertainty:}
This uncertainty is discussed in the work of Ma et al.~\cite{ma2021_self_healing_reinforcement}, which examined self-healing and recovery behaviors in the presence of faults. In this context, uncertainty encompasses not only the presence of faults but also the system's ability to reliably detect, recover from, and maintain operation following a failure. This relationship directly connects uncertainty with system dependability and confidence in fault-handling capabilities. Similarly, Yu et al.~\cite{yu2019_event_sequential_prognosis} addressed uncertainty in faults, failures, and reliability by examining degradation processes and fault progression that affect dependable system operation. The primary concern is the uncertainty about the emergence of failures over time, as well as the system's reliability in handling them.

\keyfinding{Reliability uncertainty encompasses faults or degradation and unexpected failure behavior, which requires the driving diagnosis, fault tolerance, and recovery strategies.}{44/186 (23.7\%)}{General CPS/cross-domain (26/44, 59.1\%); Software/self-adaptive/verification and Industrial/manufacturing/process (7/44 each, 15.9\%)}

\subsubsection{Category 8 -- Communication/Network/Timing Uncertainty:}\hfill\break
\noindent
\textbf{Description and Occurrences:}
 This refers to uncertainty in the cyber coordination layer, including delays, packet loss, bandwidth changes, timing jitter, asynchronous execution, congestion, and changes in communication quality that affect coordination, control, and safety. This category usually covers uncertainty about how information is passed and when it is timed in the coordination layer. This uncertainty is found in 40 out of 186 papers (21.5\%).
 


\textbf{Examples of work discussing this uncertainty:}
The strong presence in energy and power systems highlights the importance of reliable communication and timing for networked CPS infrastructure.
For example, Song et al.~\cite{song2025_quantum_particle_swarm} examined the uncertainty in wireless communication and the limited transmission resources for smart-grid control. In their work, uncertainty in the communication channel and transmission conditions directly affects the importance of timely and reliable cyber-physical coordination. Waga et al.~\cite{waga2022_model_bounded_monitoring} examined uncertainty in timing or communication behavior that influences CPS monitoring or coordination. The primary concern extends beyond plant dynamics to include the system's ability to depend on timely and stable information exchange. Consequently, network and timing behavior become a crucial source of uncertainty.

\keyfinding{Communication/timing uncertainty appears because of the delays and jitters/loss in networked CPS and affects stability and safety margins.}{40/186 (21.5\%)}{General CPS/cross-domain (27/40, 67.5\%); Energy/power systems (8/40, 20.0\%); Other (5/40, 12.5\%)}

\subsubsection{Category 9 -- Cyber/Adversarial/Security Uncertainty:}\hfill\break
\noindent
\textbf{Description and Occurrences:}
 In CPSs, this type of uncertainty involves issues like malicious behavior, attacks, adversarial effects, and unclear security states. It comprises not just keeping the information on how to respond if an attack is happening, how it might impact the CPS, or how the system should react when threat information is incomplete or unclear. Cyber, adversarial, or security uncertainty occurred in 39 out of 186 papers (21.0\%).



\textbf{Examples of work discussing this uncertainty:}
Lei et al.~\cite{lei2022_fdi_attack_edge} 
detected False Data Injection (FDI) attacks in smart grids.  The key issue is the uncertainty about whether an attack is happening and how to estimate its effects. This uncertainty affects the system's confidence in real-time detection and its ability to respond.
Hong et al.~\cite{hong2024_internal_model_cooperative} 
treated Denial-of-Service (DoS) attacks as adversarial threats that undermine distributed consensus and resilience. This shows how the same paper can fit into different categories, each highlighting a different type of uncertainty. Chen et al.~\cite{chen2023_filter_fuzzy_switched} 
examined security-relevant uncertainty in CPS operations under adversarial influence. The primary concern involves uncertainty regarding the impact of malicious interference on state estimation, filtering, or overall system behavior. Consequently, the uncertainty is explicitly associated with attacks rather than accidental factors.

\keyfinding{Adversarial/security uncertainty highlights the intentional disruptions (e.g., FDI/DoS) and mentions the need to connect the security outcomes to safety/specification assurance.}{39/186 (21.0\%)}{General CPS/cross-domain (25/39, 64.1\%); Energy/power systems (9/39, 23.1\%); Other (6/39, 15.4\%)}


\subsubsection{Category 10 --
Resource/Computational/Scheduling Uncertainty:}\hfill\break
\noindent
\textbf{Description and Occurrences:}
This category concerns the uncertainty regarding the availability, sufficiency, and allocation of multiple resources, including computational and operational resources. These include processor time, scheduling behavior, cloud or edge capacity, bandwidth or compute load, and execution-time uncertainty. The primary concern is whether the CPS has the necessary resources at the required time to ensure acceptable, safe, and reliable performance.
This category is most accurately expressed as capacity and allocation uncertainty. This type of uncertainty is identified in 24 out of 186 papers (12.9\%).



\textbf{Examples of work discussing this uncertainty:}
Panahi et al.~\cite{panahi2024_performance_automotive_efficient} 
examined control-task execution on shared platforms. In terms of Category 10, the primary concerns related to uncertainty are the availability of scheduling and computational resources for control workloads, specifically whether timing and execution resources are sufficient to maintain safe and stable control behavior. 
Cui et al.~\cite{cui2023_filtering_out_high} 
studied cloud-connected CPS environments. In these settings, unpredictable data streams introduce uncertainty in bandwidth and cloud-compute load, prompting decisions regarding workload distribution and edge-side processing. 

\keyfinding{Resource/scheduling uncertainty emphasizes the uncertain computation time availability that can suppress the need for timely control, monitoring, and decision pipelines.}{24/186 (12.9\%)}{General CPS/cross-domain (19/24, 79.2\%); Software/self-adaptive/verification (4/24, 16.7\%); Industrial/manufacturing/process (3/24, 12.5\%)}
\subsubsection{Category 11 -- Data/Distribution/Information Uncertainty:}\hfill\break
\noindent
\textbf{Description and Occurrences:}
This category encompasses uncertainty resulting from the quality, completeness, representativeness, and statistical validity of data used by the system. Relevant issues include noisy or missing data, concept drift, out-of-distribution (OOD) inputs, biased datasets, limited observations, and inconsistent reliability among multiple information sources. Notably, this form of uncertainty pertains to the information used by the CPS, rather than the physical process itself. This category is identified in 24 out of 186 papers (12.9\%).


\textbf{Examples of work discussing this uncertainty:}
Feng et al.~\cite{feng2021_improving_variational_autoencoder} 
addresses out-of-distribution uncertainty in learning-enabled safety-critical CPSs by enhancing variational autoencoder-based detection methods for embedded real-time applications. In the context of Category 11, the key observation is that changes in the input data distribution, rather than only environmental or runtime state changes, can degrade model reliability. Consequently, awareness regarding distribution is essential for dependable system operation.
Ren and Li~\cite{ren2022_interactive_learning_human} 
examined a self-evolving machine learning model within a smart manufacturing context. The connection to Data/Distribution/Information uncertainty lies in the uncertainty introduced by streaming, non-stationary, and evolving data patterns, which involves system adaptation to retain valid inference and decision support over time.

\keyfinding{Data/distribution uncertainty highlights drift and distribution shift, more specifically relevant for learning-enabled CPSs, where the validity of the evidence can degrade after deployment.}{24/186 (12.9\%)}{General CPS/cross-domain (15/24, 62.5\%); Industrial/manufacturing/process (4/24, 16.7\%); Robotics/autonomous robots and Autonomous driving (3/24 each, 12.5\%)}

\subsubsection{Category 12 -- Observability Uncertainty:}\hfill\break
\noindent
\textbf{Description and Occurrence:}
This uncertainty concerns a system's ability to observe, measure, perceive, and estimate the state of the environment. This includes noise in sensor data, missing or partial observations, imperfect measurements, perception-model errors, and uncertainty in state estimation. The primary concern is that CPS decisions often rely on uncertain observations rather than on actual system or world states.
This category of uncertainty is found in 22 out of 186 papers (11.8\%).


\textbf{Examples of work discussing this uncertainty:}
A most relevant example is provided by Wei et al.~\cite{wei2024_multiarea_probabilistic_forecasting}
which employed probabilistic forecasting-aided interval-state estimation in a power distribution context. Within this uncertainty category, the critical factor is uncertainty in state estimation due to imperfect measurements, which directly affects the reliability of monitoring and anomaly detection.
Another example is the work of Ghosh and André~\cite{ghosh2023_mouldys_monitoring_presence}, 
which addressed the challenge of uncertain logs containing noisy and missing samples in the context of safety monitoring. In this study, the monitoring process depends on imperfect and incomplete observations, making measurement uncertainty a central factor in the reliable detection of unsafe behavior.

\keyfinding{Observability uncertainty comes from noise in sensor data, missing or partial observations, imperfect measurements, perception-model
errors, and uncertainty in state estimation.}{22/186 (11.8\%)}{General CPS/cross-domain (13/22, 59.1\%); Other (4/22, 18.2\%); Industrial/manufacturing/process and Autonomous driving (2/22 each, 9.1\%)}

\subsubsection{Category 13 -- Human Interaction Uncertainty:}\hfill\break
\noindent
\textbf{Description and Occurrences:}
This category refers to uncertainty introduced by human behavior, cognitive states, user interactions, stakeholder preferences, and human-in-the-loop decision-making. The defining feature is that human actors,operators, users, supervisors, drivers, patients, or stakeholders are not fully predictable and their actions or preferences can change the system’s operational behavior. This category is best interpreted as human-linked uncertainty in cyber-physical operation.  Human/interaction/preference uncertainty occurs in 21 out of 186 papers (11.3\%). 



\textbf{Examples of work discussing this uncertainty:}
A representative example is the work of Sridhar et al.~\cite{sridhar2024mind_wandering}
in partially autonomous driving, which addressed human cognitive uncertainty, specifically mind wandering and its effect on braking/takeover behavior.  Kusherbaeva and Zhou~\cite{kusherbaeva2023_multiobjective_data_driven} 
address user-in-the-loop optimization under changing preferences. This paper focuses on the uncertainty of human preferences and interaction priorities, which affects how the CPS should adapt its optimization decisions.

\keyfinding{Human interaction uncertainty captures variability in user behavior, intent, and preferences that impacts safe control, handover, and system acceptance.}{21/186 (11.3\%)}{General CPS/cross-domain (15/21, 71.4\%); Other, Software/self-adaptive/verification, and Industrial/manufacturing/process (3/21 each, 14.3\%)}

\subsubsection{Category 14 -- Behavioral/Decision Uncertainty:}\hfill\break
\noindent
\textbf{Description and Occurrences:}
In CPSs, this uncertainty relates to unpredictable actions, policy choices, agent decisions, and the resulting system behavior. It is often seen in autonomous systems or situations where decision policies are not fully determined. 
Behavioral or decision uncertainty appears in 17 of the 186 papers we analyzed (9.1\%). 


\textbf{Examples of work discussing this uncertainty:}
Khatiri et al.~\cite{khatiri2025uav_uncertainty} 
examined autonomous UAV operations. Their work highlighted uncertainty in the UAV’s control behavior and output, where inconsistent decisions may lead to unsafe outcomes. Wang et al.~\cite{wang2025_conformal_temporal_logic} discussed and addressed uncertainty in task planning and decision-making when system conditions change. The main uncertainty-related issue in their work is choosing the correct actions or plans under incomplete or evolving information, which places the uncertainty at the system’s decision-making layer.

\keyfinding{Behavior/decision uncertainty arises because of the uncertainty in policy or action selection for autonomous components, affecting planning reliability and safety.}{17/186 (9.1\%)}{General CPS/cross-domain (8/17, 47.1\%); Robotics/autonomous robots (5/17, 29.4\%); Autonomous driving (3/17, 17.6\%)
}

\subsubsection{Category 15 -- Deployment Uncertainty:}\hfill\break
\noindent
\textbf{Description and Occurrences:}
This category can be understood as life-cycle transfer validity uncertainty and concerns whether results obtained during design, simulation, or digital-twin development remain valid when the system is deployed in the real world.
Simulation, deployment, or transfer uncertainty is identified in 16 out of 186 papers (8.6\%). 



\textbf{Examples of work discussing this uncertainty:}
Biagiola et al.~\cite{biagiola2024digital_siblings}
deal with this uncertainty by showing that different simulators can produce varying results for identical autonomous-driving test inputs. DeVries et al.~\cite{devries2021_monitoring_environmental_knowledge}
emphasize the need for runtime monitoring to detect situations that were not anticipated during design, reflecting uncertainty about whether design-time knowledge fully transfers to operational environments.

\keyfinding{Deployment uncertainty highlights the sim-to-real and test-to-deploy gap, where evidence and models may not transfer reliably to operation.}{16/186 (8.6\%)}{General CPS/cross-domain (11/16, 68.8\%); Industrial/manufacturing/process (3/16, 18.8\%); Robotics/autonomous robots and Autonomous driving (2/16 each, 12.5\%)}

\subsubsection{Category 16 -- Operational Scenarios Uncertainty:}\hfill\break
\noindent
\textbf{Description and Occurrences:}
This category refers to uncertainty in how operational scenarios are defined, selected, and covered during testing, validation, and safety analysis. The issue is not only uncertainty within individual scenarios but also uncertainty about which scenarios matter most.
 Scenario-based uncertainty is identified in 12 out of 186 papers (6.5\%).


\textbf{Examples of work discussing this uncertainty:}
Crespo-Rodriguez et al.~\cite{cresporodriguez2025_instance_space_vehicles}
examines simulation-based testing of autonomous vehicles and the challenges of identifying effective test cases when safety-critical events are infrequent. Neelofar and Aleti~\cite{neelofar2024_identifying_explaining_critical} 
focus on generating critical tests that account for interacting static and dynamic factors. The central challenge was determining whether enough relevant scenario combinations have been explored to support credible validation.

\keyfinding{Scenario uncertainty concerns whether scenario sets are sufficient to cover rare or critical cases, shaping confidence in evaluation and safety claims.}{12/186 (6.5\%)}{Autonomous driving (4/12, 33.3\%); Software/self-adaptive/verification (3/12, 25.0\%); General CPS/cross-domain (2/12, 16.7\%)}

\begin{tcolorbox}[
  colback=gray!10,
  colframe=gray!35,
  boxrule=0.6pt,
  arc=2pt,
  left=6pt,
  right=6pt,
  top=6pt,
  bottom=6pt,
  title=\textcolor{black}{\textbf{RQ1 Finding}},
  coltitle=black
]
Among the 186 primary articles, we characterized uncertainty types into 16 categories. Primarily, the literature is ruled by the uncertainty related to assurance, including safety and specification uncertainty, as well as environmental uncertainty. Model or epistemic uncertainty and probabilistic or aleatoric uncertainty are also reported frequently. Analysis of the domain–category distribution highlights that general CPS or cross-domain represent the largest proportion across most uncertainty types. In contrast, domain-specific focus was observed in safety-critical fields such as autonomous driving and robotics, where environmental variability, runtime context, and assurance evidence are more specifically significant. 
\end{tcolorbox}

\subsection{RQ2: Sources that Cause Uncertainty in CPSs}

\begin{table}[ht]
\centering
\footnotesize
\setlength{\tabcolsep}{4pt}
\renewcommand{\arraystretch}{1.15}
\caption{Major sources of uncertainty in CPSs and the uncertainty categories they induce}
\label{tab:rq2_source_groups}
\begin{tabular}{>{\raggedright\arraybackslash}p{4.0cm} >{\raggedright\arraybackslash}p{5.8cm} >{\raggedright\arraybackslash}p{4.5cm}}
\hline
\textbf{Source Group} & \textbf{Representative Sources} & \textbf{Induced Uncertainty Categories} \\

\hline
\textbf{Environmental and Physical Process Sources}
&
Weather conditions, traffic variability, terrain changes, external disturbances, nonlinear dynamics, friction, load variation, stochastic physical processes
&
Environmental,
Process/Dynamics/Disturbance,
Probabilistic/Aleatoric,
Scenario
\\
\textbf{Data, Sensing, and Perception Sources}
&
Noisy or missing data, biased datasets, concept drift, out-of-distribution inputs, sensor noise, estimation errors, partial observability
&
Data/Distribution/Information,
Sensor/Measurement/Perception,
Model/Parametric
\\
\textbf{Knowledge, Model, and Decision Sources}
&
Incomplete system knowledge, uncertain parameters, simplified assumptions, model mismatch, model drift, inaccurate abstractions, uncertain decision policies
&
Model/Parametric/Epistemic,
Behavioral/Decision,
Simulation/Deployment
\\
\textbf{Communication and Infrastructure Sources}
&
Network delays, packet loss, timing jitter, synchronization problems, bandwidth limitations, hardware failures, component degradation
&
Communication/Network/Timing,
Fault/Failure/Reliability
\\

\textbf{Human-Centered Sources}
&
Human errors, cognitive limitations, operator variability, changing preferences, trust issues, unpredictable human-system interactions
&
Human/Interaction/Preference,
Behavioral/Decision
\\

\textbf{Cybersecurity and Adversarial Sources}
&
Denial-of-service attacks, false-data injection, adversarial ML inputs, malicious behavior, privacy breaches, cyber vulnerabilities
&
Cyber/Adversarial/Security
\\

\textbf{Requirements, Specification, and Testing Sources}
&
Incomplete requirements, ambiguous specifications, insufficient safety requirements, limited verification and validation, inadequate test coverage, rare-event omission, incomplete scenario exploration
&
Safety/Specification/Testing,
Scenario,
Simulation/Deployment
\\

\textbf{Runtime Operational and Resource Sources}
&
Changing operational contexts, workload fluctuations, adaptive operating modes, execution-time variability, resource contention, energy limitations
&
Operational/Runtime,
Resource/Computational/Scheduling
\\

\hline

\end{tabular}
\end{table}

This RQ aims to link the uncertainty types from RQ1 (Table~\ref{tab:rq1_mapping_clickable}) with their sources. The relationship between different uncertainty and their sources is reported in Table~\ref{tab:rq2_source_groups}.

\begin{figure}[ht]
  \centering
  \includegraphics[width=\textwidth]{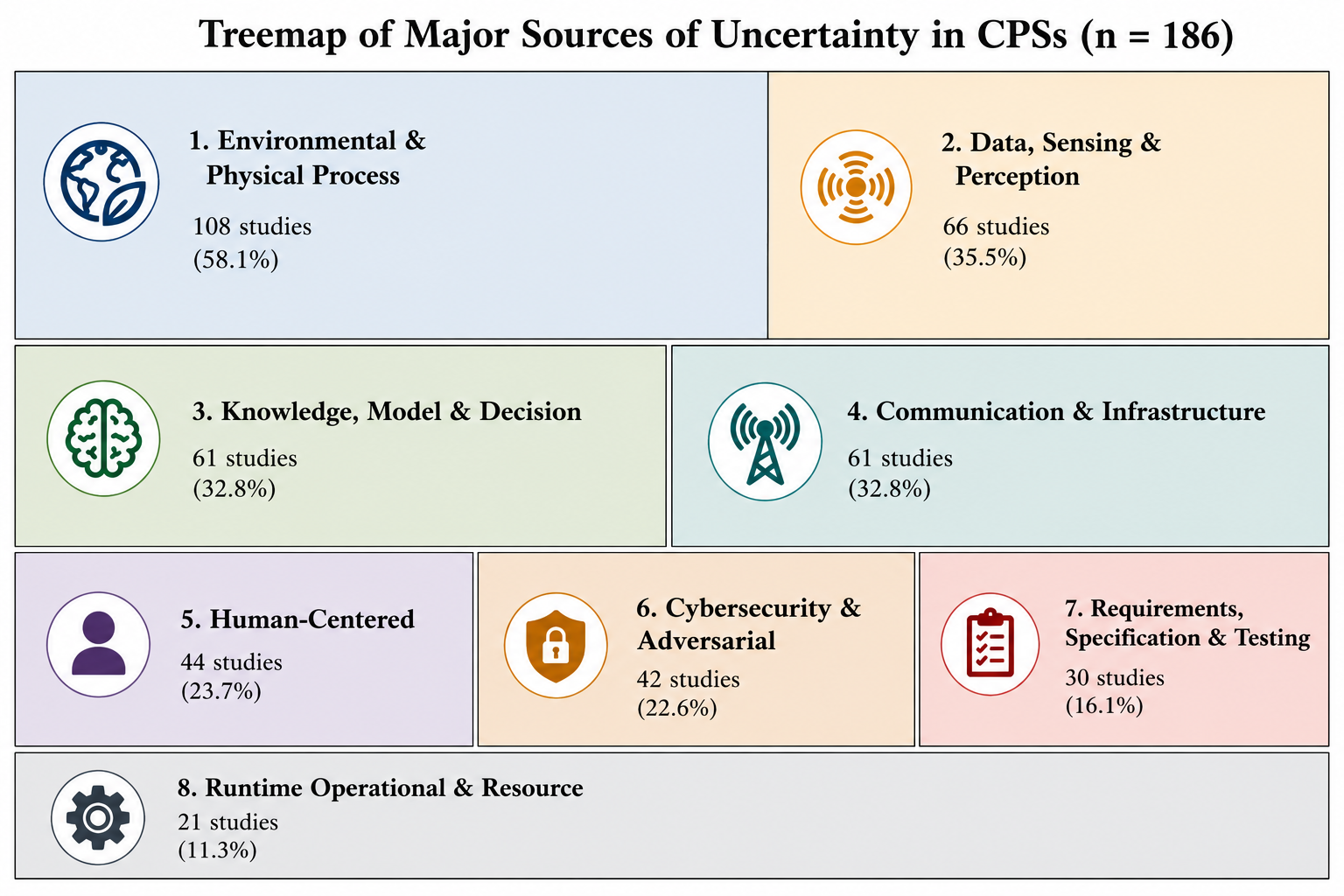}
 \caption{Treemap of the major sources of uncertainty identified across the 186 primary papers. The rectangle size is proportional to the number of articles assigned to each source group. Since multi-label coding was used, a study may contribute to more than one source group.}
\label{fig:rq2_source_treemap}
\end{figure}

Fig. \ref{fig:rq2_source_treemap} shows that uncertainty in CPSs does not originate from a single source. Instead, it comes from interactions between the environment, the system, data, humans, and adversarial threats. These sources are mutually dependent, influencing one another and propagating uncertainty throughout the system, thereby increasing the overall unpredictability. 

In the following, we describe the different sources and how they relate to the different uncertainty types.

\subsubsection{Environmental and Physical Process Sources} \hfill\break
\noindent
\textbf{Description}: 
The major reported sources of uncertainty in CPSs are the dynamic physical environment, Physical process disturbances, and inherent stochasticity. This source comes from external factors of the system \ie weather, traffic variability, terrain characteristics, obstacles, environmental disturbances, and change in operations.

\textbf{Affected uncertainty types}:
This source triggers \textit{Environmental Uncertainty} (54 studies), \textit{Process/Dynamics/Disturbance Uncertainty} (47 studies), \textit{Probabilistic/Aleatoric Uncertainty} (48 studies), and \textit{Scenario Uncertainty} (13 studies). Any change in the environment directly affects the physical conditions under which the system operates, whereas disturbances and changes in operating conditions affect the process dynamics. Furthermore, the large number of possible environmental configurations creates uncertainty, and therefore, it is important to consider scenarios during testing and validation.

\textbf{Relevant articles and challenges}: 
Evidence from the reviewed articles shows that environmental sources are specifically common in autonomous driving, UAVs, and robotics. Among the 55 articles classified under \textit{Environmental uncertainty}, autonomous driving and cross-domain CPS applications account for a significant proportion of the literature. For example, Khatiri et al.~\cite{khatiri2025uav_uncertainty} indicated that uncertainty occurring from changing environmental conditions directly impacts UAV safety during runtime operation. Zhao et al.~\cite{zhao2025_behavioral_decision_making} analyzed autonomous driving under unsafe weather and traffic conditions and showed how environmental variability affects decision-making and safety assurance. Crespo-Rodriguez et al.~\cite{cresporodriguez2025_instance_space_vehicles} also highlighted that environmental variability creates an uncertainty scenario regarding the completeness of scenario coverage, making it difficult to identify all safety-critical situations during testing. 

The main challenge associated with environmental and physical-process sources is that they are usually difficult or impossible to control. Environmental conditions always evolve during operation, while physical processes may show nonlinear behavior, stochastic disturbances, and changing operating rules. Consequently, CPS developers must design systems that remain reliable despite uncertainty rather than attempting to eliminate uncertainty. 

\sourcefinding{Environmental and physical-process conditions constitute the dominant source of uncertainty in CPSs. Investigated literature introduce uncertainty through uncontrollable operating conditions, nonlinear physical behavior, external disturbances, and inherent stochastic variation, frequently propagating into scenario selection and safety assurance.}{108/186 (58.5\%)}{Weather conditions; traffic variability; terrain changes; obstacles; external disturbances; nonlinear dynamics; friction; load variation; stochastic physical processes}

\subsubsection{Data, Sensing, and Perception Sources}\hfill\break
\noindent
\textbf{Description}: 
This source occurs when the information used by the system to sense, understand, and reason about its environment is incomplete, inaccurate, delayed, noisy, or unrepresentative of real operating requirements. These sources include noisy sensor measurements, missing observations, biased datasets, concept drift, out-of-distribution inputs, perception-model limitations, estimation errors, and partial observability of the physical conditions.

\textbf{Affected uncertainty types}:   
This source mainly causes \textit{Data/Distribution/Information Uncertainty }(24 articles, 12.8\%) and \textit{Sensor/Measurement/Perception Uncertainty} (22 articles, 11.7\%). It also contributes indirectly to \textit{Model/Parametric/Epistemic Uncertainty}, as false or incomplete observations can affect model calibration, learning, and state estimation. In many CPSs, uncertainty arising from sensing and data addition propagates throughout the entire processing chain, impacting prediction, planning, control, and, ultimately, system safety.

\textbf{Relevant articles and challenges}
Archibald et al.~\cite{archibald2025bdi}showed that noisy observations influence belief formation and action marked in autonomous agents. Similarly, Bansal et al.~\cite{bansal2024perception_simplex} show how restrictions in sensing and perception directly affect system confidence and operational safety. Furthermore, Zhu et al. \cite{zhu2025_observer_free_vehicles} showed that non-Gaussian measurement noise and uncertain observations significantly affect state estimation and collision avoidance in autonomous vehicles. Also,~\cite{yang2024_memory_distribution_shift} power-system monitoring articles revealed that uncertain measurements directly affect state estimation and anomaly detection. In autonomous-driving applications, \textit{Perception uncertainty} frequently arises from imperfect object detection and from environmental perception models.

One major challenge with this source is that CPSs rarely have direct access to the true state of the physical world. Instead, they rely on sensors and data-processing pipelines that are naturally imperfect. Sensor failures, measurement noise, communication delays, occlusions, and environmental interference can all degrade observation quality.

\sourcefinding{Data, sensing, and perception form an upstream source of uncertainty: noisy, incomplete, delayed, or distribution-shifted observations affect state estimation and subsequently propagate to prediction, decision-making, control, and safety.}{66/186 (35.6\%)}{Noisy or missing data; biased datasets; concept drift; out-of-distribution inputs; sensor noise; measurement inaccuracies; estimation errors; partial observability}

\subsubsection{Knowledge, model, and decision-related sources}\hfill\break
\noindent
\textbf{Description}:  
This source of uncertainty stems from knowledge limitations, model imperfections, and decision-making processes. This source arises when a CPS lacks complete knowledge about its environment, system dynamics, or future operating conditions. Such rules manifest as simplified system models, uncertain parameters, incomplete assumptions, model drift, and poor decision-making policies. Unlike environmental sources, which originate outside the system, this source is internal and arises from the system's representation of reality and its ability to make claims about future actions.

\textbf{Affected uncertainty types}: 
This source mainly contributes to \textit{Model/Parametric/Epistemic Uncertainty}, which was reported in 53 of the 186 reviewed articles (28.5\%), making it one of the most frequently observed uncertainty categories in the literature. It also contributes to \textit{Behavioral/Decision Uncertainty} (18 articles) and \textit{Simulation/Deployment Uncertainty} (16 articles), especially when autonomous agents make decisions with incomplete information or when deployment environments differ from those inferred during design.

\textbf{Relevant articles and challenges}:
Alfonso et al.~\cite{alfonso2020modeling} highlighted that digital twin models gradually become inaccurate as the physical system evolves, leading to increased uncertainty due to model mismatch and outdated assumptions. Zhang et al.~\cite{zhang2019_wise_test_case} showed that uncertainty-aware model-based testing is required because traditional models often fail to capture uncertain system behavior during validation. In autonomous driving systems, Zhao et al.~\cite{zhao2025_behavioral_decision_making} demonstrated that decision-making under severe scenarios depends on probabilistic risk models whose accuracy is denied by incomplete knowledge about environmental conditions and vehicle behavior. Yu et al.~\cite{yu2023_data_driven_parameterized} addressed uncertainty in nonlinear power-system dynamics, where uncertain parameters directly affect state estimation and fault diagnosis.

A main challenge associated with this source is that CPS models are naturally approximations of reality. Physical processes are usually nonlinear, time-varying, and affected by factors that cannot be fully estimated or predicted. As operating conditions evolve, models may become outdated, showing increasing differences between predicted and actual behavior. 

\sourcefinding{Knowledge, model, and decision sources reflect the limited fidelity of a CPS's internal representation of reality. Incomplete knowledge, uncertain parameters, model drift, and imperfect decision policies create epistemic uncertainty that can affect both runtime behavior and deployment validity.}{61/186 (33.0\%)}{Incomplete system knowledge; uncertain parameters; simplified assumptions; model mismatch; model drift; model obsolescence; inaccurate abstractions; uncertain decision policies}

\subsubsection{Communication, resource, and runtime-operation sources}\hfill\break
\noindent
\textbf{Description}:
Communication, resource, and runtime-operation sources constitute a major source of uncertainty in CPSs, as these systems operate in dynamic, distributed environments where system conditions continuously evolve during execution. This source arises from communication delays, packet loss, timing jitter, synchronization problems, bandwidth limitations, fluctuating workloads, changing operational contexts, resource contention, scheduling variability, and limited computational or energy resources. Unlike model-related uncertainty, which originates from incomplete knowledge, this source occurs during system operation when the cyber and physical components must coordinate under changing runtime conditions.

\textbf{Affected uncertainty types}:
This source mainly contributes to \textit{Communication/Network/Timing Uncertainty} (40 articles, 21.3\%), \textit{Operational/Runtime Context Uncertainty} (45 articles, 23.9\%), and \textit{Resource/Computational/Scheduling Uncertainty} (25 articles, 13.3\%). These uncertainty categories are closely connected because communication quality, resource availability, and runtime conditions collectively affect CPSs' ability to perform monitoring, decision-making, and control tasks reliably.

\textbf{Relevant articles and challenges}: 
Sarkar et al.~\cite{sarkar2025platooning} showed that vehicle platooning systems are affected by communication delays, dynamic interactions between vehicles, and changing operating conditions, all of which introduce uncertainty into joint control decisions. Similarly, Zhu et al.~\cite{zhu2025_observer_free_vehicles} showed that cloud-based autonomous vehicle control is influenced by network conditions, communication delays, and external disturbances, creating uncertainty in state estimation and collision-avoidance decisions.

The main challenge associated with this source is that runtime conditions are highly dynamic and difficult to predict in advance. Network quality may fluctuate, computational resources may become overloaded, workloads may change unexpectedly, and timing assumptions made during development may no longer hold during operation. Since many CPS applications are safety-critical and time-sensitive, even small delays or resource shortages can propagate throughout the system, affecting control performance, coordination, and safety.

\sourcefinding{Communication and infrastructure sources connect cyber-level variability with physical-system consequences. Delays, packet loss, timing variations, synchronization problems, hardware failures, and component degradation can directly affect coordination, control stability, and reliability.}{61/186 (33.0\%)}{Network delays; packet loss; timing jitter; synchronization problems; bandwidth limitations; hardware failures; sensor or actuator faults; component degradation}

\subsubsection{Human-Centered Sources}\hfill\break
\noindent
\textbf{Description}:
Human-centered sources of uncertainty are an important factor in CPSs because many systems interact closely with human users, operators, drivers, patients, or decision-makers. This source arises from variability in human behavior, cognitive limitations, changing preferences, fluctuations in attention, decision-making errors, trust issues, and unpredictable human-system interactions. Unlike environmental or technical sources, human-centered uncertainty arises because human actions and responses cannot always be accurately predicted or modeled.

\textbf{Affected uncertainty types}: This source mainly contributes to \textit{Human/Interaction/Preference Uncertainty} (21 articles, 11.2\%) and \textit{Behavioral/Decision Uncertainty} (17 articles, 9.1\%). In many CPS applications, uncertainty introduced by human actions directly affects system behavior, safety, adaptation strategies, and operational performance.

\textbf{Relevant articles and challenges}: 
Sridhar et al.~\cite{sridhar2024mind_wandering} showed that mind wandering and variations in driver attention, especially during braking, affect braking performance and takeover behavior in partially autonomous vehicles, introducing uncertainty into safety-critical driving situations. Similarly, Sritriratanarak et al. \cite{sritriratanarak2025_games_rational_multi} emphasized that uncertainty in human preferences and decision-making can influence the behavior of CPSs operating in collaborative or interactive environments, making it difficult to anticipate future system states and actions.

The main challenge associated with this source is that human behavior is highly dynamic and context-dependent. Factors such as exhaustion, experience, trust, situational awareness, and changing objectives can change how individuals interact with CPSs. As a result, assumptions made during design or testing may not accurately represent real-world human behavior during system operation. This challenge becomes particularly important in safety-critical applications where human decisions directly influence system outcomes.

\sourcefinding{Human-centered uncertainty confirms that CPS uncertainty is also socio-technical. Variability in attention, cognition, preferences, trust, and operator decisions can change system behavior and challenge assumptions established during design and testing.}{44/186 (23.9\%)}{Human decision errors; attention fluctuations; cognitive limitations; operator variability; changing preferences; trust issues; unpredictable human-system interactions}

\subsubsection{Cybersecurity and adversarial sources}\hfill\break
\noindent
\textbf{Description}:
Cybersecurity and adversarial sources introduce significant uncertainty in CPSs, as these systems increasingly rely on interconnected software, communication networks, cloud services, and AI-enabled components. This source originates from malicious attacks, adversarial actions, privacy breaches, false-data injection attacks, denial-of-service (DoS) attacks, adversarial machine-learning inputs, and vulnerabilities within cyber infrastructure. Unlike fault-related uncertainty, which arises from accidental failures or degradation, cybersecurity-related uncertainty stems from deliberate actions designed to disrupt, manipulate, or compromise system behavior.

\textbf{Affected uncertainty 
types}:
This source primarily contributes to \textit{Cyber/Adversarial/Security Uncertainty} (39 papers, 20.7\%). In many cases, it also indirectly influences \textit{Communication/Network Uncertainty}, \textit{Operational Uncertainty}, and \textit{Data Uncertainty}, as successful attacks often target communication channels, runtime operations, or information integrity.

\textbf{Relevant articles and challenges}: 
Wang et al.~\cite{wang2022_operating_state_reconstruction} examined denial-of-service attacks in CPS environments and showed that malicious disruption of communication channels creates uncertainty about system state, coordination, and control actions. Similarly, Liu et al.~\cite{liu2023_false_data_injection} examined false-data injection attacks in power systems and showed that adversaries can manipulate sensor and network information, causing uncertainty in state estimation and operational decision-making.

The main challenge associated with this source is that attackers are adaptive and strategic, making their actions inherently difficult to predict. Unlike environmental disturbances or hardware faults, adversarial behavior is intentional and may continuously evolve to bypass existing detection and protection mechanisms.

\sourcefinding{Cybersecurity and adversarial sources differ from naturally occurring uncertainty because they are intentionally induced. Adaptive attacks can manipulate measurements, communication, perception, and control decisions, creating uncertainty about both the true system state and the trustworthiness of available information.}{42/186 (22.3\%)}{Denial-of-service attacks; false-data injection; replay attacks; adversarial ML inputs; malicious behavior; privacy breaches; cyber vulnerabilities}

\subsubsection{Requirements, Specification, and Testing Sources}\hfill\break
\noindent
\textbf{Description}:
This source arises when system requirements are incomplete, unclear, evolving, or difficult to standardize, leading to uncertainty about the system's predicted behavior. It also arises from insufficient safety specifications, limitations in verification and validation techniques, insufficient test coverage, and the inability to predict or evaluate rare but safety-critical scenarios. Unlike environmental or sensing-related sources, this source typically appears during system development and validation activities, when uncertainty remains about whether the system has been sufficiently specified, designed, and tested.

\textbf{Affected uncertainty types}:
This source particularly contributes to \textit{Safety/Specification/Testing Uncertainty} (91 articles, 48.4\%), the most frequently identified uncertainty category in the reviewed literature. It also contributes to \textit{Scenario Uncertainty} (13 articles) and \textit{Simulation/Deployment Uncertainty} (16 articles), especially when testing environments fail to describe real-world operating conditions or when critical scenarios remain unknown.

\textbf{Relevant articles and challenges}: 
Crespo-Rodriguez et al.~\cite{cresporodriguez2025_instance_space_vehicles} showed that autonomous vehicle testing is inherently uncertain because safety-critical scenarios are rare and difficult to systematically identify, making extensive validation challenging. Similarly, Stocco et al.~\cite{stocco2023_mind_gap_transferability} showed that uncertainty can arise from limitations in simulation-based testing, where differences between simulated and real-world environments may lead to unexpected system behavior after deployment.

The main challenge with this source is that it is almost impossible to specify, verify, and test every possible operating condition a CPS may encounter. Modern CPSs operate in highly dynamic environments, creating a combinatorial explosion of possible scenarios and interactions. Consequently, developers must make assumptions regarding system behavior, environmental conditions, and safety requirements, which may later prove incomplete or inaccurate during operation.

\sourcefinding{Requirements, specification, and testing sources are mainly assurance-related: uncertainty arises when requirements, safety specifications, scenarios, simulations, and validation evidence are insufficient to establish confidence that the CPS will behave safely after deployment.}{30/186 (16.5\%)}{Incomplete requirements; ambiguous specifications; insufficient safety requirements; limited verification and validation; inadequate test coverage; rare-event omission; incomplete scenario exploration}

\subsubsection{Runtime Operational and Resource Sources}\hfill\break
\noindent
\textbf{Description}:
Runtime operational and resource sources are a major source of uncertainty in CPSs, as these systems operate in dynamic environments where conditions continuously evolve during execution. This source arises from changing operating contexts, workload fluctuations, adaptive operating modes, resource contention, scheduling variability, execution-time uncertainty, and limitations in computational, communication, or energy resources. Unlike model-related uncertainty, which arises from incomplete knowledge during design, runtime uncertainty emerges as the system actively interacts with its environment and responds to changing conditions.

\textbf{Affected uncertainty types}:
This source primarily contributes to \textit{Operational/Runtime Context Uncertainty} (45 articles, 23.9\%) and \textit{Resource/Computational/Scheduling Uncertainty} (25 articles, 13.3\%). These uncertainty categories frequently appear together because runtime decisions often depend on the availability of sufficient computational resources, predictable execution timing, and stable operating conditions.

\textbf{Relevant articles and challenges}: 
Zhu et al.~\cite{zhu2025_cloud_cruise_vehicles} showed that cloud-based autonomous vehicle control systems are influenced by runtime disturbances, communication constraints, and resource limitations, leading to uncertainty in control performance and system responsiveness.

The main challenge associated with this source is that runtime conditions cannot be fully predicted during development. Workloads may change unexpectedly, resources may become constrained, system modes may evolve, and assumptions made during design may no longer remain valid during operation. In safety-critical CPSs, these changes can directly affect the timeliness and reliability of monitoring, decision-making, and control processes.

\sourcefinding{Runtime operational and resource uncertainty emerges when execution conditions differ from design-time expectations. Changing workloads, operating modes, execution times, resource contention, and energy limitations can undermine the timeliness and dependability of monitoring and control.}{21/186 (11.7\%)}{Changing operational contexts; workload fluctuations; adaptive operating modes; execution-time variability; scheduling conflicts; resource contention; computational limitations; energy constraints}


\subsection{RQ3: Occurrence of Uncertainties in the CPS Lifecycle}
This RQ aims to map the occurrence of uncertainties onto the development and operation lifecycle of a CPS.

\subsubsection{Run-time uncertainty.}
Based on the reviewed articles mentioned in Table \ref{tab:rq3_lifecycle_distribution}, uncertainty in CPSs is most frequently reported during runtime/operation time. However, a significant portion of the literature also reports that uncertainty also occurs during development. 91 articles (48.7\%) analyzed discuss uncertainty during the runtime or operation stage; 35 articles (18.9\%) discuss it mainly with the development or testing stage; and 59 articles (31.9\%) discuss its presence in both development and runtime.
Therefore, while CPS uncertainty may also become evident at development time, it mostly occurs at run-time. 

\begin{table}[H]
\centering
\caption{Lifecycle stage at which uncertainty occurs in CPS articles}
\label{tab:rq3_lifecycle_distribution}
\begin{tabular}{lcc}
\toprule
\textbf{Lifecycle stage of uncertainty occurrence} & \textbf{No. of articles} & \textbf{\% of coded articles} \\
\midrule
Runtime / operation & 91 & 48.7\% \\
Development / testing & 35 & 18.9\% \\
Both development and runtime & 59 & 31.9\% \\
\midrule
\textbf{Total} & \textbf{185} & \textbf{100.0\%} \\
\bottomrule
\end{tabular}
\end{table}

More specifically, various articles show that although uncertainties arise during real-world operation, they can be anticipated, modeled, bounded, or introduced during earlier phases such as design, simulation, verification, and testing.
Real-world interactions with CPSs introduce sensor noise, disturbances, operator behavior, and unexpected environmental changes.  Khatiri et al.~\cite{khatiri2025uav_uncertainty} show that uncertainty is associated with faulty control signals during real-time flights. The study proposes a runtime detector to identify such behaviors early. This work describes runtime uncertainty, as observed during runtime rather than during the UAV development cycle.  Sarkar et al.~\cite{sarkar2025platooning} deal with vehicle slips, external disturbances, delays, and dynamic vehicle interactions during adaptive cruise control mode. These occur during live operation, requiring the control system to adapt to continuously changing conditions. Sridhar et al.~\cite{sridhar2024mind_wandering} showed that, in partially autonomous driving, uncertainty may arise when driver attention fluctuates during handover and braking situations. This operational uncertainty occurs because system behavior depends on the driver's real-time readiness to assume control. In this context, uncertainty is operational because it depends on the user’s real-time attention span, which directly influences the effectiveness of safety-critical control handovers. Similarly, Shao et al.~\cite{shao2025secure_control} investigated secure control of nonlinear CPSs subject to uncertain mode transitions, cyber-attacks, and nonlinear disturbances. This case demonstrates that runtime uncertainty may originate not only from the physical environment or human behavior, but also from unstable system modes, adversarial inputs, and uncertain nonlinear dynamics.

\subsubsection{Development-time uncertainty.}
Several articles also report uncertainty during the development stage. In these articles, uncertainty is discussed as a challenge resulting from modeling, simulation, safety analysis, and verification. Such uncertainties arise because developers cannot conceptualize and analyze all possible scenarios. Developers find it infeasible to consider all possible test cases, and, therefore, the simulators they develop cannot represent all real-world conditions. Tan et al.~\cite{tan2025remedy} analyzed the discovery of unknown and risky behaviors arising from model evolution and simulation in CARLA, and address the uncertainty developed during development to uncover behaviors that have not yet been adequately modeled or tested.  Biagiola et al.~\cite{biagiola2024digital_siblings} investigated uncertainty in simulation and behavioral variability across simulators, focusing on whether test results from one simulator can be generalized to other platforms. This represents a development and testing uncertainty. Bansal et al.~\cite{bansal2024perception_simplex} explicitly explored the unverifiability of DNN-based perception systems in autonomous vehicles, contrasting the challenges of verifying DNN-based perception with the methods available for classical LiDAR-based detection.

\subsubsection{Cross-lifecycle uncertainty.}
The literature also comprises a large body of articles in which uncertainty cannot be meaningfully assigned to a single life-cycle stage. These works discuss uncertainty that arises at both development time and runtime~\cite{wang2023_novel_model_dynamic,archibald2025bdi}, either because design-time assumptions are later challenged during operation or because runtime behavior can only be managed through models and mechanisms developed during development. This is an important finding because it shows that uncertainty in CPSs is often not phase-specific; rather, it spans the lifecycle. Related to that Sartaj et al.~\cite{sartaj2025envdt} showed that testing with digital twins without modeling uncertain environmental factors leaves important behaviors untested, while real devices operating in the field may drift out of sync with their digital twins. This means the uncertainty is addressed during development and simulation, but it also remains active during real operation.  Archibald et al.~\cite{archibald2025bdi} addressed runtime uncertainty due to sensor noise, probabilistic action outcomes, and nondeterministic plan selection, while also emphasizing model correctness, simulation, and verification during development. The uncertainty is therefore both operational and design-related. This kind of work clearly shows that, in AI-enabled CPSs, uncertainty cannot be cleanly separated into pre- or post-deployment concerns.

Overall, the findings shown in Table \ref{tab:rq3_mapping_normalized_clickable} emerging from the analyzed papers indicate that the life-cycle stage at which uncertainty is addressed depends on the specific problem under investigation. In the presence of disturbances, sensor noise, control instability, human state variability, communication delays, concept drift, or cyber-attacks, uncertainty is predominantly managed as a runtime phenomenon. In contrast, when the focus is on scenario exploration, simulation fidelity, formal verification, architecture design, controller synthesis, or learning model assurance, uncertainty is more often considered a development or testing concern. Numerous articles on CPSs ~\cite{an2020_modeling_verification_vehicles,ma2021_self_healing_reinforcement,caldas2024_verification_field_ros,ayerdi2025_marmot_metamorphic_monitoring,xu2023_anomaly_curriculum_learning} indicate that these perspectives are interconnected: development-phase assumptions influence what can be managed at runtime, while runtime observations reveal the limitations of design-time models and tests. Consequently, the evidence from the reviewed literature does not support the restrictive view that uncertainty is confined to a single lifecycle stage. Rather, it substantiates the broader conclusion that uncertainty in CPSs primarily arises at run-time/operation and is driven or influenced by development-phase decisions.

\begin{table}[ht]
\centering
\scriptsize
\setlength{\tabcolsep}{3pt}
\renewcommand{\arraystretch}{1.05}
\caption{Mapping of normalized lifecycle-stage categories to primary articles.}
\label{tab:rq3_mapping_normalized_clickable}
\begin{tabularx}{\textwidth}{@{}
>{\vspace{0pt}\raggedright\arraybackslash}p{0.28\textwidth}
>{\vspace{0pt}\raggedright\arraybackslash}X@{}}
\toprule
\textbf{Lifecycle-stage category} & \textbf{Used by} \\
\midrule

Runtime/operation uncertainty
&~\cite{khatiri2025uav_uncertainty}, \cite{vermaelen2025_evaluation_formal_versus}, \cite{sarkar2025platooning}, \cite{alsaedi2026rdsad}, \cite{shao2025secure_control}, \cite{wen2025_distributed_event_triggered}, \cite{castiglioni2024_stark_tool_cpss}, \cite{sridhar2024mind_wandering}, \cite{xu2024_pretrain_prompt_transfer}, \cite{lu2024_new_saturation_uncertain}, \cite{liu2024_probabilistic_dynamic_programming}, \cite{wu2024_mixed_textbraceleft_textbraceright}, \cite{yang2024_memory_distribution_shift}, \cite{mortlock2024_castnet_context_spatio}, \cite{wang2024_resilient_cpss_mixed}, \cite{qie2024_self_trajectory_prediction}, \cite{wei2024_multiarea_probabilistic_forecasting}, \cite{wang2024_quality_pharmaceutical_data}, \cite{kafunah2024_ensemble_combination_quality}, \cite{hong2024_internal_model_cooperative}, \cite{caldas2024_verification_field_ros}, \cite{ren2024_lateral_ground_vehicles}, \cite{chen2023_fast_online_planning}, \cite{langford2023_modalas_addressing_assurance}, \cite{vierhauser2023_grum_flexible_model}, \cite{pinciroli2023_performance_fault_tolerant}, \cite{weiss2022uncertainty}, \cite{mahmood2023_distributed_trust_management}, \cite{liu2023_distributed_antittack_fault}, \cite{han2023_robustness_assessment_industrial}, \cite{outkin2023_defender_policy_evaluation}, \cite{kaur2023_intelligent_fault_tolerant}, \cite{liu2023_sliding_mode_security}, \cite{mao2023_textsubscript1_simplex_velocity}, \cite{pal2022_emphhow_emphhard_risk}, \cite{liu2023_false_data_injection}, \cite{chai2023_multiphase_overtaking_maneuver}, \cite{balzereit2023_emphautoconf_emph_new}, \cite{lei2022_fdi_attack_edge}, \cite{pettet2022_hierarchical_planning_dynamic}, \cite{waga2022_model_bounded_monitoring}, \cite{lin2022_bayesian_deep_learning}, \cite{ren2022_interactive_learning_human}, \cite{nie2022_hypergraphical_real_multirobot}, \cite{rodrigues2022_architecture_mission_coordination}, \cite{zhang2022_adp_remote_networked}, \cite{liu2022_guided_optimal_automated}, \cite{ebrahimi2022_real_estimation_frameworks}, \cite{wang2022_operating_state_reconstruction}, \cite{chen2022_stability_assessment_secondary}, \cite{ramakrishna2022_efficient_out_distribution}, \cite{zhang2022_economic_driven_hierarchical}, \cite{myllyaho2022_misbehaviour_fault_tolerance}, \cite{zhou2021_dependable_real_workflows}, \cite{song2021_smart_collaborative_balancing}, \cite{chowdhury2021_trustworthiness_self_vehicles}, \cite{esterle2020_think_therefore_you}, \cite{fang2019_job_shop_smart}, \cite{rahimilarki2019_neural_network_fault}, \cite{khayatian2020_crossroads_intersection_management}, \cite{li2019_sustainable_user_home}, \cite{ma2019_modeling_foundations_executable}, \cite{chaves2019_fragility_dsverifier}, \cite{zhang2019_wise_test_modeling}, \cite{heinzemann2019_transactional_execution_hierarchical}, \cite{zhang2019_national_scale_traffic}, \cite{gerostathopoulos2019_tuning_self_adaptation}, \cite{usamentiaga2018_temperature_monitoring_electrical}, \cite{camara2018_reasoning_about_sensing}, \cite{du2018_pcssl_stochastic_extension}, \cite{zhang2018specifying}, \cite{ma2019_holistic_management_dependable}, \cite{martin2018_slats_simultaneous_localization}, \cite{wen2018_consensus_strategy_economic}, \cite{jiang2018_recursive_total_principle}, \cite{lu2017_distributed_secondary_voltage}, \cite{gerostathopoulos2017_strengthening_adaptation_meta}, \cite{martini2017_real_energy}, \cite{sanchez2024_automated_planning_temporal}, \cite{hashimoto2022_collaborative_rover_copter}, \cite{wang2025_conformal_temporal_logic}, \cite{valle2025_defining_generating_multi}, \cite{shin2018_hitecs_uml_profile}, \cite{bandyszak2020_orthogonal_modeling_engineering}, \cite{devries2021_monitoring_environmental_knowledge}, \cite{gerostathopoulos2016_self_adaptation_intensive}, \cite{bao2025_quantification_physics_informed}, \cite{biglari2022_model_validity_tolerance}, \cite{foldiak2024_probabilistic_graph_queries}, \cite{ali2015_test_evolving_modelling}, \cite{das2024_interpretable_latent_space} \\
\\
Development/testing uncertainty
&~\cite{tan2025remedy}, \cite{cresporodriguez2025_instance_space_vehicles}, \cite{yu2025_kernel_latent_state}, \cite{bansal2024perception_simplex}, \cite{biagiola2024digital_siblings}, \cite{neelofar2024_identifying_explaining_critical}, \cite{kochanthara2024_perception_automated_case}, \cite{manfredi2024_consensus_uncertain_multiagent}, \cite{wang2024_data_guaranteed_trajectory}, \cite{weiss2023_generating_detecting_true}, \cite{laurent2023_parameter_coverage}, \cite{yu2023_data_driven_parameterized}, \cite{stocco2023_mind_gap_transferability}, ~\cite{islam2023_intelligent_privacy_preservation}, \cite{cornejo2022_mutation_scalable_solutions}, \cite{mandrioli2022_self_probabilistic_guarantees}, \cite{alasmari2022_quantitative_verification_reduction}, \cite{bento2022_hybrid_particle_swarm}, \cite{belzner2021_synthesizing_policies_probabilistic}, \cite{wang2021_data_driven_quality}, \cite{kaigom2021_value_driven_robotic}, \cite{shin2021_specification_loop}, \cite{bertoa2020_incorporating_measurement_ocl}, \cite{wang2020_resilient_rmpc_polytopic}, \cite{wang2020_achieving_personalized_anonymity}, \cite{medina2019_designing_controller_image}, \cite{wang2019_dynamic_data_injection}, \cite{zhang2019_wise_test_case}, \cite{ma2019_self_healing_fragility}, \cite{joerger2018_improve_management_large}, \cite{galarraga2021_genetic_algorithm_industrial}, \cite{zhang2017_wise_test_ready}, \cite{ayerdi2025_marmot_metamorphic_monitoring}, \cite{li2017_stochastic_contracts_probabilistic}, \cite{cai2020_real_out_distribution} \\
\\
Both development and runtime uncertainty
&~\cite{zhao2025_behavioral_decision_making}, \cite{sartaj2025envdt}, \cite{archibald2025bdi}, \cite{sritriratanarak2025_games_rational_multi}, \cite{song2025_quantum_particle_swarm}, \cite{zhu2025_cloud_cruise_vehicles}, \cite{chen2025_neural_network_reinforcement}, \cite{zhu2025_observer_free_vehicles}, \cite{pinciroli2024_modeling_more_performance}, \cite{calinescu2024_controller_synthesis_deep}, \cite{tan2024_multistep_exploration_exploitation}, \cite{panahi2024_performance_automotive_efficient}, \cite{wang2024_mutual_active_learning}, \cite{vandinten2024_slow_furious_performance}, \cite{liu2024_leader_follower_formation}, \cite{song2024_credibility_adversarial_examples}, \cite{ahmed2024_spatio_temporal_deep}, \cite{huang2024_validating_emulation_cybersecurity}, \cite{mandrioli2024_stress_loops}, \cite{wang2023_novel_model_dynamic}, \cite{mani2023_machine_learning_resilience}, \cite{chen2023_filter_fuzzy_switched}, \cite{ghosh2023_mouldys_monitoring_presence}, \cite{chawla2023_deep_learning_denial}, \cite{xu2023_anomaly_curriculum_learning}, \cite{dong2023_non_line_sight}, \cite{guo2023_path_tracking_vehicles}, \cite{lian2023_resilient_cyberattack_input}, \cite{zampetti2023_continuous_integration_delivery}, \cite{kusherbaeva2023_multiobjective_data_driven}, \cite{hu2023_enhancing_output_feedback}, \cite{lu2023_learning_configurations_operating}, \cite{ruan2022_inertia_data_recovery}, \cite{camara2022_interaction_problem_self}, \cite{catak2022_prediction_validator_deep}, \cite{duman2020_factor_security_fos}, \cite{liu2022_resilient_distributed_optimal}, \cite{zhao2021_spatiotemporal_prediction_energy}, \cite{burevs2021_targeting_smart_confidence}, \cite{ma2021_self_healing_reinforcement}, \cite{afrasiabi2021_advanced_deep_learning}, \cite{an2020_modeling_verification_vehicles}, \cite{camilli2020_model_hypothesis_uncertain}, \cite{yu2019_event_sequential_prognosis}, \cite{kar2019_event_triggered_neural}, \cite{wang2018_deep_learning_interval}, \cite{zhang2017_heterogeneous_model_integration}, \cite{talusan2025_end_end_solution}, \cite{lu2023_evolutionary_computation_rl}, \cite{shin2018_test_case_prioritization}, \cite{wang2018_rsimplex_architecture_failures}, \cite{hartsell2021_resonate_risk_assessment}, \cite{zhou2024_building_nvidia_isaac}, \cite{cui2023_filtering_out_high}, \cite{lin2025_monitoring_learning_enabled}, \cite{ali2017_evaluation_mutation_crossover}, \cite{feng2021_improving_variational_autoencoder}, \cite{camilli2023_enforcing_resilience_equilibrium} \\
\\
\bottomrule
\end{tabularx}
\end{table}

\begin{tcolorbox}[
  colback=gray!10,
  colframe=gray!35,
  boxrule=0.6pt,
  arc=2pt,
  left=6pt,
  right=6pt,
  top=6pt,
  bottom=6pt,
  title=\textcolor{black}{\textbf{RQ3 Finding}},
  coltitle=black
]
Uncertainty in CPSs is not limited to a single life cycle stage. The majority of reviewed articles address uncertainty during runtime or operation (49.2\%), while approximately one-third of the literature considers both development and runtime stages (31.9\%). These findings suggest that although uncertainty is most frequently encountered during operation, it is often influenced, anticipated, or constrained earlier through development-time models, simulation, verification, and testing.
\end{tcolorbox}

\FloatBarrier

\subsection{RQ4: Mitigation Strategies for Uncertainties in CPS}
This RQ discusses the uncertainty-mitigating strategies documented in the literature. 
\label{subsec:RQ4}
Based on the papers we analyzed, uncertainty in CPSs can be addressed through several complementary mitigation strategies rather than a single approach. The literature presents methods that are active throughout the CPS life-cycle. Normalized coding of the reviewed papers shows some common categories of mitigation, \ie it include concepts such as optimization and resource management, simulation and digital twin-based analysis, security and resilience mechanisms, as mentioned in detail in Table~\ref{tab:rq4_clickable_mapping}.

\begin{table}[ht]
\centering
\scriptsize
\setlength{\tabcolsep}{3pt}
\renewcommand{\arraystretch}{1.05}
\caption{Mapping of normalized mitigation-strategy categories to primary articles.}
\label{tab:rq4_clickable_mapping}
\begin{tabularx}{\textwidth}{@{}
>{\vspace{0pt}\raggedright\arraybackslash}p{0.28\textwidth}
>{\vspace{0pt}\raggedright\arraybackslash}X@{}}
\toprule
\textbf{Mitigation strategy category} & \textbf{Used by} \\
\bottomrule
\end{tabularx}
\end{table}
\FloatBarrier

\textbf{Optimization and resource management.} This is the most common mitigation measure in the coded data set. Given this prevalence (72/186; 38.7\%), many papers have sought to reduce uncertainty by improving the allocation of system resources, communication, computation, and control actions in constrained and dynamic environments. Uncertainty is not always removed or eliminated directly, but its operational effects are minimized by increasing system selectivity, prioritizing and efficiency. Zhu et al.~\cite{zhu2025_cloud_cruise_vehicles} introduced prioritized data transmission, observer-based control, and a collision-avoidance constraint strategy to ensure safe vehicle operation in the presence of disturbances and network-related uncertainty. Crespo-Rodriguez et al.~\cite{cresporodriguez2025_instance_space_vehicles} addressed uncertainty by identifying critical scenario features and prioritizing high-value test cases, thereby directing testing efforts toward areas where uncertainty is most likely to reveal failures.
Collectively, these works achieveed mitigation by managing limited resources under uncertain conditions.

\textbf{Application of simulation, digital twins, and runtime models.} We found such mitigation strategies in 58 out of 186 studied papers (31.2\%). These approaches are specifically recommended when direct testing of real-world uncertainty is impractical, costly, or unsafe. Researchers employ simulated environments and model-based replicas to investigate uncertain conditions, anticipate hazardous behaviors, and enhance system preparedness prior to deployment or during operation. For example, Khatiri et al.~\cite{khatiri2025uav_uncertainty} used over 5,000 simulated UAV flights to quantify behavioral uncertainty and develop a runtime uncertainty detector that predicts unsafe states up to 50 seconds in advance. Sartaj et al.~\cite{sartaj2025envdt} presented an environment digital twin methodology that models uncertain environmental factors, simulates a range of uncertain scenarios, and improves test coverage for IoT healthcare applications while maintaining calibration with the physical device. Collectively, these works suggest that simulation-based mitigation strategies are used to identify uncertainty before it leads to operational failure.

\textbf{Security, privacy, and attack resilience.} 
Security, privacy, and attack-resilience mechanisms were identified in 43 of 186 reviewed papers (23.1\%). These approaches address uncertainty by detecting attacks, filtering manipulated data, securing communication, preserving privacy, and ensuring reliable CPS operation in adversarial environments. For example, Alsaedi et al.~\cite{alsaedi2026rdsad} addressed uncertainty through dynamic deviation detection and shift-aware model adaptation, which enables the detection mechanism to remain effective as the data stream evolves.  Zhu et al.~\cite{zhu2025_observer_free_vehicles} proposed a solution that integrates a dynamic encryption mechanism, an observer-based controller, and a chance-constrained collision-avoidance index to ensure safe operation in the presence of privacy leakage, non-Gaussian noise, and uncertain obstacles. Collectively, these works suggest that a substantial portion of the mitigation literature is dedicated to preserving CPS trustworthiness in adversarial or compromised environments.

\textbf{Probabilistic and Bayesian modeling} (40/186,21.5\%), frequently combined with uncertainty quantification and risk estimation, constitutes a recurring mitigation strategy. Rather than seeking to eliminate uncertainty, these approaches explicitly represent it and leverage this representation to inform safer decision-making. For example, Zhao et al.~\cite{zhao2025_behavioral_decision_making}
addressed uncertainty by using a Bayesian network for hazard-probability inference, a Bayesian decision-making model for selecting driving behavior, and formal safety verification. Archibald et al.~\cite{archibald2025bdi} used Markov Decision Processes, employing the quantitative verification and error checking. These cases illustrate that a central mitigation strategy in the CPS literature is to render uncertainty measurable, analyzable, and relevant to decision-making rather than to disregard it.


\textbf{Robust and adaptive control} (37/186, 19.9\%) has been considered an effective strategy for addressing uncertain dynamics in the literature. The primary objective is not to determine the precise source of uncertainty in advance, but rather to design controllers that maintain acceptable performance despite imperfect knowledge of the environment, vehicle dynamics, or process parameters.  Sarkar et al.~\cite{sarkar2025platooning} aimed at mitigating uncertainty through an adaptive time-delay control framework and a dual-rate adaptation law, which corrects switching-gain estimation and ensures bounded inter-vehicular states using Lyapunov stability analysis. Wen et al.~\cite{wen2025_distributed_event_triggered} employed an event-triggered scheme, neuro-adaptive control, a sparse-sample learning mechanism, and a low-pass filter to stabilize platooning under uncertain nonlinear vehicle dynamics and limited communication resources. Collectively, these works show that robust control remains among the most practical mitigation strategies in CPS, particularly when uncertainty cannot be eliminated and must instead be managed.

\textbf{State estimation and observer design} (26/186,14.0\%)  constitute a closely related category that frequently appears in the literature as a supporting mitigation mechanism. In CPSs affected by uncertainty, direct observation of all system states is often unreliable, necessitating the reconstruction of hidden or noisy states in a stable manner. Zhu et al.~\cite{zhu2025_observer_free_vehicles} employed an observer to manage unmeasurable states under uncertainty and communication constraints. Zhu et al.~\cite{zhu2025_cloud_cruise_vehicles} introduced a novel observer structure for noise-robust state estimation as part of its comprehensive safety mechanism. These examples demonstrate that observer design is widely adopted as an enabling mitigation strategy to support control, monitoring, and decision-making under imperfect information.

\textbf{Runtime monitoring and detection} (37/186,19.9\%). These strategies acknowledge that uncertainty in CPSs frequently arises during operation and cannot always be addressed entirely at the design stage. The corresponding mitigation approach involves continuous monitoring of system behavior, early detection of deviations, and initiation of supervisory actions to prevent uncertainty from leading to safety failures. Khatiri et al.~\cite{khatiri2025uav_uncertainty} introduced an early-warning supervisor that identifies uncertain behavior and predicts unsafe states in advance.  Yu et al.~\cite{yu2025_kernel_latent_state} proposed a kernel latent state-space model that enhances fault detection and identification in uncertain nonlinear industrial processes by capturing the most predictable dynamics in a concise representation. Collectively, these works indicate that runtime monitoring is a fundamental mitigation strategy in CPS, enabling systems to respond effectively when uncertainty becomes apparent during operation.

\textbf{Formal verification.} At the assurance level, several articles address uncertainty by employing formal verification, logic-based assurance, runtime assurance, and simplex-style architectures(34/186,18.3\%). These approaches are adopted when stronger guarantees are required beyond those provided by empirical testing. For example, Archibald et al.~\cite{archibald2025bdi} used model checking to verify mission and safety properties under uncertain beliefs and action outcomes. Bansal et al.~\cite{bansal2024perception_simplex} introduced an architecture that monitors deep neural network outputs, compares them with a verifiable classical detector, identifies inconsistencies, and initiates safe fail-over or corrective action as needed. Such mitigation strategies are significant because they recognize that uncertainty in learning-enabled CPSs may never be entirely eliminated, necessitating a reliable fallback mechanism to ensure safety.

\textbf{Safe planning, decision-making, uncertainty modeling, and traceability }(18/186,9.7\%).
Rather than emphasizing immediate control actions, these approaches prioritize system structuring to ensure explicit representation of uncertainty in both design and strategic behavior. Sritriratanarak and Garcia~\cite{sritriratanarak2025_games_rational_multi} proposed Rational Multi-Agent Decision-Making in Temporally Non-Deterministic Environments, which models the uncertain evolution of a CPS using probabilistic finite-state automata and applies strategy optimization in both collaborative and adversarial contexts, thereby enhancing decision-making safety under temporal non-determinism.

Overall, the evidence from the reviewed papers suggests that mitigation strategies for uncertainty in CPS are most effective when implemented as a layered combination rather than as a single technique. 

\begin{tcolorbox}[
  colback=gray!10,
  colframe=gray!35,
  boxrule=0.6pt,
  arc=2pt,
  left=6pt,
  right=6pt,
  top=6pt,
  bottom=6pt,
  title=\textcolor{black}{\textbf{RQ4 Finding}},
  coltitle=black
]
The literature highlights that uncertainty in CPSs is dealt with through a combination of mitigation strategies rather than a single solution. The most prevalent approaches include optimization and resource management (72/186, 38.7\%), simulation and digital twin methodologies (58/186, 31.2\%), and security and resilience mechanisms (43/186, 23.1\%). Overall, our findings suggest that uncertainty is typically managed, monitored, and bounded, rather than being entirely removed from the system.
\end{tcolorbox}

%% file: sections/discussion.tex
\section{Discussion and Implications}
\label{sec:discussion}
This section discusses common patterns observed across research question results and identifies open challenges and future research directions. The analysis emphasizes (i) the grouping of uncertainty types across domains, (ii) the evolution of uncertainty throughout different lifecycle stages, and (iii) the degree to which mitigation strategies conform to the identified uncertainty landscape.

\subsection{Uncertainty preferences and Intersections}
Within the reviewed literature, a consistent pattern is that uncertainty is addressed primarily when it undermines confidence in CPS behavior or poses risks to safety assurance. Specifically, studies examine uncertainty in the context of environmental variability, imperfect sensing, incomplete models, or runtime changes that complicate the prediction, verification, or control of system behavior. This observation highlights why the most dominant uncertainty type is safety and specification uncertainty. It also shows why it often co-occurs with uncertainties that destabilize assurance arguments in practice. For example, environmental variability and shifts in runtime or operational context. Thus, uncertainty is not simply a modeling challenge; it becomes most vital when it undermines the evidence supporting safe deployment.

A second key pattern is the coupling of uncertainty sources rather than their isolation. Multiple papers address various uncertainties simultaneously, highlighting that mitigation should be evaluated as a pipeline (\ie environment $\rightarrow$ sensing/data $\rightarrow$ model $\rightarrow$ runtime context $\rightarrow$ assurance claim) rather than as a series of independent fixes. The strongest co-occurrences in the dataset support this perspective: safety and specification uncertainty frequently co-occur with environmental and runtime context uncertainty, suggesting that assurance most often fails at the boundary between controlled development assumptions and uncontrolled operational conditions. This finding supports reporting practices that explicitly document uncertainty dependencies, rather than listing uncertainty types as independent factors.

Types of uncertainty are not spread evenly across different domains. Most articles focus on general CPS or cross-domain topics, often serving as “method carriers” that address uncertainty broadly and span several categories. On the other hand, articles on specific CPS domains, such as autonomous driving, energy systems, or industrial CPS, exhibit more focused uncertainty clusters. This suggests that uncertainty taxonomies are most helpful when combined with domain-specific reporting, such as identifying which uncertainties are most important in each domain and why.

\begin{tcolorbox}[
  colback=gray!10,
  colframe=gray!35,
  boxrule=0.6pt,
  arc=2pt,
  left=6pt,
  right=6pt,
  top=6pt,
  bottom=6pt,
  title=\textcolor{black}{\textbf{Implication}},
  coltitle=black
]

This survey supports future research on CPS uncertainty by presenting a multi-label mapping of uncertainty types, sources, lifecycle stages, and mitigation strategies across 186 studies. This work enables researchers to compare how uncertainties manifest individually or in combination, and to assess whether mitigation strategies target isolated uncertainties or interacting sources. Future studies should report uncertainty combinations more explicitly to enable more comparable and realistic evaluations.

\end{tcolorbox}

\subsection{Lifecycle placement: why uncertainty at runtime prevails and why it matters}
Lifecycle coding shows that uncertainty is most often addressed at runtime or in operations, with a noticeable proportion also spanning development, testing, and operational phases. This trend highlights that many CPS assurance methods cannot rely solely on pre-deployment validation, as fundamental assumptions may be invalidated in operation by environmental changes, data shifts, component degradation, or adversarial conditions.

The focus on runtime considerations clarifies why assurance-oriented uncertainty is predominant in RQ1. Safety and specification claims are not established only at design time; they are continually challenged at runtime as operating conditions change. Consequently, future research should conceptualize assurance as a continuous process that integrates pre-deployment and runtime evidence, monitoring mechanisms, and safety envelopes. Approaches that focus exclusively on improving modeling fidelity without incorporating runtime observability frequently fail to address the critical juncture at which uncertainty becomes operationally hazardous.

\begin{tcolorbox}[
  colback=gray!10,
  colframe=gray!35,
  boxrule=0.6pt,
  arc=2pt,
  left=6pt,
  right=6pt,
  top=6pt,
  bottom=6pt,
  title=\textcolor{black}{\textbf{Implication}},
  coltitle=black
]

Integrating runtime assurance artifacts, such as monitors, confidence estimators, safety envelopes, and anomaly detectors, into development-time verification and testing pipelines represents a promising approach. This integration is particularly important for learning-enabled CPSs, where distribution shifts frequently occur during deployment.

\end{tcolorbox}

\subsection{Mitigation landscape: alignment, fragmentation, and gaps}
The mitigation strategies found in the corpus fall into several main groups: runtime monitoring and supervision, robust or adaptive control and estimation, simulation or scenario-based testing, probabilistic or Bayesian reasoning, formal verification or model checking, learning-based mitigation, and security or resilience mechanisms. Two issues are especially important.

First, here is a partial alignment between types of uncertainty and their corresponding mitigation strategies. For instance, scenario or environment-driven uncertainty typically corresponds to scenario-based testing and simulation expansion. Dynamics or disturbance uncertainty aligns with robust or adaptive control, while adversarial or security uncertainty is addressed through resilience mechanisms. Safety or specification uncertainty is often managed using formal verification and monitoring.
The alignment discussed above highlights that the field comprises established tool-chains for addressing common uncertainty categories.

Second, there is an important fragmentation in the reported mitigation strategies. Numerous articles present mitigation measures in highly specialized terminology, which complicates cross-domain comparisons. Consequently, synthesis tables and road maps are valuable because they apply a standardized framework to diverse solutions.

Several infrequently reported uncertainties can have a significant impact in real-world deployments, particularly those related to simulation-to-deployment transfer and data or distribution uncertainty in learning-enabled CPSs. Although these uncertainties are often recognized, mitigation strategies are typically articulated in general terms, such as ``improve generalization'' or ``expand testing,'' rather than operationalized through measurable evidence-transfer criteria, deployment monitoring, and updated governance.

\begin{tcolorbox}[
  colback=gray!10,
  colframe=gray!35,
  boxrule=0.6pt,
  arc=2pt,
  left=6pt,
  right=6pt,
  top=6pt,
  bottom=6pt,
  title=\textcolor{black}{\textbf{Limitations}},
  coltitle=black
]

The field requires a more robust approach to ``end-to-end'' mitigation framing. Mitigation should be explicitly described in terms of the (i) specific uncertainty it addresses, (ii) the stage in the lifecycle at which it is applied, and (iii) the evidence it generates to substantiate assurance claims.

\end{tcolorbox} 

\subsection{Cross-cutting gaps and underexplored directions}
There are, more specifically, two cross-cutting gaps across domains:

The first one is Under-reporting of architecture and traceability. Most of the articles address uncertainty without specifying the architectural context. For example, the locations where uncertainty is sensed or estimated at the architectural level, decisions are made, and mitigation is implemented. This lack of clarity reduces reproducibility and complicates the interpretation of mitigation boundaries.

Secondly, Human-related uncertainty remains comparatively underrepresented in contrast to safety, reliability, and robustness, even though real-world CPS deployments are frequently socio-technical. This observation indicates a need to systematically integrate human uncertainty into assurance arguments and runtime supervision, rather than considering it a peripheral factor. 

Table~\ref{tab:next_steps_future_research} presents future research directions that are being identified through cross-RQ synthesis. Each item highlights a gap identified by linking (i) dominant uncertainty types and domain concentration (RQ1), (ii) the lifecycle stage where uncertainty arises (RQ3), and (iii) mitigation families and reporting practices documented in the literature (RQ4). The resulting roadmap prioritizes actions to enhance comparability (taxonomy, metrics, reporting), reinforce assurance under runtime uncertainty (continuous assurance, evidence transfer), and address interconnected uncertainty chains common in real CPS deployments (security/safety, timing/control, data shift).

\begin{table}[ht]
\centering
\footnotesize
\setlength{\tabcolsep}{5pt}
\renewcommand{\arraystretch}{1.00}
\caption{Next steps for future research (RQ1--RQ4 synthesis).}
\label{tab:next_steps_future_research}
\begin{tabularx}{\textwidth}{@{}>{\raggedright\arraybackslash}p{0.28\textwidth} X@{}}
\toprule
\multicolumn{2}{@{}l@{}}{\textbf{Next Steps for Future Research}}\\
\midrule

\textbf{(1) Report uncertainty intersections} &
Needs to move beyond documenting only marginal category counts and instead explicitly document co-occurrence chains, \ie environment  $\rightarrow$ sensing/data $\rightarrow$ model $\rightarrow$ runtime context $\rightarrow$ assurance. Require articles to explicitly mention, whether uncertainties are handled jointly vs. in isolation and why. \textit{(RQ1, RQ4)} \\
\midrule

\textbf{(2) Standardize uncertainty taxonomies and boundary rules} &
Adopt a shared glossary of normalized uncertainty categories and publish boundary criteria (e.g., environmental vs. runtime context; epistemic vs. aleatoric). Include a short mapping from paper terminology to normalized labels to improve comparability. \textit{(RQ1)} \\
\midrule
\textbf{(3) Standardize uncertainty-source identification and reporting} &	Establish a common taxonomy for sources of uncertainty and require articles to explicitly document the originating source, affected system component, induced uncertainty categories, and supporting evidence.\textit{(RQ2)} \\
\midrule

\textbf{(4) Investigate multi-source uncertainty propagation across CPS layers} & Modeling how sources interact and propagate through perception, decision-making, control, and assurance mechanisms. Future evaluations will quantify cascading effects and assess mitigation strategies under multiple simultaneous sources of uncertainty. \textit{(RQ2, RQ4)} \\
\midrule

\textbf{(5) Make lifecycle placement explicit and evaluation-aligned} &
For each addressed uncertainty, it is required to explicitly declare  whether it arises during development/testing, runtime/operation, or both, and align evaluation evidence accordingly. In simple terms, it is required to explicitly declare the life-cycle placement as a design choice. \textit{(RQ3, RQ4)} \\
\midrule

\textbf{(6) Shift from one-shot assurance to continuous assurance} &
Prioritize approaches that combine pre-deployment V\&V with runtime monitoring, safety envelopes, and fallback strategies. Define what runtime evidence is collected, how it updates confidence, and what triggers mitigation. \textit{(RQ3, RQ4)} \\
\midrule

\textbf{(7) Operationalize evidence transfer (simulation/testing $\rightarrow$ deployment)} &
Introduce measurable transfer criteria for sim-to-real and testing-to-deployment validity (e.g., coverage, distribution shift detection, assumption monitoring). Report deployment mismatch explicitly and provide governance for post-deployment updates. \textit{(RQ3, RQ4)} \\
\midrule

\textbf{(8) Treat data/distribution shift as a first-class safety concern in learning-enabled CPS} &
Standardize how out-of-distribution (OOD) conditions are detected and handled (monitoring + fallback) and report failure modes under shift. Link data uncertainty to assurance artifacts rather than reporting it only as a generic limitation. \textit{(RQ1, RQ4)} \\
\midrule

\textbf{(9) Strengthen coupled cyber--physical assurance (security $\leftrightarrow$ safety)} &
Evaluate security-driven uncertainty (e.g., FDI/DoS effects) in terms of its impact on safety/specification compliance, not only detection accuracy. Promote joint security--safety mitigation reporting (threat model, monitoring, recovery, safety degradation). \textit{(RQ1, RQ4)} \\
\midrule

\textbf{(10) Co-design communication/timing resilience with control objectives} &
Move from treating network/timing uncertainty as a constraint to explicit co-design of scheduling, communication, and control robustness. Report how timing losses propagate into physical risk and how mitigation bounds that risk. \textit{(RQ1, RQ4)} \\
\midrule

\textbf{(11) Expand human-in-the-loop uncertainty integration} &
Increase coverage of human preference, cognition, and interaction uncertainty and connect it to assurance (e.g., safe handover, operator trust, controllability). Use measurable human-factor signals and define how they influence runtime decisions. \textit{(RQ1, RQ4)} \\
\midrule

\textbf{(12) Standardize uncertainty-aware evaluation metrics and disclosures} &
Report uncertainty handling with comparable metrics (e.g., calibration, risk bounds, safety margin violations, runtime monitor performance) and disclose assumptions. Encourage reporting of effect sizes, confidence intervals, and evaluation logs for reproducibility. \textit{(RQ4)} \\
\midrule

\textbf{(13) Improve artifact and tooling support for reproducibility} &
Release coded datasets, mapping tables, and scripts (e.g., bubble chart generation) and publish analysis-ready files enabling independent replication of RQ counts and domain intersections. \textit{(RQ1, RQ3, RQ4)} \\
\bottomrule
\end{tabularx}
\end{table}

%% file: sections/threats.tex
\section{Threats to Validity}
\label{sec:threats}

This section discusses the study threats to validity.

Threats to \textbf{construct validity} concern the relationship between theory and observation. 
One threat is related to how articles have been classified along the investigated dimensions, namely, (1) how uncertainty types are defined, (2) how their data sources are identified, (3) the life-cycle manifestation, and (4) the mitigation strategies. As described in Section \ref{sec:synthesis}, all these dimensions were characterized by carefully scrutinizing the selected articles. The process of creating a taxonomy for each of these dimensions and classifying papers along the taxonomies may be larger, subjective, and error-prone.

Threats to \textbf{internal validity} concern factors internal to our study that may affect the identification and selection of primary studies. One possible threat is our reliance on a limited pool of digital libraries. Scopus was used as the primary bibliographic source because of its broad cross-publisher coverage, while the ACM Digital Library and IEEE Xplore were searched as complementary publisher-specific sources. Since these databases differ in their search syntax, indexing mechanisms, and retrieval behavior, the corresponding search strings were adapted to each platform and were therefore not strictly equivalent. This may have caused some relevant studies to be retrieved from one database but not from another. We mitigated this threat by using Scopus as the main source, merging the results from all databases, applying the same inclusion and exclusion criteria to all retrieved records, and focusing on software engineering venues whose publications are largely indexed by these sources. In addition, forward and backward snowballing were performed to identify potentially relevant studies missed by the database searches. Nevertheless, citation chasing may itself be subject to imprecision, since its effectiveness depends on the selected seed papers and on the completeness of the citation indexes. CitationChaser~\cite{Haddaway2022Citationchaser} was used to support the retrieval of both references and citing papers, after which the resulting candidates were manually assessed according to the same selection criteria.
Snowballing could also be subject to imprecision. For the forward snowballing, one author examined all references of the papers obtained from the query and selection criteria, while for the backward snowballing CitationChaser~\cite{Haddaway2022Citationchaser} was used. 

Threats to \textbf{external validity} concern the generalizability of our findings. They primarily stem from the fact that our study focuses on articles published in software engineering venues, as specified in Section \ref{sec:results}. While, as explained, this is intentional because we want to focus on software engineering aspects, it may exclude some relevant software engineering contributions that, given the interdisciplinary nature of CPS, appeared in other venues not considered. We have mitigated this threat by including some relevant CPS-specific and interdisciplinary venues, i.e., those in the last row of Table \ref{tab:selected_venues}.
Another threat concerns the limited timespan we considered, which, again, may have excluded some older yet still relevant contributions. However, the topic we studied is very technology-oriented, and therefore, considering older articles would have yielded irrelevant/obsolete findings, especially regarding the analysis of uncertainty sources (RQ2) and mitigation strategies (RQ4).

%% file: sections/related.tex
\section{Related Work}
\label{sec:related}

This section discusses other secondary studies about (i) software engineering for CPSs and (ii) uncertainty in CPSs.

The work most closely related to ours is the systematic literature review by Asmat \etal~\cite{asmat2023uncertainty}, which examines uncertainty handling approaches, tools, root causes, and evaluation metrics in CPSs. The review identifies uncertainty arising from environmental variability, imperfect plant models, hardware and software issues, communication limitations, and resource constraints. It also groups techniques for managing uncertainty, including robust design and uncertainty-aware modeling and analysis tools. 
Our work differs by starting from the uncertainty phenomena reported in primary studies and developing a multi-dimensional characterization of their types, sources, and affected CPS layers. Instead of focusing mainly on handling technologies and metrics, we provide a broader lifecycle-oriented mapping of CPS uncertainty. In particular, RQ$_3$ examines uncertainty identification and mitigation strategies at both development time and runtime, while RQ$_4$ investigates whether uncertainty is considered during normal operation or mainly in exceptional situations, such as failures, attacks, environmental or process disturbances, and safety-critical events reported in the reviewed studies.







Makelburg \etal~\cite{makelburg2026surveying} surveyed uncertainty representations in CPSs and proposed a unified conceptual model. Their model distinguishes uncertainty across cyber, physical, and platform components, considers autonomous and non-autonomous subsystems, and maps uncertainty to representation paradigms, including probabilistic, interval-based, fuzzy, and qualitative models. They also highlight inconsistent terminology and the limited treatment of autonomy in existing research.
Their work focuses primarily on how uncertainty is represented, whereas ours examines how uncertainty originates, propagates, is detected, and is mitigated across CPS domains. Thus, representation techniques constitute one dimension of our analysis rather than its principal outcome. 

Pinto~\cite{pinto2023analysis} provides a broad conceptual treatment of uncertainty in CPS analysis and design. The study covers uncertainty in the physical plants, sensors and actuators, software, communication, human interaction, and AI-enabled components. It also discusses design-time, verification-time, and runtime approaches.
This work is closely related to ours because it connects the control and software engineering perspectives and discusses uncertainty propagation across CPS layers and life-cycle phases. However, Pinto's work is a conceptual analysis of uncertainty rather than a synthesis of primary studies, as was done in our work. Therefore, our study complements Pinto's work by identifying and quantifying, on the existing literature, uncertainty types, sources, affected layers, propagation paths, and development and runtime management mechanisms.

Other secondary studies in the field of software engineering for CPSs exist, e.g., those related to testing~\cite{abbaspour2015survey,sadri2022survey}, anomaly detection and attacks~\cite{moriano2025adaptive}, or model-driven engineering~\cite{mohamed2021model}. All these papers, however, do not cover anything related to uncertainty.

%% file: sections/conclusion.tex
\section{Conclusion}
\label{sec:conclusion}

This literature survey synthesizes evidence from 186 primary studies to characterize how uncertainty is handled in CPSs. In particular, we (i) characterize the uncertainty types, (ii) identify what are the sources of CPS uncertainty and map them to the types, (iii) map uncertainty types across the CPS life-cycle, and (iv) identify uncertainty mitigation strategies foreseen by the analyzed articles. More specifically, the main findings of this survey can be summarized as follows:
\begin{itemize}
    \item Uncertainty in CPSs is multi-dimensional. The reviewed studies addressed 16 uncertainty categories, with prominent attention to safety/specification/testing uncertainty, environmental uncertainty, and model/epistemic uncertainty.
    \item The primary sources of uncertainty are environmental and physical-process factors, followed by data/sensing/perception and knowledge/model/or decision-related factors. These findings suggest that CPS uncertainty is significantly influenced by the interplay among the physical environment, incomplete information, and software-driven decision-making processes.
    \item Uncertainty is present throughout the CPS life cycle, with runtime uncertainty being particularly significant. Numerous studies focus on operational uncertainty, while others examine development-time uncertainty or both. These findings underscore the need for methodologies that integrate design-time verification with runtime monitoring and adaptation.
    \item Mitigation strategies in the literature are diverse yet remain fragmented. Approaches include optimization and resource management, simulation and digital twin methodology, and security/resilience mechanisms. These strategies are typically applied to specific types of uncertainty or particular domains, rather than being integrated into a comprehensive CPS assurance process.
    \end{itemize}

In contrast to previous conceptual analyses of uncertainty in CPSs, this survey offers an evidence-based synthesis of how uncertainty is addressed in the primary study literature. It extends earlier work by quantifying uncertainty types, sources, affected life-cycle phases, propagation concerns, and mitigation mechanisms. Consequently, the contribution encompasses not only a taxonomy but also a structured map that identifies areas of concentrated CPS uncertainty research and significant gaps. 

As a next step, we intend to expand the SLR findings by conducting semi-structured interviews with CPS researchers and practitioners. The aim of these interviews will be to validate the uncertainty categories, assess whether the identified mitigation strategies meet practical requirements, and identify uncertainty-handling challenges that may be underrepresented in the literature.

\section{Data Availability}
\label{sec:data}
A replication package containing the analyzed studies, working data sets, and scripts used to perform the data extraction is available online~\citep{replication}.

\section*{Acknowledgments}
This work is supported by the InnoGuard Marie Skłodowska-Curie Doctoral Network (Grant Agreement No. 101169233). Aitor Arrieta is part of the Systems and Software Engineering research group of Mondragon Unibertsitatea (IT1919-26), supported by the Department of Education, Universities and Research of the Basque Country. Aitor Arrieta is supported by the Spanish Ministry of Science, Innovation and Universities (project PID2023-152979OA-I00), funded by MCIU /AEI.

%% file: Biblio.bib
@article{Haddaway2022PRISMA2020,
  author  = {Haddaway, Neal R. and Page, Matthew J. and Pritchard, Chris C. and McGuinness, Luke A.},
  title   = {PRISMA2020: An R package and Shiny app for producing PRISMA 2020-compliant flow diagrams, with interactivity for optimised digital transparency and Open Synthesis},
  journal = {Campbell Systematic Reviews},
  year    = {2022},
  volume  = {18},
  number  = {2},
  pages   = {e1230},
  doi     = {10.1002/cl2.1230},
  url     = {https://doi.org/10.1002/cl2.1230}
}

@inproceedings{Wohlin14,
  author       = {Claes Wohlin},
  title        = {Guidelines for snowballing in systematic literature studies and a
                  replication in software engineering},
  booktitle    = {18th International Conference on Evaluation and Assessment in Software
                  Engineering, {EASE} '14, London, England, United Kingdom, May 13-14,
                  2014},
  pages        = {38:1--38:10},
  publisher    = {{ACM}},
  year         = {2014}
}

@inproceedings{abbaspour2015survey,
  title={A survey on testing for cyber physical system},
  author={Abbaspour Asadollah, Sara and Inam, Rafia and Hansson, Hans},
  booktitle={IFIP International Conference on Testing Software and Systems},
  pages={194--207},
  year={2015},
  organization={Springer}
}

@article{mohamed2021model,
  title={Model-driven engineering tools and languages for cyber-physical systems--a systematic literature review},
  author={Mohamed, Mustafa Abshir and Kardas, Geylani and Challenger, Moharram},
  journal={IEEE Access},
  volume={9},
  pages={48605--48630},
  year={2021},
  publisher={IEEE}
}

@article{sadri2022survey,
  title={Survey on test case generation, selection and prioritization for cyber-physical systems},
  author={Sadri-Moshkenani, Zahra and Bradley, Justin and Rothermel, Gregg},
  journal={Software Testing, Verification and Reliability},
  volume={32},
  number={1},
  pages={e1794},
  year={2022},
  publisher={Wiley Online Library}
}

@article{moriano2025adaptive,
  title={Adaptive anomaly detection for identifying attacks in cyber-physical systems: A systematic literature review},
  author={Moriano, Pablo and Hespeler, Steven C and Li, Mingyan and Mahbub, Maria},
  journal={Artificial Intelligence Review},
  volume={58},
  number={9},
  pages={283},
  year={2025},
  publisher={Springer}
}

@article{derler2011modeling,
  title={Modeling cyber--physical systems},
  author={Derler, Patricia and Lee, Edward A and Vincentelli, Alberto Sangiovanni},
  journal={Proceedings of the IEEE},
  volume={100},
  number={1},
  pages={13--28},
  year={2011},
  publisher={IEEE}
}

@article{KitchenhamBBTBL09,
  author       = {Barbara A. Kitchenham and
                  Pearl Brereton and
                  David Budgen and
                  Mark Turner and
                  John Bailey and
                  Stephen G. Linkman},
  title        = {Systematic literature reviews in software engineering - {A} systematic
                  literature review},
  journal      = {Inf. Softw. Technol.},
  volume       = {51},
  number       = {1},
  pages        = {7--15},
  year         = {2009}
}

@inproceedings{4519604,
  author={Lee, Edward A.},
  booktitle={2008 11th IEEE International Symposium on Object and Component-Oriented Real-Time Distributed Computing (ISORC)}, 
  title={Cyber Physical Systems: Design Challenges}, 
  year={2008},
  volume={},
  number={},
  pages={363-369},
  doi={10.1109/ISORC.2008.25}}

@inbook{inbook,
author = {Jadidi, Nastaran and Varmazyar, Mohsen},
year = {2023},
month = {08},
pages = {2089-2117},
title = {A Survey of Cyber-physical Systems Applications (2017–2022)},
isbn = {978-3-030-97939-3},
doi = {10.1007/978-3-030-97940-9_145}
}

@article{art,
author = {Webster, Jane and Watson, Richard},
year = {2002},
month = {06},
pages = {},
title = {Analyzing the Past to Prepare for the Future: Writing a Literature Review},
volume = {26},
journal = {MIS Quarterly},
doi = {10.2307/4132319}
}

@inproceedings{rajkumar2010cps,
  title={Cyber-Physical Systems: The Next Computing Revolution},
  author={Rajkumar, Ragunathan and Lee, Insup and Sha, Lui and Stankovic, John},
  booktitle={Proceedings of the 47th Design Automation Conference},
  pages={731--736},
  year={2010}
}

@incollection{baheti2011cps,
  title={Cyber-physical Systems},
  author={Baheti, Radhakisan and Gill, Helen},
  booktitle={The Impact of Control Technology},
  pages={161--166},
  year={2011},
  publisher={IEEE Control Systems Society}
}

@book{lee2017introduction,
  title={Introduction to Embedded Systems: A Cyber-Physical Systems Approach},
  author={Lee, Edward A. and Seshia, Sanjit A.},
  year={2017},
  publisher={MIT Press}
}

@article{derkiureghian2009aleatory,
  title={Aleatory or epistemic? Does it matter?},
  author={Der Kiureghian, Armen and Ditlevsen, Ove},
  journal={Structural Safety},
  volume={31},
  number={2},
  pages={105--112},
  year={2009},
  doi={10.1016/j.strusafe.2008.06.020}
}

@article{hullermeier2021aleatoric,
  title={Aleatoric and epistemic uncertainty in machine learning: an introduction to concepts and methods},
  author={H{\"u}llermeier, Eyke and Waegeman, Willem},
  journal={Machine Learning},
  volume={110},
  pages={457--506},
  year={2021},
  doi={10.1007/s10994-021-05946-3}
}

@article{zhang2018specifying,
  author  = {Zhang, M. and Yue, T. and Ali, S. and Selic, B. and Okariz, O. and Norgren, R. and Intxausti, K.},
  title   = {Specifying Uncertainty in Use Case Models},
  journal = {Journal of Systems and Software},
  year    = {2018},
  doi     = {10.1016/j.jss.2018.06.075}
}

@article{alfonso2020modeling,
  author = {Alfonso, I. and Sottet, J.-S. and Brimont, P. and Cabot, J.},
  title = {Modeling the obsolescence of models},
  journal = {Unknown},
  year = {2020}
}

@article{weiss2022uncertainty,
  author = {Weiss, M. and Tonella, P.},
  title = {Uncertainty quantification for deep neural networks: An empirical comparison and usage guidelines},
  journal = {Unknown},
  year = {2022}
}

@article{al2022guide,
  author = {Al, S. and others},
  title = {A Guide to Handling Uncertainty in Software Engineering},
  journal = {ACM Computing Surveys},
  year = {2022}
}

@article{khatiri2025uav_uncertainty,
  author  = {Khatiri, Sajad and Mohammadi Amin, Fatemeh and Panichella, Sebastiano and Tonella, Paolo},
  title   = {When uncertainty leads to unsafety: Empirical insights into the role of uncertainty in unmanned aerial vehicle safety},
  journal = {Empirical Software Engineering},
  year    = {2025},
  volume  = {30},
  number  = {6},
  pages   = {166},
  doi     = {10.1007/s10664-025-10697-z}
}

@article{tan2025remedy,
  author  = {Tan, Chao and Wang, Tiexin and Zhang, Man and Yue, Tao},
  title   = {Safety behavior abstraction and model evolution in autonomous driving},
  journal = {Software and Systems Modeling},
  year    = {2025},
  volume  = {24},
  number  = {3},
  pages   = {843--863},
  doi     = {10.1007/s10270-024-01261-2}
}

@article{sartaj2025envdt,
  author  = {Sartaj, Hassan and Ali, Shaukat and Gj{\o}by, Julie Marie},
  title   = {Uncertainty-aware environment simulation of medical devices digital twins},
  journal = {Software and Systems Modeling},
  year    = {2025},
  volume  = {24},
  number  = {3},
  pages   = {651--677},
  doi     = {10.1007/s10270-024-01223-8}
}

@article{archibald2025bdi,
  author  = {Archibald, Blair and Sevegnani, Michele and Xu, Mengwei},
  title   = {Modelling and verifying {BDI} agents under uncertainty},
  journal = {Science of Computer Programming},
  year    = {2025},
  volume  = {242},
  pages   = {103254},
  doi     = {10.1016/j.scico.2024.103254}
}

@article{sridhar2024mind_wandering,
  author  = {Sridhar, Harini and Huang, Gaojian and Thorpe, Adam and Oishi, Meeko M. K. and Pitts, Brandon J.},
  title   = {Characterizing the Effect of Mind Wandering on Braking Dynamics in Partially Autonomous Vehicles},
  journal = {ACM Transactions on Cyber-Physical Systems},
  year    = {2024},
  volume  = {8},
  number  = {3},
  pages   = {1--21},
  doi     = {10.1145/3653678}
}

@article{biagiola2024digital_siblings,
  author  = {Biagiola, Matteo and Stocco, Andrea and Riccio, Vincenzo and Tonella, Paolo},
  title   = {Two is better than one: Digital siblings to improve autonomous driving testing},
  journal = {Empirical Software Engineering},
  year    = {2024},
  volume  = {29},
  number  = {4},
  pages   = {72},
  doi     = {10.1007/s10664-024-10458-4}
}

@article{sarkar2025platooning,
  author = {Sarkar, Rajasree and Banerjee, Arunava and Hashemi, Ehsan},
  title = {Adaptive {{Time-Delay Control}} for {{Cooperative Platooning Considering Slip}}},
  journal = {ACM Transactions on Cyber-Physical Systems},
  year = 2025,
  month = jan,
  volume = {9},
  number = {1},
  pages = {1--22},
  issn = {2378-962X, 2378-9638},
  doi = {10.1145/3704738},
}

@article{alsaedi2026rdsad,
  author = {Alsaedi, Abdullah and Tari, Zahir and Mahmud, Redowan},
  title = {{{RDSAD}}: {{Robust Threat Detection}} in {{Evolving Data Streams}} via {{Adaptive Latent Dynamics}}},
  journal = {IEEE Transactions on Dependable and Secure Computing},
  year = 2026,
  month = jan,
  volume = {23},
  number = {1},
  pages = {1063--1080},
  issn = {1545-5971, 1941-0018, 2160-9209},
  doi = {10.1109/TDSC.2025.3614294},
}

@article{shao2025secure_control,
  author = {Shao, Xingchen and Xie, Xiang-Peng and Luan, Xiaoli},
  title = {Asynchronous {{Gain-Scheduling Secure Control}} of {{Nonlinear Cyber-Physical Systems Under Complex Transition Probabilities}}: {{A Dual-Domain Polynomial Framework}}},
  journal = {IEEE Transactions on Industrial Informatics},
  year = 2025,
  month = nov,
  volume = {21},
  number = {11},
  pages = {8258--8269},
  issn = {1551-3203, 1941-0050},
  doi = {10.1109/TII.2025.3586061},
}

@article{bansal2024perception_simplex,
  author = {Bansal, Ayoosh and Kim, Hunmin and Yu, Simon and Li, Bo and Hovakimyan, Naira and Caccamo, Marco and Sha, Lui},
  title = {Perception Simplex: {{Verifiable}} Collision Avoidance in Autonomous Vehicles amidst Obstacle Detection Faults},
  journal = {Software Testing, Verification and Reliability},
  year = 2024,
  month = sep,
  volume = {34},
  number = {6},
  pages = {e1879},
  issn = {0960-0833, 1099-1689},
  doi = {10.1002/stvr.1879},
}

@article{zhao2025_behavioral_decision_making,
  author = {Zhao, Ying and Zhu, Yi and Zhao, Li and Huang, Junge and Zhi, Qiang},
  title = {Behavioral Decision-Making and Safety Verification Approaches for Autonomous Driving System in Extreme Scenarios},
  journal = {Journal of Systems and Software},
  year = 2025,
  month = aug,
  volume = {226},
  pages = {112385},
  issn = {01641212},
  doi = {10.1016/j.jss.2025.112385},
}

@article{sritriratanarak2025_games_rational_multi,
  author = {Sritriratanarak, Warisa and Garcia, Paulo},
  title = {Cyber {{Physical Games}}: {{Rational Multi-Agent Decision-Making}} in {{Temporally Non-Deterministic Environments}}},
  journal = {ACM Transactions on Cyber-Physical Systems},
  year = 2025,
  month = apr,
  volume = {9},
  number = {2},
  pages = {1--21},
  issn = {2378-962X, 2378-9638},
  doi = {10.1145/3719659},
}

@article{vermaelen2025_evaluation_formal_versus,
  author = {Vermaelen, Jan and Holvoet, Tom},
  title = {An Empirical Evaluation of a Formal Approach versus Ad Hoc Implementations in Robot Behavior Planning},
  journal = {Science of Computer Programming},
  year = 2025,
  month = apr,
  volume = {241},
  pages = {103226},
  issn = {01676423},
  doi = {10.1016/j.scico.2024.103226},
}

@article{cresporodriguez2025_instance_space_vehicles,
  author = {{Crespo-Rodriguez}, Victor and {Neelofar} and Aleti, Aldeida and Turhan, Burak},
  title = {Instance {{Space Analysis}} of {{Testing}} of {{Autonomous Vehicles}} in {{Critical Scenarios}}},
  journal = {ACM Transactions on Software Engineering and Methodology},
  year = 2025,
  month = mar,
  volume = {34},
  number = {3},
  pages = {1--36},
  issn = {1049-331X, 1557-7392},
  doi = {10.1145/3699596},
}

@article{song2025_quantum_particle_swarm,
  author = {Song, Chaobin and Yue, Dong and Zhang, Bo and Hancke, Gerhard P. and Dou, Chunxia and Wang, Haiwen},
  title = {Quantum {{Particle Swarm Optimization-Based Robust Relay Power Allocation Strategy}} for {{Cyber-Physical Power System}}},
  journal = {IEEE Transactions on Industrial Informatics},
  year = 2025,
  month = nov,
  volume = {21},
  number = {11},
  pages = {8327--8338},
  issn = {1551-3203, 1941-0050},
  doi = {10.1109/TII.2025.3609044},
}

@article{zhu2025_cloud_cruise_vehicles,
  author = {Zhu, Kaiqun and Wang, Zidong and Ding, Derui and Hu, Jun and Dong, Hongli},
  title = {Cloud-{{Based Collision Avoidance Adaptive Cruise Control}} for {{Autonomous Vehicles Under External Disturbances With Token Bucket Shapers}}},
  journal = {IEEE Transactions on Industrial Informatics},
  year = 2025,
  month = nov,
  volume = {21},
  number = {11},
  pages = {8759--8769},
  issn = {1551-3203, 1941-0050},
  doi = {10.1109/TII.2025.3586068},
}

@article{yu2025_kernel_latent_state,
  author = {Yu, Jiaxin and Dong, Yining and Qin, S. Joe},
  title = {Kernel {{Latent State-Space Modeling}} for {{Nonlinear Dynamic Process Monitoring}}},
  journal = {IEEE Transactions on Industrial Informatics},
  year = 2025,
  month = sep,
  volume = {21},
  number = {9},
  pages = {6701--6711},
  issn = {1551-3203, 1941-0050},
  doi = {10.1109/TII.2025.3567382},
}

@article{chen2025_neural_network_reinforcement,
  author = {Chen, Zhengquan and Zhang, Zhiheng and Yan, Jiayuan and Zhong, Maiying and Lv, Lingling and Hou, Yandong},
  title = {Neural {{Network}}--{{Based Reinforcement Iterative Learning Fault Estimation Scheme}} for {{Nonlinear Uncertain Manipulator Systems With Time-Delay}}},
  journal = {IEEE Transactions on Industrial Informatics},
  year = 2025,
  month = sep,
  volume = {21},
  number = {9},
  pages = {7287--7298},
  issn = {1551-3203, 1941-0050},
  doi = {10.1109/TII.2025.3575123},
}

@article{zhu2025_observer_free_vehicles,
  author = {Zhu, Kaiqun and Wang, Zidong and Li, Zhenning and Xu, Cheng-Zhong},
  title = {Secure {{Observer-Based Collision-Free Control}} for {{Autonomous Vehicles Under Non-Gaussian Noises}}},
  journal = {IEEE Transactions on Industrial Informatics},
  year = 2025,
  month = mar,
  volume = {21},
  number = {3},
  pages = {2184--2193},
  issn = {1551-3203, 1941-0050},
  doi = {10.1109/TII.2024.3495763},
}

@article{wen2025_distributed_event_triggered,
  author = {Wen, Guanghui and Wan, Ying and Zhou, Jialing and Zheng, Dezhi and Chen, C. L. Philip},
  title = {Distributed {{Robust Event-Triggered Platooning Control}} of {{Connected Vehicles With Uncertain Dynamics}}: {{A Neuro-adaptive Approach}}},
  journal = {IEEE Transactions on Industrial Informatics},
  year = 2025,
  month = mar,
  volume = {21},
  number = {3},
  pages = {2570--2579},
  issn = {1551-3203, 1941-0050},
  doi = {10.1109/TII.2024.3508842},
}

@article{castiglioni2024_stark_tool_cpss,
  author = {Castiglioni, Valentina and Loreti, Michele and Tini, Simone},
  title = {Stark: {{A}} Tool for the Analysis of {{CPSs}} Robustness},
  journal = {Science of Computer Programming},
  year = 2024,
  month = sep,
  volume = {236},
  pages = {103134},
  issn = {01676423},
  doi = {10.1016/j.scico.2024.103134},
}

@article{pinciroli2024_modeling_more_performance,
  author = {Pinciroli, Riccardo and Smith, Connie U. and Trubiani, Catia},
  title = {Modeling More Software Performance Antipatterns in Cyber-Physical Systems},
  journal = {Software and Systems Modeling},
  year = 2024,
  month = aug,
  volume = {23},
  number = {4},
  pages = {1003--1023},
  issn = {1619-1366, 1619-1374},
  doi = {10.1007/s10270-023-01137-x},
}

@article{xu2024_pretrain_prompt_transfer,
  author = {Xu, Qinghua and Yue, Tao and Ali, Shaukat and Arratibel, Maite},
  title = {Pretrain, {{Prompt}}, and {{Transfer}}: {{Evolving Digital Twins}} for {{Time-to-Event Analysis}} in {{Cyber-Physical Systems}}},
  journal = {IEEE Transactions on Software Engineering},
  year = 2024,
  month = jun,
  volume = {50},
  number = {6},
  pages = {1464--1477},
  issn = {0098-5589, 1939-3520, 2326-3881},
  doi = {10.1109/TSE.2024.3388572},
}

@article{calinescu2024_controller_synthesis_deep,
  author = {Calinescu, Radu and Imrie, Calum and Mangal, Ravi and Rodrigues, Gena{\'i}na Nunes and P{\u a}s{\u a}reanu, Corina and Santana, Misael Alpizar and V{\'a}zquez, Gricel},
  title = {Controller {{Synthesis}} for {{Autonomous Systems With Deep-Learning Perception Components}}},
  journal = {IEEE Transactions on Software Engineering},
  year = 2024,
  month = jun,
  volume = {50},
  number = {6},
  pages = {1374--1395},
  issn = {0098-5589, 1939-3520, 2326-3881},
  doi = {10.1109/TSE.2024.3385378},
}

@article{lu2024_new_saturation_uncertain,
  author = {Lu, Sirong and Pan, Muxuan and Cheng, Ke and Sun, Qinqin},
  title = {A {{New Saturation Control}} for {{Uncertain System Using Diffeomorphism Approach}}: {{Application}} to {{Turbofan Engine}}},
  journal = {IEEE Transactions on Industrial Informatics},
  year = 2024,
  month = jun,
  volume = {20},
  number = {6},
  pages = {8861--8872},
  issn = {1551-3203, 1941-0050},
  doi = {10.1109/TII.2024.3376775},
}

@article{liu2024_probabilistic_dynamic_programming,
  author = {Liu, Jincan and Xie, Zhengchao and Zhao, Jing and Wong, Pak Kin},
  title = {Probabilistic {{Adaptive Dynamic Programming}} for {{Optimal Reliability-Critical Control With Fault Interruption Estimation}}},
  journal = {IEEE Transactions on Industrial Informatics},
  year = 2024,
  month = jun,
  volume = {20},
  number = {6},
  pages = {8524--8535},
  issn = {1551-3203, 1941-0050},
  doi = {10.1109/TII.2024.3369714},
}

@article{tan2024_multistep_exploration_exploitation,
  author = {Tan, Yuan and Yang, Jun and Chen, Wen-Hua and Li, Shihua},
  title = {Multistep {{Dual Control}} for {{Exploration}} and {{Exploitation}} in {{Autonomous Search With Convergence Guarantee}}},
  journal = {IEEE Transactions on Industrial Informatics},
  year = 2024,
  month = jun,
  volume = {20},
  number = {6},
  pages = {8207--8217},
  issn = {1551-3203, 1941-0050},
  doi = {10.1109/TII.2024.3366979},
}

@article{wu2024_mixed_textbraceleft_textbraceright,
  author = {Wu, Yuying and Zhang, Langwen and Xie, Wei and Wang, Bohui},
  title = {Robust {{Mixed}} \${{H}}\_\textbraceleft 2\textbraceright\$/\${{H}}\_\textbraceleft\textbackslash infty \textbraceright\$ {{Model Predictive Control}} for {{Cyber-Physical Systems With Input Saturation}} and {{Energy-Bounded Disturbance}}},
  journal = {IEEE Transactions on Industrial Informatics},
  year = 2024,
  month = jun,
  volume = {20},
  number = {6},
  pages = {8328--8337},
  issn = {1551-3203, 1941-0050},
  doi = {10.1109/TII.2024.3372036},
}

@article{yang2024_memory_distribution_shift,
  author = {Yang, Yahan and Kaur, Ramneet and Dutta, Souradeep and Lee, Insup},
  title = {Memory-Based {{Distribution Shift Detection}} for {{Learning Enabled Cyber-Physical Systems}} with {{Statistical Guarantees}}},
  journal = {ACM Transactions on Cyber-Physical Systems},
  year = 2024,
  month = apr,
  volume = {8},
  number = {2},
  pages = {1--28},
  issn = {2378-962X, 2378-9638},
  doi = {10.1145/3643892},
}

@article{mortlock2024_castnet_context_spatio,
  author = {Mortlock, Trier and Malawade, Arnav and Tsujio, Kohei and Al Faruque, Mohammad},
  title = {{{CASTNet}}: {{A Context-Aware}}, {{Spatio-Temporal Dynamic Motion Prediction Ensemble}} for {{Autonomous Driving}}},
  journal = {ACM Transactions on Cyber-Physical Systems},
  year = 2024,
  month = apr,
  volume = {8},
  number = {2},
  pages = {1--20},
  issn = {2378-962X, 2378-9638},
  doi = {10.1145/3648622},
}

@article{panahi2024_performance_automotive_efficient,
  author = {Panahi, Vahid and Kargahi, Mehdi and Faghih, Fathiyeh},
  title = {Control {{Performance Analysis}} of {{Automotive Cyber-physical Systems}}: {{A Study}} on {{Efficient Formal Verification}}},
  journal = {ACM Transactions on Cyber-Physical Systems},
  year = 2024,
  month = apr,
  volume = {8},
  number = {2},
  pages = {1--19},
  issn = {2378-962X, 2378-9638},
  doi = {10.1145/3576046},
}

@article{wang2024_resilient_cpss_mixed,
  author = {Wang, Hui-Ting and Zhang, Kunwu and He, Yong and Shi, Yang},
  title = {Resilient {{Control}} of {{CPSs Under Mixed DoS Attacks}}: {{A Type-Dependent ADT Approach}}},
  journal = {IEEE Transactions on Industrial Informatics},
  year = 2024,
  month = may,
  volume = {20},
  number = {5},
  pages = {7832--7841},
  issn = {1551-3203, 1941-0050},
  doi = {10.1109/TII.2024.3363091},
}

@article{neelofar2024_identifying_explaining_critical,
  author = {Neelofar, Neelofar and Aleti, Aldeida},
  title = {Identifying and {{Explaining Safety-critical Scenarios}} for {{Autonomous Vehicles}} via {{Key Features}}},
  journal = {ACM Transactions on Software Engineering and Methodology},
  year = 2024,
  month = may,
  volume = {33},
  number = {4},
  pages = {1--32},
  issn = {1049-331X, 1557-7392},
  doi = {10.1145/3640335},
}

@article{qie2024_self_trajectory_prediction,
  author = {Qie, Tianqi and Wang, Weida and Yang, Chao and Li, Ying},
  title = {A {{Self-Trajectory Prediction Approach}} for {{Autonomous Vehicles Using Distributed Decouple LSTM}}},
  journal = {IEEE Transactions on Industrial Informatics},
  year = 2024,
  month = apr,
  volume = {20},
  number = {4},
  pages = {6708--6717},
  issn = {1551-3203, 1941-0050},
  doi = {10.1109/TII.2024.3352231},
}

@article{wang2024_mutual_active_learning,
  author = {Wang, Lening and Wang, Xi and Ji, Qinglei and Wang, Lihui and Jin, Ran},
  title = {Mutual {{Active Learning}} for {{Engineering Regulated Statistical Digital Twin Models}}},
  journal = {IEEE Transactions on Industrial Informatics},
  year = 2024,
  month = apr,
  volume = {20},
  number = {4},
  pages = {6167--6176},
  issn = {1551-3203, 1941-0050},
  doi = {10.1109/TII.2023.3344134},
}

@article{vandinten2024_slow_furious_performance,
  author = {Van Dinten, Imara and Derakhshanfar, Pouria and Panichella, Annibale and Zaidman, Andy},
  title = {The Slow and the Furious? {{Performance}} Antipattern Detection in {{Cyber}}--{{Physical Systems}}},
  journal = {Journal of Systems and Software},
  year = 2024,
  month = apr,
  volume = {210},
  pages = {111904},
  issn = {01641212},
  doi = {10.1016/j.jss.2023.111904},
}

@article{kochanthara2024_perception_automated_case,
  author = {Kochanthara, Sangeeth and Singh, Tajinder and Forrai, Alexandru and Cleophas, Loek},
  title = {Safety of {{Perception Systems}} for {{Automated Driving}}: {{A Case Study}} on {{Apollo}}},
  journal = {ACM Transactions on Software Engineering and Methodology},
  year = 2024,
  month = mar,
  volume = {33},
  number = {3},
  pages = {1--28},
  issn = {1049-331X, 1557-7392},
  doi = {10.1145/3631969},
}

@article{liu2024_leader_follower_formation,
  author = {Liu, Zhao-Qing and Ge, Xiaohua and Xie, Hao and Han, Qing-Long and Zheng, Jinchuan and Wang, Yu-Long},
  title = {Secure {{Leader}}--{{Follower Formation Control}} of {{Networked Mobile Robots Under Replay Attacks}}},
  journal = {IEEE Transactions on Industrial Informatics},
  year = 2024,
  month = mar,
  volume = {20},
  number = {3},
  pages = {4149--4159},
  issn = {1551-3203, 1941-0050},
  doi = {10.1109/TII.2023.3313651},
}

@article{wei2024_multiarea_probabilistic_forecasting,
  author = {Wei, Shuheng and Wu, Zaijun and Xu, Junjun and Hu, Qinran},
  title = {Multiarea {{Probabilistic Forecasting-Aided Interval State Estimation}} for {{FDIA Identification}} in {{Power Distribution Networks}}},
  journal = {IEEE Transactions on Industrial Informatics},
  year = 2024,
  month = mar,
  volume = {20},
  number = {3},
  pages = {4271--4282},
  issn = {1551-3203, 1941-0050},
  doi = {10.1109/TII.2023.3321098},
}

@article{wang2024_quality_pharmaceutical_data,
  author = {Wang, Zhengsong and Tang, Shengnan and Guo, Ge and Yang, Yanqiu and He, Dakuo and Yang, Le and Han, Meng and Hou, Yue},
  title = {Adaptive {{Quality Control With Uncertainty}} for a {{Pharmaceutical Cyber-Physical System Based}} on {{Data}} and {{Knowledge Integration}}},
  journal = {IEEE Transactions on Industrial Informatics},
  year = 2024,
  month = mar,
  volume = {20},
  number = {3},
  pages = {3339--3350},
  issn = {1551-3203, 1941-0050},
  doi = {10.1109/TII.2023.3306355},
}

@article{song2024_credibility_adversarial_examples,
  author = {Song, Qun and Tan, Rui and Ren, Chao and Xu, Yan and Lou, Yang and Wang, Jianping and Gooi, Hoay Beng},
  title = {On {{Credibility}} of {{Adversarial Examples Against Learning-Based Grid Voltage Stability Assessment}}},
  journal = {IEEE Transactions on Dependable and Secure Computing},
  year = 2024,
  month = mar,
  volume = {21},
  number = {2},
  pages = {585--599},
  issn = {1545-5971, 1941-0018, 2160-9209},
  doi = {10.1109/TDSC.2022.3213012},
}

@article{ahmed2024_spatio_temporal_deep,
  author = {Ahmed, Arman and Basumallik, Sagnik and Gholami, Amir and Sadanandan, Sajan K. and Namaki, Mohammad H. N. and Srivastava, Anurag K. and Wu, Yinghui},
  title = {Spatio-{{Temporal Deep Graph Network}} for {{Event Detection}}, {{Localization}}, and {{Classification}} in {{Cyber-Physical Electric Distribution System}}},
  journal = {IEEE Transactions on Industrial Informatics},
  year = 2024,
  month = feb,
  volume = {20},
  number = {2},
  pages = {2397--2407},
  issn = {1551-3203, 1941-0050},
  doi = {10.1109/TII.2023.3290942},
}

@article{kafunah2024_ensemble_combination_quality,
  author = {Kafunah, Jefkine and Ali, Muhammad Intizar and Breslin, John G.},
  title = {Uncertainty-{{Aware Ensemble Combination Method}} for {{Quality Monitoring Fault Diagnosis}} in {{Safety-Related Products}}},
  journal = {IEEE Transactions on Industrial Informatics},
  year = 2024,
  month = feb,
  volume = {20},
  number = {2},
  pages = {1975--1986},
  issn = {1551-3203, 1941-0050},
  doi = {10.1109/TII.2023.3280566},
}

@article{manfredi2024_consensus_uncertain_multiagent,
  author = {Manfredi, Sabato},
  title = {Robust {{Consensus Design}} of {{Uncertain Multiagent Systems With Bounded Gains}} and {{Incremental Nonlinear Interactions}}},
  journal = {IEEE Transactions on Industrial Informatics},
  year = 2024,
  month = oct,
  volume = {20},
  number = {10},
  pages = {11844--11853},
  issn = {1551-3203, 1941-0050},
  doi = {10.1109/TII.2024.3413319},
}

@article{hong2024_internal_model_cooperative,
  author = {Hong, Yaxian and Su, Youfeng and Cai, He},
  title = {Internal {{Model Based Cooperative Robust Resilient Control Under DoS Attacks With Application}} to {{Vehicles Formation}}},
  journal = {IEEE Transactions on Industrial Informatics},
  year = 2024,
  month = nov,
  volume = {20},
  number = {11},
  pages = {13124--13134},
  issn = {1551-3203, 1941-0050},
  doi = {10.1109/TII.2024.3431094},
}

@article{caldas2024_verification_field_ros,
  author = {Caldas, Ricardo and Garc{\'i}a, Juan Antonio Pi{\~n}era and Schiopu, Matei and Pelliccione, Patrizio and Rodrigues, Gena{\'i}na and Berger, Thorsten},
  title = {Runtime {{Verification}} and {{Field-Based Testing}} for {{ROS-Based Robotic Systems}}},
  journal = {IEEE Transactions on Software Engineering},
  year = 2024,
  month = oct,
  volume = {50},
  number = {10},
  pages = {2544--2567},
  issn = {0098-5589, 1939-3520, 2326-3881},
  doi = {10.1109/TSE.2024.3444697},
}

@article{ren2024_lateral_ground_vehicles,
  author = {Ren, Yunshuai and Xie, Xiangpeng and Li, Yongfu},
  title = {Lateral {{Control}} of {{Autonomous Ground Vehicles}} via a {{New Homogeneous Polynomial Parameter Dependent-Type Fuzzy Controller}}},
  journal = {IEEE Transactions on Industrial Informatics},
  year = 2024,
  month = sep,
  volume = {20},
  number = {9},
  pages = {11233--11241},
  issn = {1551-3203, 1941-0050},
  doi = {10.1109/TII.2024.3403269},
}

@article{wang2024_data_guaranteed_trajectory,
  author = {Wang, Xudong and Zhang, Hui and Wang, Yuan and Wang, Yaonan and Dong, Hanlin and Li, Jitao},
  title = {Data-{{Based Guaranteed Trajectory Estimation}} for {{Unmanned Surface Vehicles}}},
  journal = {IEEE Transactions on Industrial Informatics},
  year = 2024,
  month = jul,
  volume = {20},
  number = {7},
  pages = {9793--9802},
  issn = {1551-3203, 1941-0050},
  doi = {10.1109/TII.2024.3390442},
}

@article{huang2024_validating_emulation_cybersecurity,
  author = {Huang, Hao and Wlazlo, Patrick and Sahu, Abhijeet and Walker, Adele and Goulart, Ana E. and Davis, Katherine R. and Swiler, Laura and Tarman, Thomas D. and Vugrin, Eric},
  title = {Validating an {{Emulation-Based Cybersecurity Model With}} a {{Physical Testbed}}},
  journal = {IEEE Transactions on Dependable and Secure Computing},
  year = 2024,
  month = jul,
  volume = {21},
  number = {4},
  pages = {2997--3011},
  issn = {1545-5971, 1941-0018, 2160-9209},
  doi = {10.1109/TDSC.2023.3321176},
}

@article{mandrioli2024_stress_loops,
  author = {Mandrioli, Claudio and Shin, Seung Yeob and Maggio, Martina and Bianculli, Domenico and Briand, Lionel},
  title = {Stress {{Testing Control Loops}} in {{Cyber-physical Systems}}},
  journal = {ACM Transactions on Software Engineering and Methodology},
  year = 2024,
  month = feb,
  volume = {33},
  number = {2},
  pages = {1--58},
  issn = {1049-331X, 1557-7392},
  doi = {10.1145/3624742},
}

@article{wang2023_novel_model_dynamic,
  author = {Wang, Lei and Luo, Zhengda and Tang, Hongtao and Guo, Shunsheng and Li, Xixing},
  title = {A {{Novel Model}} for {{Dynamic Manufacturing Service Collaboration}} on {{Industrial Internet}}},
  journal = {IEEE Transactions on Industrial Informatics},
  year = 2023,
  month = dec,
  volume = {19},
  number = {12},
  pages = {11788--11799},
  issn = {1551-3203, 1941-0050},
  doi = {10.1109/TII.2023.3252408},
}

@article{chen2023_fast_online_planning,
  author = {Chen, Yanjie and Liu, Jiangjiang and Lan, Limin and Zhang, Hui and Miao, Zhiqiang and Wang, Yaonan},
  title = {A {{Fast Online Planning Under Partial Observability Using Information Entropy Rewards}}},
  journal = {IEEE Transactions on Industrial Informatics},
  year = 2023,
  month = dec,
  volume = {19},
  number = {12},
  pages = {11596--11607},
  issn = {1551-3203, 1941-0050},
  doi = {10.1109/TII.2023.3248086},
}

@article{weiss2023_generating_detecting_true,
  author = {Weiss, Michael and G{\'o}mez, Andr{\'e} Garc{\'i}a and Tonella, Paolo},
  title = {Generating and Detecting True Ambiguity: A Forgotten Danger in {{DNN}} Supervision Testing},
  journal = {Empirical Software Engineering},
  year = 2023,
  month = nov,
  volume = {28},
  number = {6},
  pages = {146},
  issn = {1382-3256, 1573-7616},
  doi = {10.1007/s10664-023-10393-w},
}

@article{mani2023_machine_learning_resilience,
  author = {Mani, Ganapathy and Haliem, Marina and Bhargava, Bharat and Manickam, Indu and Kochpatcharin, Kevin and Kim, Myeongsu and Vugrin, Eric and Wang, Weichao and Jenkins, Chris and Angin, Pelin and Yu, Meng},
  title = {Machine {{Learning Based Resilience Testing}} of an {{Address Randomization Cyber Defense}}},
  journal = {IEEE Transactions on Dependable and Secure Computing},
  year = 2023,
  month = nov,
  volume = {20},
  number = {6},
  pages = {4853--4867},
  issn = {1545-5971, 1941-0018, 2160-9209},
  doi = {10.1109/TDSC.2023.3234561},
}

@article{langford2023_modalas_addressing_assurance,
  author = {Langford, Michael Austin and Chan, Kenneth H. and Fleck, Jonathon Emil and McKinley, Philip K. and Cheng, Betty H. C.},
  title = {{{MoDALAS}}: Addressing Assurance for Learning-Enabled Autonomous Systems in the Face of Uncertainty},
  journal = {Software and Systems Modeling},
  year = 2023,
  month = oct,
  volume = {22},
  number = {5},
  pages = {1543--1563},
  issn = {1619-1366, 1619-1374},
  doi = {10.1007/s10270-023-01090-9},
}

@article{chen2023_filter_fuzzy_switched,
  author = {Chen, Haiyang and Zong, Guangdeng and Zhao, Xudong and Gao, Fangzheng and Shi, Kaibo},
  title = {Secure {{Filter Design}} of {{Fuzzy Switched CPSs With Mismatched Modes}} and {{Application}}: {{A Multidomain Event-Triggered Strategy}}},
  journal = {IEEE Transactions on Industrial Informatics},
  year = 2023,
  month = oct,
  volume = {19},
  number = {10},
  pages = {10034--10044},
  issn = {1551-3203, 1941-0050},
  doi = {10.1109/TII.2022.3232768},
}

@article{vierhauser2023_grum_flexible_model,
  author = {Vierhauser, Michael and Garmendia, Antonio and Stadler, Marco and Wimmer, Manuel and {Cleland-Huang}, Jane},
  title = {{{GRuM}} --- {{A}} Flexible Model-Driven Runtime Monitoring Framework and Its Application to Automated Aerial and Ground Vehicles},
  journal = {Journal of Systems and Software},
  year = 2023,
  month = sep,
  volume = {203},
  pages = {111733},
  issn = {01641212},
  doi = {10.1016/j.jss.2023.111733},
}

@article{pinciroli2023_performance_fault_tolerant,
  author = {Pinciroli, Riccardo and Trubiani, Catia},
  title = {Performance {{Analysis}} of {{Fault-Tolerant Multiagent Coordination Mechanisms}}},
  journal = {IEEE Transactions on Industrial Informatics},
  year = 2023,
  month = sep,
  volume = {19},
  number = {9},
  pages = {9821--9832},
  issn = {1551-3203, 1941-0050},
  doi = {10.1109/TII.2023.3234606},
}

@article{ghosh2023_mouldys_monitoring_presence,
  author = {Ghosh, Bineet and Andr{\'e}, {\'E}tienne},
  title = {{{MoULDyS}}: {{Monitoring}} of Autonomous Systems in the Presence of Uncertainties},
  journal = {Science of Computer Programming},
  year = 2023,
  month = aug,
  volume = {230},
  pages = {102976},
  issn = {01676423},
  doi = {10.1016/j.scico.2023.102976},
}

@article{chawla2023_deep_learning_denial,
  author = {Chawla, Astha and Panigrahi, Bijaya Ketan and Bhalja, Bhavesh R.},
  title = {Deep-{{Learning-Based Denial-of-Service Resilient Framework}} for {{Wide Area Situational Awareness}} of {{Power Systems}}},
  journal = {IEEE Transactions on Industrial Informatics},
  year = 2023,
  month = aug,
  volume = {19},
  number = {8},
  pages = {9204--9216},
  issn = {1551-3203, 1941-0050},
  doi = {10.1109/TII.2022.3227726},
}

@article{xu2023_anomaly_curriculum_learning,
  author = {Xu, Qinghua and Ali, Shaukat and Yue, Tao},
  title = {Digital {{Twin-based Anomaly Detection}} with {{Curriculum Learning}} in {{Cyber-physical Systems}}},
  journal = {ACM Transactions on Software Engineering and Methodology},
  year = 2023,
  month = sep,
  volume = {32},
  number = {5},
  pages = {1--32},
  issn = {1049-331X, 1557-7392},
  doi = {10.1145/3582571},
}

@article{mahmood2023_distributed_trust_management,
  author = {Mahmood, Adnan and Sheng, Quan Z. and Zhang, Wei Emma and Wang, Yan and Sagar, Subhash},
  title = {Toward a {{Distributed Trust Management System}} for {{Misbehavior Detection}} in the {{Internet}} of {{Vehicles}}},
  journal = {ACM Transactions on Cyber-Physical Systems},
  year = 2023,
  month = jul,
  volume = {7},
  number = {3},
  pages = {1--25},
  issn = {2378-962X, 2378-9638},
  doi = {10.1145/3594637},
}

@article{dong2023_non_line_sight,
  author = {Dong, Mengyao and Qi, Yihong and Wang, Xianbin and Liu, Yiming},
  title = {A {{Non-Line-of-Sight Mitigation Method}} for {{Indoor Ultra-Wideband Localization With Multiple Walls}}},
  journal = {IEEE Transactions on Industrial Informatics},
  year = 2023,
  month = jul,
  volume = {19},
  number = {7},
  pages = {8183--8195},
  issn = {1551-3203, 1941-0050},
  doi = {10.1109/TII.2022.3217533},
}

@article{liu2023_distributed_antittack_fault,
  author = {Liu, Chun and Zhao, Jing and Patton, Ron J.},
  title = {Distributed {{Antittack Fault-Tolerant Tracking Control}} for {{Vehicle Platoon Systems Under Cyber-Physical Threats}}},
  journal = {IEEE Transactions on Industrial Informatics},
  year = 2023,
  month = jun,
  volume = {19},
  number = {6},
  pages = {7825--7834},
  issn = {1551-3203, 1941-0050},
  doi = {10.1109/TII.2022.3171343},
}

@article{han2023_robustness_assessment_industrial,
  author = {Han, Liping and Ali, Shaukat and Yue, Tao and Arrieta, Aitor and Arratibel, Maite},
  title = {Uncertainty-{{Aware Robustness Assessment}} of {{Industrial Elevator Systems}}},
  journal = {ACM Transactions on Software Engineering and Methodology},
  year = 2023,
  month = oct,
  volume = {32},
  number = {4},
  pages = {1--51},
  issn = {1049-331X, 1557-7392},
  doi = {10.1145/3576041},
}

@article{guo2023_path_tracking_vehicles,
  author = {Guo, Jinghua and Li, Lubin and Wang, Jingyao and Li, Keqiang},
  title = {Cyber-{{Physical System-Based Path Tracking Control}} of {{Autonomous Vehicles Under Cyber-Attacks}}},
  journal = {IEEE Transactions on Industrial Informatics},
  year = 2023,
  month = may,
  volume = {19},
  number = {5},
  pages = {6624--6635},
  issn = {1551-3203, 1941-0050},
  doi = {10.1109/TII.2022.3206354},
}

@article{lian2023_resilient_cyberattack_input,
  author = {Lian, Zhi and Shi, Peng and Lim, Cheng-Chew},
  title = {Adaptive {{Resilient Control}} for {{Cyber-Physical Systems Under Cyberattack}} and {{Input Saturation}}},
  journal = {IEEE Transactions on Industrial Informatics},
  year = 2023,
  month = may,
  volume = {19},
  number = {5},
  pages = {6513--6524},
  issn = {1551-3203, 1941-0050},
  doi = {10.1109/TII.2022.3198699},
}

@article{outkin2023_defender_policy_evaluation,
  author = {Outkin, Alexander V. and Schulz, Patricia V. and Schulz, Timothy and Tarman, Thomas D. and Pinar, Ali},
  title = {Defender {{Policy Evaluation}} and {{Resource Allocation With MITRE ATT}}\&{{CK Evaluations Data}}},
  journal = {IEEE Transactions on Dependable and Secure Computing},
  year = 2023,
  month = may,
  volume = {20},
  number = {3},
  pages = {1909--1926},
  issn = {1545-5971, 1941-0018, 2160-9209},
  doi = {10.1109/TDSC.2022.3165624},
}

@article{zampetti2023_continuous_integration_delivery,
  author = {Zampetti, Fiorella and Tamburri, Damian and Panichella, Sebastiano and Panichella, Annibale and Canfora, Gerardo and Di Penta, Massimiliano},
  title = {Continuous {{Integration}} and {{Delivery Practices}} for {{Cyber-Physical Systems}}: {{An Interview-Based Study}}},
  journal = {ACM Transactions on Software Engineering and Methodology},
  year = 2023,
  month = may,
  volume = {32},
  number = {3},
  pages = {1--44},
  issn = {1049-331X, 1557-7392},
  doi = {10.1145/3571854},
}

@article{laurent2023_parameter_coverage,
  author = {Laurent, Thomas and Klikovits, Stefan and Arcaini, Paolo and Ishikawa, Fuyuki and Ventresque, Anthony},
  title = {Parameter {{Coverage}} for {{Testing}} of {{Autonomous Driving Systems}} under {{Uncertainty}}},
  journal = {ACM Transactions on Software Engineering and Methodology},
  year = 2023,
  month = jul,
  volume = {32},
  number = {3},
  pages = {1--31},
  issn = {1049-331X, 1557-7392},
  doi = {10.1145/3550270},
}

@article{yu2023_data_driven_parameterized,
  author = {Yu, Handi and Li, Xin},
  title = {Data-{{Driven Parameterized Corner Synthesis}} for {{Efficient Validation}} of {{Perception Systems}} for {{Autonomous Driving}}},
  journal = {ACM Transactions on Cyber-Physical Systems},
  year = 2023,
  month = apr,
  volume = {7},
  number = {2},
  pages = {1--24},
  issn = {2378-962X, 2378-9638},
  doi = {10.1145/3571286},
}

@article{kaur2023_intelligent_fault_tolerant,
  author = {Kaur, Gagandeep and Chanak, Prasenjit},
  title = {An {{Intelligent Fault Tolerant Data Routing Scheme}} for {{Wireless Sensor Network-Assisted Industrial Internet}} of {{Things}}},
  journal = {IEEE Transactions on Industrial Informatics},
  year = 2023,
  month = apr,
  volume = {19},
  number = {4},
  pages = {5543--5553},
  issn = {1551-3203, 1941-0050},
  doi = {10.1109/TII.2022.3204560},
}

@article{stocco2023_mind_gap_transferability,
  author = {Stocco, Andrea and Pulfer, Brian and Tonella, Paolo},
  title = {Mind the {{Gap}}! {{A Study}} on the {{Transferability}} of {{Virtual Versus Physical-World Testing}} of {{Autonomous Driving Systems}}},
  journal = {IEEE Transactions on Software Engineering},
  year = 2023,
  month = apr,
  volume = {49},
  number = {4},
  pages = {1928--1940},
  issn = {0098-5589, 1939-3520, 2326-3881},
  doi = {10.1109/TSE.2022.3202311},
}

@article{kusherbaeva2023_multiobjective_data_driven,
  author = {Kusherbaeva, Victoria and Zhou, Nianjun},
  title = {Multiobjective {{Data-Driven Production Optimization With}} a {{Feedback Mechanism}}},
  journal = {IEEE Transactions on Industrial Informatics},
  year = 2023,
  month = apr,
  volume = {19},
  number = {4},
  pages = {5456--5464},
  issn = {1551-3203, 1941-0050},
  doi = {10.1109/TII.2022.3193286},
}

@article{liu2023_sliding_mode_security,
  author = {Liu, Zhen and Chen, Xinkai and Yu, Jinpeng},
  title = {Adaptive {{Sliding Mode Security Control}} for {{Stochastic Markov Jump Cyber-Physical Nonlinear Systems Subject}} to {{Actuator Failures}} and {{Randomly Occurring Injection Attacks}}},
  journal = {IEEE Transactions on Industrial Informatics},
  year = 2023,
  month = mar,
  volume = {19},
  number = {3},
  pages = {3155--3165},
  issn = {1551-3203, 1941-0050},
  doi = {10.1109/TII.2022.3181274},
}

@article{hu2023_enhancing_output_feedback,
  author = {Hu, Jianchen and Ding, Baocang and Zhang, Meng and Zhao, Jun and Xu, Zuhua and Pan, Hongguang},
  title = {Enhancing {{Output Feedback Robust MPC}} via {{Lexicographic Optimization}}},
  journal = {IEEE Transactions on Industrial Informatics},
  year = 2023,
  month = mar,
  volume = {19},
  number = {3},
  pages = {3068--3078},
  issn = {1551-3203, 1941-0050},
  doi = {10.1109/TII.2022.3174717},
}

@article{mao2023_textsubscript1_simplex_velocity,
  author = {Mao, Yanbing and Gu, Yuliang and Hovakimyan, Naira and Sha, Lui and Voulgaris, Petros},
  title = {{{S$L$}}{\textsubscript{1}} -{{Simplex}}: {{Safe Velocity Regulation}} of {{Self-Driving Vehicles}} in {{Dynamic}} and {{Unforeseen Environments}}},
  journal = {ACM Transactions on Cyber-Physical Systems},
  year = 2023,
  month = jan,
  volume = {7},
  number = {1},
  pages = {1--24},
  issn = {2378-962X, 2378-9638},
  doi = {10.1145/3564273},
}

@article{islam2023_intelligent_privacy_preservation,
  author = {Islam, Shafkat and Badsha, Shahriar and Sengupta, Shamik and Khalil, Ibrahim and Atiquzzaman, Mohammed},
  title = {An {{Intelligent Privacy Preservation Scheme}} for {{EV Charging Infrastructure}}},
  journal = {IEEE Transactions on Industrial Informatics},
  year = 2023,
  month = feb,
  volume = {19},
  number = {2},
  pages = {1238--1247},
  issn = {1551-3203, 1941-0050},
  doi = {10.1109/TII.2022.3203707},
}

@article{pal2022_emphhow_emphhard_risk,
  author = {Pal, Ranjan and Liu, Peihan and Lu, Taoan and Hua, Ed},
  title = {{\emph{How }}{{{\emph{Hard Is Cyber-risk Management}}}}{\emph{ in }}{{{\emph{IT}}}}{\emph{/}}{{{\emph{OT Systems}}}}{\emph{?}} {{A Theory}} to {{Classify}} and {{Conquer Hardness}} of {{Insuring ICSs}}},
  journal = {ACM Transactions on Cyber-Physical Systems},
  year = 2022,
  month = oct,
  volume = {6},
  number = {4},
  pages = {1--31},
  issn = {2378-962X, 2378-9638},
  doi = {10.1145/3568399},
}

@article{liu2023_false_data_injection,
  author = {Liu, Chensheng and He, Wangli and Deng, Ruilong and Tian, Yu-Chu and Du, Wenli},
  title = {False-{{Data-Injection-Enabled Network Parameter Modifications}} in {{Power Systems}}: {{Attack}} and {{Detection}}},
  journal = {IEEE Transactions on Industrial Informatics},
  year = 2023,
  month = jan,
  volume = {19},
  number = {1},
  pages = {177--188},
  issn = {1551-3203, 1941-0050},
  doi = {10.1109/TII.2022.3172688},
}

@article{chai2023_multiphase_overtaking_maneuver,
  author = {Chai, Runqi and Tsourdos, Antonios and Chai, Senchun and Xia, Yuanqing and Savvaris, Al and Chen, C. L. Philip},
  title = {Multiphase {{Overtaking Maneuver Planning}} for {{Autonomous Ground Vehicles Via}} a {{Desensitized Trajectory Optimization Approach}}},
  journal = {IEEE Transactions on Industrial Informatics},
  year = 2023,
  month = jan,
  volume = {19},
  number = {1},
  pages = {74--87},
  issn = {1551-3203, 1941-0050},
  doi = {10.1109/TII.2022.3168434},
}

@article{lu2023_learning_configurations_operating,
  author = {Lu, Chengjie and Shi, Yize and Zhang, Huihui and Zhang, Man and Wang, Tiexin and Yue, Tao and Ali, Shaukat},
  title = {Learning {{Configurations}} of {{Operating Environment}} of {{Autonomous Vehicles}} to {{Maximize}} Their {{Collisions}}},
  journal = {IEEE Transactions on Software Engineering},
  year = 2023,
  month = jan,
  volume = {49},
  number = {1},
  pages = {384--402},
  issn = {0098-5589, 1939-3520, 2326-3881},
  doi = {10.1109/TSE.2022.3150788},
}

@article{balzereit2023_emphautoconf_emph_new,
  author = {Balzereit, Kaja and Niggemann, Oliver},
  title = {{{{\emph{AutoConf}}}}{\emph{:}} {{New Algorithm}} for {{Reconfiguration}} of {{Cyber-Physical Production Systems}}},
  journal = {IEEE Transactions on Industrial Informatics},
  year = 2023,
  month = jan,
  volume = {19},
  number = {1},
  pages = {739--749},
  issn = {1551-3203, 1941-0050},
  doi = {10.1109/TII.2022.3146940},
}

@article{lei2022_fdi_attack_edge,
  author = {Lei, Wenxin and Pang, Zhibo and Wen, Hong and Hou, Wenjing and Han, Wen},
  title = {{{FDI Attack Detection}} at the {{Edge}} of {{Smart Grids Based}} on {{Classification}} of {{Predicted Residuals}}},
  journal = {IEEE Transactions on Industrial Informatics},
  year = 2022,
  month = dec,
  volume = {18},
  number = {12},
  pages = {9302--9311},
  issn = {1551-3203, 1941-0050},
  doi = {10.1109/TII.2022.3174159},
}

@article{pettet2022_hierarchical_planning_dynamic,
  author = {Pettet, Geoffrey and Mukhopadhyay, Ayan and Kochenderfer, Mykel J. and Dubey, Abhishek},
  title = {Hierarchical {{Planning}} for {{Dynamic Resource Allocation}} in {{Smart}} and {{Connected Communities}}},
  journal = {ACM Transactions on Cyber-Physical Systems},
  year = 2022,
  month = oct,
  volume = {6},
  number = {4},
  pages = {1--26},
  issn = {2378-962X, 2378-9638},
  doi = {10.1145/3502869},
}

@article{waga2022_model_bounded_monitoring,
  author = {Waga, Masaki and Andr{\'e}, {\'E}tienne and Hasuo, Ichiro},
  title = {Model-Bounded {{Monitoring}} of {{Hybrid Systems}}},
  journal = {ACM Transactions on Cyber-Physical Systems},
  year = 2022,
  month = oct,
  volume = {6},
  number = {4},
  pages = {1--26},
  issn = {2378-962X, 2378-9638},
  doi = {10.1145/3529095},
}

@article{ruan2022_inertia_data_recovery,
  author = {Ruan, Jiaqi and Liang, Gaoqi and Zhao, Junhua and Qiu, Jing and Dong, Zhao Yang},
  title = {An {{Inertia-Based Data Recovery Scheme}} for {{False Data Injection Attack}}},
  journal = {IEEE Transactions on Industrial Informatics},
  year = 2022,
  month = nov,
  volume = {18},
  number = {11},
  pages = {7814--7823},
  issn = {1551-3203, 1941-0050},
  doi = {10.1109/TII.2022.3146859},
}

@article{lin2022_bayesian_deep_learning,
  author = {Lin, Yan-Hui and Li, Gang-Hui},
  title = {A {{Bayesian Deep Learning Framework}} for {{RUL Prediction Incorporating Uncertainty Quantification}} and {{Calibration}}},
  journal = {IEEE Transactions on Industrial Informatics},
  year = 2022,
  month = oct,
  volume = {18},
  number = {10},
  pages = {7274--7284},
  issn = {1551-3203, 1941-0050},
  doi = {10.1109/TII.2022.3156965},
}

@article{ren2022_interactive_learning_human,
  author = {Ren, Yutian and Li, Guann-Pyng},
  title = {An {{Interactive}} and {{Adaptive Learning Cyber Physical Human System}} for {{Manufacturing With}} a {{Case Study}} in {{Worker Machine Interactions}}},
  journal = {IEEE Transactions on Industrial Informatics},
  year = 2022,
  month = oct,
  volume = {18},
  number = {10},
  pages = {6723--6732},
  issn = {1551-3203, 1941-0050},
  doi = {10.1109/TII.2022.3150795},
}

@article{cornejo2022_mutation_scalable_solutions,
  author = {Cornejo, Oscar and Pastore, Fabrizio and Briand, Lionel C.},
  title = {Mutation {{Analysis}} for {{Cyber-Physical Systems}}: {{Scalable Solutions}} and {{Results}} in the {{Space Domain}}},
  journal = {IEEE Transactions on Software Engineering},
  year = 2022,
  month = oct,
  volume = {48},
  number = {10},
  pages = {3913--3939},
  issn = {0098-5589, 1939-3520, 2326-3881},
  doi = {10.1109/TSE.2021.3107680},
}

@article{nie2022_hypergraphical_real_multirobot,
  author = {Nie, Zixiang and Chen, Kwang-Cheng},
  title = {Hypergraphical {{Real-Time Multirobot Task Allocation}} in a {{Smart Factory}}},
  journal = {IEEE Transactions on Industrial Informatics},
  year = 2022,
  month = sep,
  volume = {18},
  number = {9},
  pages = {6047--6056},
  issn = {1551-3203, 1941-0050},
  doi = {10.1109/TII.2021.3135297},
}

@article{rodrigues2022_architecture_mission_coordination,
  author = {Rodrigues, Gabriel and Caldas, Ricardo and Araujo, Gabriel and De Moraes, Vicente and Rodrigues, Gena{\'i}na and Pelliccione, Patrizio},
  title = {An Architecture for Mission Coordination of Heterogeneous Robots},
  journal = {Journal of Systems and Software},
  year = 2022,
  month = sep,
  volume = {191},
  pages = {111363},
  issn = {01641212},
  doi = {10.1016/j.jss.2022.111363},
}

@article{zhang2022_adp_remote_networked,
  author = {Zhang, Longjie and Chen, Yong and Li, Meng},
  title = {{{ADP-Based Remote Secure Control}} for {{Networked Control Systems Under Unknown Nonlinear Attacks}} in {{Sensors}} and {{Actuators}}},
  journal = {IEEE Transactions on Industrial Informatics},
  year = 2022,
  month = sep,
  volume = {18},
  number = {9},
  pages = {6003--6014},
  issn = {1551-3203, 1941-0050},
  doi = {10.1109/TII.2021.3137816},
}

@article{mandrioli2022_self_probabilistic_guarantees,
  author = {Mandrioli, Claudio and Maggio, Martina},
  title = {Testing {{Self-Adaptive Software With Probabilistic Guarantees}} on {{Performance Metrics}}: {{Extended}} and {{Comparative Results}}},
  journal = {IEEE Transactions on Software Engineering},
  year = 2022,
  month = sep,
  volume = {48},
  number = {9},
  pages = {3554--3572},
  issn = {0098-5589, 1939-3520, 2326-3881},
  doi = {10.1109/TSE.2021.3101130},
}

@article{camara2022_interaction_problem_self,
  author = {C{\'a}mara, Javier and Troya, Javier and Vallecillo, Antonio and Bencomo, Nelly and Calinescu, Radu and Cheng, Betty H. C. and Garlan, David and Schmerl, Bradley},
  title = {The Uncertainty Interaction Problem in Self-Adaptive Systems},
  journal = {Software and Systems Modeling},
  year = 2022,
  month = aug,
  volume = {21},
  number = {4},
  pages = {1277--1294},
  issn = {1619-1366, 1619-1374},
  doi = {10.1007/s10270-022-01037-6},
}

@article{liu2022_guided_optimal_automated,
  author = {Liu, Wei and Zhu, Huainian and Yuan, Haoliang},
  title = {Uncertainty-{{Guided Optimal Scheduling}} of {{Automated Energy Internet}}},
  journal = {IEEE Transactions on Industrial Informatics},
  year = 2022,
  month = aug,
  volume = {18},
  number = {8},
  pages = {5662--5669},
  issn = {1551-3203, 1941-0050},
  doi = {10.1109/TII.2021.3126325},
}

@article{catak2022_prediction_validator_deep,
  author = {Catak, Ferhat Ozgur and Yue, Tao and Ali, Shaukat},
  title = {Uncertainty-Aware {{Prediction Validator}} in {{Deep Learning Models}} for {{Cyber-physical System Data}}},
  journal = {ACM Transactions on Software Engineering and Methodology},
  year = 2022,
  month = oct,
  volume = {31},
  number = {4},
  pages = {1--31},
  issn = {1049-331X, 1557-7392},
  doi = {10.1145/3527451},
}

@article{ebrahimi2022_real_estimation_frameworks,
  author = {Ebrahimi, Mehrdad and Rastegar, Mohammad and Arefi, Mohammad Mehdi},
  title = {Real-{{Time Estimation Frameworks}} for {{Feeder-Level Load Disaggregation}} and {{PEVs}}' {{Charging Behavior Characteristics Extraction}}},
  journal = {IEEE Transactions on Industrial Informatics},
  year = 2022,
  month = jul,
  volume = {18},
  number = {7},
  pages = {4715--4724},
  issn = {1551-3203, 1941-0050},
  doi = {10.1109/TII.2021.3118101},
}

@article{alasmari2022_quantitative_verification_reduction,
  author = {Alasmari, Naif and Calinescu, Radu and Paterson, Colin and Mirandola, Raffaela},
  title = {Quantitative Verification with Adaptive Uncertainty Reduction},
  journal = {Journal of Systems and Software},
  year = 2022,
  month = jun,
  volume = {188},
  pages = {111275},
  issn = {01641212},
  doi = {10.1016/j.jss.2022.111275},
}

@article{wang2022_operating_state_reconstruction,
  author = {Wang, Huaizhi and Wen, Xichang and Xu, Yinliang and Zhou, Bin and Peng, Jianchun and Liu, Wenxin},
  title = {Operating {{State Reconstruction}} in {{Cyber Physical Smart Grid}} for {{Automatic Attack Filtering}}},
  journal = {IEEE Transactions on Industrial Informatics},
  year = 2022,
  month = may,
  volume = {18},
  number = {5},
  pages = {2909--2922},
  issn = {1551-3203, 1941-0050},
  doi = {10.1109/TII.2020.3000172},
}

@article{chen2022_stability_assessment_secondary,
  author = {Chen, Chunyu and Zhang, Xiao and Cui, Mingjian and Zhang, Kaifeng and Zhao, Junbo and Li, Fangxing},
  title = {Stability {{Assessment}} of {{Secondary Frequency Control System With Dynamic False Data Injection Attacks}}},
  journal = {IEEE Transactions on Industrial Informatics},
  year = 2022,
  month = may,
  volume = {18},
  number = {5},
  pages = {3224--3234},
  issn = {1551-3203, 1941-0050},
  doi = {10.1109/TII.2021.3107539},
}

@article{ramakrishna2022_efficient_out_distribution,
  author = {Ramakrishna, Shreyas and Rahiminasab, Zahra and Karsai, Gabor and Easwaran, Arvind and Dubey, Abhishek},
  title = {Efficient {{Out-of-Distribution Detection Using Latent Space}} of {\emph{{$\beta$}}} -{{VAE}} for {{Cyber-Physical Systems}}},
  journal = {ACM Transactions on Cyber-Physical Systems},
  year = 2022,
  month = apr,
  volume = {6},
  number = {2},
  pages = {1--34},
  issn = {2378-962X, 2378-9638},
  doi = {10.1145/3491243},
}

@article{zhang2022_economic_driven_hierarchical,
  author = {Zhang, Zhijun and Zhang, Yudi and Yue, Dong and Dou, Chunxia and Ding, Xiaohua and Zhang, Huifeng},
  title = {Economic-{{Driven Hierarchical Voltage Regulation}} of {{Incremental Distribution Networks}}: {{A Cloud-Edge Collaboration Based Perspective}}},
  journal = {IEEE Transactions on Industrial Informatics},
  year = 2022,
  month = mar,
  volume = {18},
  number = {3},
  pages = {1746--1757},
  issn = {1551-3203, 1941-0050},
  doi = {10.1109/TII.2021.3085670},
}

@article{duman2020_factor_security_fos,
  author = {Duman, Onur and Zhang, Mengyuan and Wang, Lingyu and Debbabi, Mourad and Atallah, Ribal and Lebel, Bernard},
  title = {Factor of {{Security}} ({{FoS}}): {{Quantifying}} the {{Security Effectiveness}} of {{Redundant Smart Grid Subsystems}}},
  journal = {IEEE Transactions on Dependable and Secure Computing},
  year = 2020,
  pages = {1--1},
  issn = {1545-5971, 1941-0018, 2160-9209},
  doi = {10.1109/TDSC.2020.3009931},
}

@article{myllyaho2022_misbehaviour_fault_tolerance,
  author = {Myllyaho, Lalli and Raatikainen, Mikko and M{\"a}nnist{\"o}, Tomi and Nurminen, Jukka K. and Mikkonen, Tommi},
  title = {On Misbehaviour and Fault Tolerance in Machine Learning Systems},
  journal = {Journal of Systems and Software},
  year = 2022,
  month = jan,
  volume = {183},
  pages = {111096},
  issn = {01641212},
  doi = {10.1016/j.jss.2021.111096},
}

@article{liu2022_resilient_distributed_optimal,
  author = {Liu, Yun and Li, Yuanzheng and Wang, Yu and Zhang, Xian and Gooi, Hoay Beng and Xin, Huanhai},
  title = {Robust and {{Resilient Distributed Optimal Frequency Control}} for {{Microgrids Against Cyber Attacks}}},
  journal = {IEEE Transactions on Industrial Informatics},
  year = 2022,
  month = jan,
  volume = {18},
  number = {1},
  pages = {375--386},
  issn = {1551-3203, 1941-0050},
  doi = {10.1109/TII.2021.3071753},
}

@article{bento2022_hybrid_particle_swarm,
  author = {Bento, Murilo E. C.},
  title = {A {{Hybrid Particle Swarm Optimization Algorithm}} for the {{Wide-Area Damping Control Design}}},
  journal = {IEEE Transactions on Industrial Informatics},
  year = 2022,
  month = jan,
  volume = {18},
  number = {1},
  pages = {592--599},
  issn = {1551-3203, 1941-0050},
  doi = {10.1109/TII.2021.3054846},
}

@article{zhao2021_spatiotemporal_prediction_energy,
  author = {Zhao, Jun and Liu, Yang and Pedrycz, Witold and Wang, Wei},
  title = {Spatiotemporal {{Prediction}} for {{Energy System}} of {{Steel Industry}} by {{Generalized Tensor Granularity Based Evolving Type-2 Fuzzy Neural Network}}},
  journal = {IEEE Transactions on Industrial Informatics},
  year = 2021,
  month = dec,
  volume = {17},
  number = {12},
  pages = {7933--7945},
  issn = {1551-3203, 1941-0050},
  doi = {10.1109/TII.2021.3062036},
}

@article{burevs2021_targeting_smart_confidence,
  author = {Bure{\v s}, Tom{\'a}{\v s} and Hn{\v e}tynka, Petr and Pl{\'a}{\v s}il, Franti{\v s}ek and {\v S}koda, Dominik and Kofro{\v n}, Jan and Al Ali, Rima and Gerostathopoulos, Ilias},
  title = {Targeting Uncertainty in Smart {{CPS}} by Confidence-Based Logic},
  journal = {Journal of Systems and Software},
  year = 2021,
  month = nov,
  volume = {181},
  pages = {111065},
  issn = {01641212},
  doi = {10.1016/j.jss.2021.111065},
}

@article{zhou2021_dependable_real_workflows,
  author = {Zhou, Junlong and Sun, Jin and Zhang, Mingyue and Ma, Yue},
  title = {Dependable {{Scheduling}} for {{Real-Time Workflows}} on {{Cyber}}--{{Physical Cloud Systems}}},
  journal = {IEEE Transactions on Industrial Informatics},
  year = 2021,
  month = nov,
  volume = {17},
  number = {11},
  pages = {7820--7829},
  issn = {1551-3203, 1941-0050},
  doi = {10.1109/TII.2020.3011506},
}

@article{song2021_smart_collaborative_balancing,
  author = {Song, Fei and Ai, Zhengyang and Zhang, Haowei and You, Ilsun and Li, Shiyong},
  title = {Smart {{Collaborative Balancing}} for {{Dependable Network Components}} in {{Cyber-Physical Systems}}},
  journal = {IEEE Transactions on Industrial Informatics},
  year = 2021,
  month = oct,
  volume = {17},
  number = {10},
  pages = {6916--6924},
  issn = {1551-3203, 1941-0050},
  doi = {10.1109/TII.2020.3029766},
}

@article{belzner2021_synthesizing_policies_probabilistic,
  author = {Belzner, Lenz and Wirsing, Martin},
  title = {Synthesizing Safe Policies under Probabilistic Constraints with Reinforcement Learning and {{Bayesian}} Model Checking},
  journal = {Science of Computer Programming},
  year = 2021,
  month = jun,
  volume = {206},
  pages = {102620},
  issn = {01676423},
  doi = {10.1016/j.scico.2021.102620},
}

@article{ma2021_self_healing_reinforcement,
  author = {Ma, Tao and Ali, Shaukat and Yue, Tao},
  title = {Testing Self-Healing Cyber-Physical Systems under Uncertainty with Reinforcement Learning: An Empirical Study},
  journal = {Empirical Software Engineering},
  year = 2021,
  month = may,
  volume = {26},
  number = {3},
  pages = {52},
  issn = {1382-3256, 1573-7616},
  doi = {10.1007/s10664-021-09941-z},
}

@article{wang2021_data_driven_quality,
  author = {Wang, Zhengsong and He, Dakuo and Hou, Yue},
  title = {Data-{{Driven Adaptive Quality Control Under Uncertain Conditions}} for a {{Cyber-Pharmaceutical-Development System}}},
  journal = {IEEE Transactions on Industrial Informatics},
  year = 2021,
  month = may,
  volume = {17},
  number = {5},
  pages = {3165--3175},
  issn = {1551-3203, 1941-0050},
  doi = {10.1109/TII.2020.3004132},
}

@article{kaigom2021_value_driven_robotic,
  author = {Kaigom, Eric Guiffo and Rosmann, Jurgen},
  title = {Value-{{Driven Robotic Digital Twins}} in {{Cyber}}--{{Physical Applications}}},
  journal = {IEEE Transactions on Industrial Informatics},
  year = 2021,
  month = may,
  volume = {17},
  number = {5},
  pages = {3609--3619},
  issn = {1551-3203, 1941-0050},
  doi = {10.1109/TII.2020.3011062},
}

@article{chowdhury2021_trustworthiness_self_vehicles,
  author = {Chowdhury, Abdullahi and Karmakar, Gour and Kamruzzaman, Joarder and Islam, Syed},
  title = {Trustworthiness of {{Self-Driving Vehicles}} for {{Intelligent Transportation Systems}} in {{Industry Applications}}},
  journal = {IEEE Transactions on Industrial Informatics},
  year = 2021,
  month = feb,
  volume = {17},
  number = {2},
  pages = {961--970},
  issn = {1551-3203, 1941-0050},
  doi = {10.1109/TII.2020.2987431},
}

@article{afrasiabi2021_advanced_deep_learning,
  author = {Afrasiabi, Mousa and Mohammadi, Mohammad and Rastegar, Mohammad and Afrasiabi, Shahabodin},
  title = {Advanced {{Deep Learning Approach}} for {{Probabilistic Wind Speed Forecasting}}},
  journal = {IEEE Transactions on Industrial Informatics},
  year = 2021,
  month = jan,
  volume = {17},
  number = {1},
  pages = {720--727},
  issn = {1551-3203, 1941-0050},
  doi = {10.1109/TII.2020.3004436},
}

@article{shin2021_specification_loop,
  author = {Shin, Seung Yeob and Chaouch, Karim and Nejati, Shiva and Sabetzadeh, Mehrdad and Briand, Lionel C. and Zimmer, Frank},
  title = {Uncertainty-Aware Specification and Analysis for Hardware-in-the-Loop Testing of Cyber-Physical Systems},
  journal = {Journal of Systems and Software},
  year = 2021,
  month = jan,
  volume = {171},
  pages = {110813},
  issn = {01641212},
  doi = {10.1016/j.jss.2020.110813},
}

@article{an2020_modeling_verification_vehicles,
  author = {An, Dongdong and Liu, Jing and Zhang, Min and Chen, Xiaohong and Chen, Mingsong and Sun, Haiying},
  title = {Uncertainty Modeling and Runtime Verification for Autonomous Vehicles Driving Control: {{A}} Machine Learning-Based Approach},
  journal = {Journal of Systems and Software},
  year = 2020,
  month = sep,
  volume = {167},
  pages = {110617},
  issn = {01641212},
  doi = {10.1016/j.jss.2020.110617},
}

@article{bertoa2020_incorporating_measurement_ocl,
  author = {Bertoa, Manuel F. and Burgue{\~n}o, Loli and Moreno, Nathalie and Vallecillo, Antonio},
  title = {Incorporating Measurement Uncertainty into {{OCL}}/{{UML}} Primitive Datatypes},
  journal = {Software and Systems Modeling},
  year = 2020,
  month = sep,
  volume = {19},
  number = {5},
  pages = {1163--1189},
  issn = {1619-1366, 1619-1374},
  doi = {10.1007/s10270-019-00741-0},
}

@article{esterle2020_think_therefore_you,
  author = {Esterle, Lukas and Brown, John N. A.},
  title = {I {{Think Therefore You Are}}: {{Models}} for {{Interaction}} in {{Collectives}} of {{Self-aware Cyber-physical Systems}}},
  journal = {ACM Transactions on Cyber-Physical Systems},
  year = 2020,
  month = oct,
  volume = {4},
  number = {4},
  pages = {1--25},
  issn = {2378-962X, 2378-9638},
  doi = {10.1145/3375403},
}

@article{wang2020_resilient_rmpc_polytopic,
  author = {Wang, Jianhua and Song, Yan},
  title = {Resilient {{RMPC}} for {{Cyber-Physical Systems With Polytopic Uncertainties}} and {{State Saturation Under TOD Scheduling}}: {{An ADT Approach}}},
  journal = {IEEE Transactions on Industrial Informatics},
  year = 2020,
  month = jul,
  volume = {16},
  number = {7},
  pages = {4900--4908},
  issn = {1551-3203, 1941-0050},
  doi = {10.1109/TII.2019.2938889},
}

@article{wang2020_achieving_personalized_anonymity,
  author = {Wang, Jinbao and Cai, Zhipeng and Yu, Jiguo},
  title = {Achieving {{Personalized}} \$k\$-{{Anonymity-Based Content Privacy}} for {{Autonomous Vehicles}} in {{CPS}}},
  journal = {IEEE Transactions on Industrial Informatics},
  year = 2020,
  month = jun,
  volume = {16},
  number = {6},
  pages = {4242--4251},
  issn = {1551-3203, 1941-0050},
  doi = {10.1109/TII.2019.2950057},
}

@article{camilli2020_model_hypothesis_uncertain,
  author = {Camilli, Matteo and Gargantini, Angelo and Scandurra, Patrizia},
  title = {Model-based Hypothesis Testing of Uncertain Software Systems},
  journal = {Software Testing, Verification and Reliability},
  year = 2020,
  month = mar,
  volume = {30},
  number = {2},
  pages = {e1730},
  issn = {0960-0833, 1099-1689},
  doi = {10.1002/stvr.1730},
}

@article{fang2019_job_shop_smart,
  author = {Fang, Yilin and Peng, Chao and Lou, Ping and Zhou, Zude and Hu, Jianmin and Yan, Junwei},
  title = {Digital-{{Twin-Based Job Shop Scheduling Toward Smart Manufacturing}}},
  journal = {IEEE Transactions on Industrial Informatics},
  year = 2019,
  month = dec,
  volume = {15},
  number = {12},
  pages = {6425--6435},
  issn = {1551-3203, 1941-0050},
  doi = {10.1109/TII.2019.2938572},
}

@article{rahimilarki2019_neural_network_fault,
  author = {Rahimilarki, Reihane and Gao, Zhiwei and Zhang, Aihua and Binns, Richard},
  title = {Robust {{Neural Network Fault Estimation Approach}} for {{Nonlinear Dynamic Systems With Applications}} to {{Wind Turbine Systems}}},
  journal = {IEEE Transactions on Industrial Informatics},
  year = 2019,
  month = dec,
  volume = {15},
  number = {12},
  pages = {6302--6312},
  issn = {1551-3203, 1941-0050},
  doi = {10.1109/TII.2019.2893845},
}

@article{khayatian2020_crossroads_intersection_management,
  author = {Khayatian, Mohammad and Lou, Yingyan and Mehrabian, Mohammadreza and Shirvastava, Aviral},
  title = {Crossroads+: {{A Time-aware Approach}} for {{Intersection Management}} of {{Connected Autonomous Vehicles}}},
  journal = {ACM Transactions on Cyber-Physical Systems},
  year = 2020,
  month = apr,
  volume = {4},
  number = {2},
  pages = {1--28},
  issn = {2378-962X, 2378-9638},
  doi = {10.1145/3364182},
}

@article{medina2019_designing_controller_image,
  author = {Medina, R{\'o}binson and Valencia, Juan and Stuijk, Sander and Goswami, Dip and Basten, Twan},
  title = {Designing a {{Controller}} with {{Image-based Pipelined Sensing}} and {{Additive Uncertainties}}},
  journal = {ACM Transactions on Cyber-Physical Systems},
  year = 2019,
  month = jul,
  volume = {3},
  number = {3},
  pages = {1--26},
  issn = {2378-962X, 2378-9638},
  doi = {10.1145/3326067},
}

@article{li2019_sustainable_user_home,
  author = {Li, Wei and Chang, Xiaomin and Cao, Junwei and Yang, Ting and Sun, Yaojie and Zomaya, Albert Y.},
  title = {A {{Sustainable}} and {{User-Behavior-Aware Cyber-Physical System}} for {{Home Energy Management}}},
  journal = {ACM Transactions on Cyber-Physical Systems},
  year = 2019,
  month = oct,
  volume = {3},
  number = {4},
  pages = {1--24},
  issn = {2378-962X, 2378-9638},
  doi = {10.1145/3320075},
}

@article{wang2019_dynamic_data_injection,
  author = {Wang, Huaizhi and Ruan, Jiaqi and Zhou, Bin and Li, Canbing and Wu, Qiuwei and Raza, Muhammad Qamar and Cao, Guang-Zhong},
  title = {Dynamic {{Data Injection Attack Detection}} of {{Cyber Physical Power Systems With Uncertainties}}},
  journal = {IEEE Transactions on Industrial Informatics},
  year = 2019,
  month = oct,
  volume = {15},
  number = {10},
  pages = {5505--5518},
  issn = {1551-3203, 1941-0050},
  doi = {10.1109/TII.2019.2902163},
}

@article{ma2019_modeling_foundations_executable,
  author = {Ma, Tao and Ali, Shaukat and Yue, Tao},
  title = {Modeling Foundations for Executable Model-Based Testing of Self-Healing Cyber-Physical Systems},
  journal = {Software \& Systems Modeling},
  year = 2019,
  month = oct,
  volume = {18},
  number = {5},
  pages = {2843--2873},
  issn = {1619-1366, 1619-1374},
  doi = {10.1007/s10270-018-00703-y},
}

@article{yu2019_event_sequential_prognosis,
  author = {Yu, Ming and Lan, Dun and Huang, Yunzhi and Wang, Hai and Jiang, Canghua and Zhao, Linfeng},
  title = {Event-{{Based Sequential Prognosis}} for {{Uncertain Hybrid Systems With Intermittent Faults}}},
  journal = {IEEE Transactions on Industrial Informatics},
  year = 2019,
  month = aug,
  volume = {15},
  number = {8},
  pages = {4455--4468},
  issn = {1551-3203, 1941-0050},
  doi = {10.1109/TII.2018.2889883},
}

@article{chaves2019_fragility_dsverifier,
  author = {Chaves, Lennon C. and Ismail, Hussama I. and Bessa, Iury V. and Cordeiro, Lucas C. and {de Lima Filho}, Eddie B.},
  title = {Verifying Fragility in Digital Systems with Uncertainties Using {{DSVerifier}} v2.0},
  journal = {Journal of Systems and Software},
  year = 2019,
  month = jul,
  volume = {153},
  pages = {22--43},
  issn = {01641212},
  doi = {10.1016/j.jss.2019.03.015},
}

@article{zhang2019_wise_test_case,
  author = {Zhang, Man and Ali, Shaukat and Yue, Tao},
  title = {Uncertainty-Wise Test Case Generation and Minimization for {{Cyber-Physical Systems}}},
  journal = {Journal of Systems and Software},
  year = 2019,
  month = jul,
  volume = {153},
  pages = {1--21},
  issn = {01641212},
  doi = {10.1016/j.jss.2019.03.011},
}

@article{ma2019_self_healing_fragility,
  author = {Ma, Tao and Ali, Shaukat and Yue, Tao and Elaasar, Maged},
  title = {Testing Self-Healing Cyber-Physical Systems under Uncertainty: A Fragility-Oriented Approach},
  journal = {Software Quality Journal},
  year = 2019,
  month = jun,
  volume = {27},
  number = {2},
  pages = {615--649},
  issn = {0963-9314, 1573-1367},
  doi = {10.1007/s11219-018-9437-3},
}

@article{zhang2019_wise_test_modeling,
  author = {Zhang, Man and Ali, Shaukat and Yue, Tao and Norgren, Roland and Okariz, Oscar},
  title = {Uncertainty-{{Wise Cyber-Physical System}} Test Modeling},
  journal = {Software \& Systems Modeling},
  year = 2019,
  month = apr,
  volume = {18},
  number = {2},
  pages = {1379--1418},
  issn = {1619-1366, 1619-1374},
  doi = {10.1007/s10270-017-0609-6},
}

@article{kar2019_event_triggered_neural,
  author = {Kar, Aniket K. and Dhar, Narendra Kumar and Verma, Nishchal K.},
  title = {Event-{{Triggered Adaptive Neural Network Controller}} in a {{Cyber}}--{{Physical Framework}}},
  journal = {IEEE Transactions on Industrial Informatics},
  year = 2019,
  month = apr,
  volume = {15},
  number = {4},
  pages = {2101--2111},
  issn = {1551-3203, 1941-0050},
  doi = {10.1109/TII.2018.2861904},
}

@article{heinzemann2019_transactional_execution_hierarchical,
  author = {Heinzemann, Christian and Becker, Steffen and Volk, Andreas},
  title = {Transactional Execution of Hierarchical Reconfigurations in Cyber-Physical Systems},
  journal = {Software \& Systems Modeling},
  year = 2019,
  month = feb,
  volume = {18},
  number = {1},
  pages = {157--189},
  issn = {1619-1366, 1619-1374},
  doi = {10.1007/s10270-017-0583-z},
}

@article{zhang2019_national_scale_traffic,
  author = {Zhang, Desheng and He, Tian and Zhang, Fan},
  title = {National-Scale {{Traffic Model Calibration}} in {{Real Time}} with {{Multi-source Incomplete Data}}},
  journal = {ACM Transactions on Cyber-Physical Systems},
  year = 2019,
  month = apr,
  volume = {3},
  number = {2},
  pages = {1--26},
  issn = {2378-962X, 2378-9638},
  doi = {10.1145/3300186},
}

@article{gerostathopoulos2019_tuning_self_adaptation,
  author = {Gerostathopoulos, Ilias and Skoda, Dominik and Plasil, Frantisek and Bures, Tomas and Knauss, Alessia},
  title = {Tuning Self-Adaptation in Cyber-Physical Systems through Architectural Homeostasis},
  journal = {Journal of Systems and Software},
  year = 2019,
  month = feb,
  volume = {148},
  pages = {37--55},
  issn = {01641212},
  doi = {10.1016/j.jss.2018.10.051},
}

@article{usamentiaga2018_temperature_monitoring_electrical,
  author = {Usamentiaga, Ruben and Fernandez, Miguel Angel and Villan, Alberto Fernandez and Carus, Juan Luis},
  title = {Temperature {{Monitoring}} for {{Electrical Substations Using Infrared Thermography}}: {{Architecture}} for {{Industrial Internet}} of {{Things}}},
  journal = {IEEE Transactions on Industrial Informatics},
  year = 2018,
  month = dec,
  volume = {14},
  number = {12},
  pages = {5667--5677},
  issn = {1551-3203, 1941-0050},
  doi = {10.1109/TII.2018.2868452},
}

@article{camara2018_reasoning_about_sensing,
  author = {C{\'a}mara, Javier and Peng, Wenxin and Garlan, David and Schmerl, Bradley},
  title = {Reasoning about Sensing Uncertainty and Its Reduction in Decision-Making for Self-Adaptation},
  journal = {Science of Computer Programming},
  year = 2018,
  month = dec,
  volume = {167},
  pages = {51--69},
  issn = {01676423},
  doi = {10.1016/j.scico.2018.07.002},
}

@article{du2018_pcssl_stochastic_extension,
  author = {Du, Dehui and Huang, Ping and Jiang, Kaiqiang and Mallet, Fr{\'e}d{\'e}ric},
  title = {{{pCSSL}}: {{A}} Stochastic Extension to {{MARTE}}/{{CCSL}} for Modeling Uncertainty in {{Cyber Physical Systems}}},
  journal = {Science of Computer Programming},
  year = 2018,
  month = nov,
  volume = {166},
  pages = {71--88},
  issn = {01676423},
  doi = {10.1016/j.scico.2018.05.005},
}

@article{wang2018_deep_learning_interval,
  author = {Wang, Huaizhi and Ruan, Jiaqi and Wang, Guibin and Zhou, Bin and Liu, Yitao and Fu, Xueqian and Peng, Jianchun},
  title = {Deep {{Learning-Based Interval State Estimation}} of {{AC Smart Grids Against Sparse Cyber Attacks}}},
  journal = {IEEE Transactions on Industrial Informatics},
  year = 2018,
  month = nov,
  volume = {14},
  number = {11},
  pages = {4766--4778},
  issn = {1551-3203, 1941-0050},
  doi = {10.1109/TII.2018.2804669},
}

@article{joerger2018_improve_management_large,
  author = {Joerger, Guillaume and Rambourg, Juliette and {Gaspard-Boulinc}, Helene and Conversy, Stephane and Bass, Barbara L. and Dunkin, Brian J. and Garbey, Marc},
  title = {A {{Cyber-Physical System}} to {{Improve}} the {{Management}} of a {{Large Suite}} of {{Operating Rooms}}},
  journal = {ACM Transactions on Cyber-Physical Systems},
  year = 2018,
  month = oct,
  volume = {2},
  number = {4},
  pages = {1--24},
  issn = {2378-962X, 2378-9638},
  doi = {10.1145/3140234},
}

@article{ma2019_holistic_management_dependable,
  author = {Ma, Yehan and Gunatilaka, Dolvara and Li, Bo and Gonzalez, Humberto and Lu, Chenyang},
  title = {Holistic {{Cyber-Physical Management}} for {{Dependable Wireless Control Systems}}},
  journal = {ACM Transactions on Cyber-Physical Systems},
  year = 2019,
  month = jan,
  volume = {3},
  number = {1},
  pages = {1--25},
  issn = {2378-962X, 2378-9638},
  doi = {10.1145/3185510},
}

@article{martin2018_slats_simultaneous_localization,
  author = {Martin, Paul and Symington, Andrew and Srivastava, Mani},
  title = {{{SLATS}}: {{Simultaneous Localization}} and {{Time Synchronization}}},
  journal = {ACM Transactions on Cyber-Physical Systems},
  year = 2018,
  month = jul,
  volume = {2},
  number = {3},
  pages = {1--25},
  issn = {2378-962X, 2378-9638},
  doi = {10.1145/3134841},
}

@article{wen2018_consensus_strategy_economic,
  author = {Wen, Guanghui and Yu, Xinghuo and Liu, Zhi-Wei and Yu, Wenwu},
  title = {Adaptive {{Consensus-Based Robust Strategy}} for {{Economic Dispatch}} of {{Smart Grids Subject}} to {{Communication Uncertainties}}},
  journal = {IEEE Transactions on Industrial Informatics},
  year = 2018,
  month = jun,
  volume = {14},
  number = {6},
  pages = {2484--2496},
  issn = {1551-3203, 1941-0050},
  doi = {10.1109/TII.2017.2772088},
}

@article{jiang2018_recursive_total_principle,
  author = {Jiang, Yuchen and Yin, Shen},
  title = {Recursive {{Total Principle Component Regression Based Fault Detection}} and {{Its Application}} to {{Vehicular Cyber-Physical Systems}}},
  journal = {IEEE Transactions on Industrial Informatics},
  year = 2018,
  month = apr,
  volume = {14},
  number = {4},
  pages = {1415--1423},
  issn = {1551-3203, 1941-0050},
  doi = {10.1109/TII.2017.2752709},
}

@article{lu2017_distributed_secondary_voltage,
  author = {Lu, Xiaoqing and Yu, Xinghuo and Lai, Jingang and Guerrero, Josep M. and Zhou, Hong},
  title = {Distributed {{Secondary Voltage}} and {{Frequency Control}} for {{Islanded Microgrids With Uncertain Communication Links}}},
  journal = {IEEE Transactions on Industrial Informatics},
  year = 2017,
  month = apr,
  volume = {13},
  number = {2},
  pages = {448--460},
  issn = {1551-3203, 1941-0050},
  doi = {10.1109/TII.2016.2603844},
}

@article{gerostathopoulos2017_strengthening_adaptation_meta,
  author = {Gerostathopoulos, Ilias and Bures, Tomas and Hnetynka, Petr and Hujecek, Adam and Plasil, Frantisek and Skoda, Dominik},
  title = {Strengthening {{Adaptation}} in {{Cyber-Physical Systems}} via {{Meta-Adaptation Strategies}}},
  journal = {ACM Transactions on Cyber-Physical Systems},
  year = 2017,
  month = jul,
  volume = {1},
  number = {3},
  pages = {1--25},
  issn = {2378-962X, 2378-9638},
  doi = {10.1145/2823345},
}

@article{martini2017_real_energy,
  author = {Martini, Daniele De and Benetti, Guido and Vedova, Marco L. Della and Facchinetti, Tullio},
  title = {Adaptive {{Real-Time Scheduling}} of {{Cyber-Physical Energy Systems}}},
  journal = {ACM Transactions on Cyber-Physical Systems},
  year = 2017,
  month = oct,
  volume = {1},
  number = {4},
  pages = {1--25},
  issn = {2378-962X, 2378-9638},
  doi = {10.1145/3047412},
}

@article{zhang2017_heterogeneous_model_integration,
  author = {Zhang, Desheng and Zhao, Juanjuan and Zhang, Fan and He, Tian and Lee, Haengju and Son, Sang H.},
  title = {Heterogeneous {{Model Integration}} for {{Multi-Source Urban Infrastructure Data}}},
  journal = {ACM Transactions on Cyber-Physical Systems},
  year = 2017,
  month = jan,
  volume = {1},
  number = {1},
  pages = {1--26},
  issn = {2378-962X, 2378-9638},
  doi = {10.1145/2967503},
}

@article{talusan2025_end_end_solution,
  author = {Talusan, Jose Paolo and Han, Chaeeun and Rogers, David and Mukhopadhyay, Ayan and Laszka, Aron and Freudberg, Dan and Dubey, Abhishek},
  title = {An {{End-to-End Solution}} for {{Public Transit Stationing}} and {{Dispatch Problem}}},
  journal = {ACM Transactions on Cyber-Physical Systems},
  year = 2025,
  month = oct,
  volume = {9},
  number = {4},
  pages = {1--25},
  issn = {2378-962X, 2378-9638},
  doi = {10.1145/3754454},
}

@inproceedings{sanchez2024_automated_planning_temporal,
  author = {Sanchez, Raquel and Troya, Javier and Camara, Javier},
  title = {Automated {{Planning}} for {{Adaptive Cyber-Physical Systems}} under {{Uncertainty}} in {{Temporal Availability Constraints}}},
  booktitle = {Proceedings of the 19th {{International Symposium}} on {{Software Engineering}} for {{Adaptive}} and {{Self-Managing Systems}}},
  year = 2024,
  month = apr,
  pages = {14--24},
  publisher = {ACM},
  address = {Lisbon AA Portugal},
  isbn = {979-8-4007-0585-4},
  doi = {10.1145/3643915.3644083},
}

@article{hashimoto2022_collaborative_rover_copter,
  author = {Hashimoto, Kazumune and Tsumagari, Natsuko and Ushio, Toshimitsu},
  title = {Collaborative {{Rover-copter Path Planning}} and {{Exploration}} with {{Temporal Logic Specifications Based}} on {{Bayesian Update Under Uncertain Environments}}},
  journal = {ACM Transactions on Cyber-Physical Systems},
  year = 2022,
  month = apr,
  volume = {6},
  number = {2},
  pages = {1--24},
  issn = {2378-962X, 2378-9638},
  doi = {10.1145/3470453},
}

@article{wang2025_conformal_temporal_logic,
  author = {Wang, Jun and Tong, Jiaming and Tan, Kaiyuan and Vorobeychik, Yevgeniy and Kantaros, Yiannis},
  title = {Conformal {{Temporal Logic Planning}} Using {{Large Language Models}}},
  journal = {ACM Transactions on Cyber-Physical Systems},
  year = 2025,
  month = sep,
  pages = {3769111},
  issn = {2378-962X, 2378-9638},
  doi = {10.1145/3769111},
}

@article{valle2025_defining_generating_multi,
  author = {Valle, Pablo and Arrieta, Aitor and Han, Liping and Ali, Shaukat and Yue, Tao},
  title = {Defining and Generating Multi-Level and Uncertainty-Wise Test Oracles for Cyber-Physical Systems},
  journal = {Software and Systems Modeling},
  year = 2025,
  month = jun,
  volume = {24},
  number = {3},
  pages = {679--704},
  issn = {1619-1366, 1619-1374},
  doi = {10.1007/s10270-025-01271-8},
}

@inproceedings{galarraga2021_genetic_algorithm_industrial,
  author = {Galarraga, Joritz and Arrieta, Aitor and Ali, Shaukat and Sagardui, Goiuria and Arratibel, Maite},
  title = {Genetic {{Algorithm-based Testing}} of {{Industrial Elevators}} under {{Passenger Uncertainty}}},
  booktitle = {2021 {{IEEE International Symposium}} on {{Software Reliability Engineering Workshops}} ({{ISSREW}})},
  year = 2021,
  month = oct,
  pages = {353--358},
  publisher = {IEEE},
  address = {Wuhan, China},
  isbn = {978-1-6654-2603-9},
  doi = {10.1109/ISSREW53611.2021.00101},
}

@article{zhang2017_wise_test_ready,
  author = {Zhang, Man and Ali, Shaukat and Yue, Tao and Norgre, Roland},
  title = {Uncertainty-Wise Evolution of Test Ready Models},
  journal = {Information and Software Technology},
  year = 2017,
  month = jul,
  volume = {87},
  pages = {140--159},
  issn = {09505849},
  doi = {10.1016/j.infsof.2017.03.003},
}

@inproceedings{shin2018_test_case_prioritization,
  author = {Shin, Seung Yeob and Nejati, Shiva and Sabetzadeh, Mehrdad and Briand, Lionel C. and Zimmer, Frank},
  title = {Test Case Prioritization for Acceptance Testing of Cyber Physical Systems: A Multi-Objective Search-Based Approach},
  booktitle = {Proceedings of the 27th {{ACM SIGSOFT International Symposium}} on {{Software Testing}} and {{Analysis}}},
  year = 2018,
  month = jul,
  pages = {49--60},
  publisher = {ACM},
  address = {Amsterdam Netherlands},
  isbn = {978-1-4503-5699-2},
  doi = {10.1145/3213846.3213852},
}

@article{ayerdi2025_marmot_metamorphic_monitoring,
  author = {Ayerdi, Jon and Iriarte, Asier and Valle, Pablo and Roman, Ibai and Illarramendi, Miren and Arrieta, Aitor},
  title = {{{MarMot}}: {{Metamorphic Runtime Monitoring}} of {{Autonomous Driving Systems}}},
  journal = {ACM Transactions on Software Engineering and Methodology},
  year = 2025,
  month = jan,
  volume = {34},
  number = {1},
  pages = {1--35},
  issn = {1049-331X, 1557-7392},
  doi = {10.1145/3678171},
}

@inproceedings{shin2018_hitecs_uml_profile,
  author = {Shin, Seung Yeob and Chaouch, Karim and Nejati, Shiva and Sabetzadeh, Mehrdad and Briand, Lionel C. and Zimmer, Frank},
  title = {{{HITECS}}: {{A UML Profile}} and {{Analysis Framework}} for {{Hardware-in-the-Loop Testing}} of {{Cyber Physical Systems}}},
  booktitle = {Proceedings of the 21th {{ACM}}/{{IEEE International Conference}} on {{Model Driven Engineering Languages}} and {{Systems}}},
  year = 2018,
  month = oct,
  pages = {357--367},
  publisher = {ACM},
  address = {Copenhagen Denmark},
  isbn = {978-1-4503-4949-9},
  doi = {10.1145/3239372.3239382},
}

@article{bandyszak2020_orthogonal_modeling_engineering,
  author = {Bandyszak, Torsten and Daun, Marian and Tenbergen, Bastian and Kuhs, Patrick and Wolf, Stefanie and Weyer, Thorsten},
  title = {Orthogonal {{Uncertainty Modeling}} in the {{Engineering}} of {{Cyber-Physical Systems}}},
  journal = {IEEE Transactions on Automation Science and Engineering},
  year = 2020,
  pages = {1--16},
  issn = {1545-5955, 1558-3783},
  doi = {10.1109/TASE.2020.2980726},
}

@article{wang2018_rsimplex_architecture_failures,
  author = {Wang, Xiaofeng and Hovakimyan, Naira and Sha, Lui},
  title = {{{RSimplex}}: {{A Robust Control Architecture}} for {{Cyber And Physical Failures}}},
  journal = {ACM Transactions on Cyber-Physical Systems},
  year = 2018,
  month = oct,
  volume = {2},
  number = {4},
  pages = {1--26},
  issn = {2378-962X, 2378-9638},
  doi = {10.1145/3121428},
}

@inproceedings{devries2021_monitoring_environmental_knowledge,
  author = {DeVries, Byron and Fredericks, Erik M. and Cheng, Betty H.C.},
  title = {Analysis and {{Monitoring}} of {{Cyber-Physical Systems}} via {{Environmental Domain Knowledge}} \& {{Modeling}}},
  booktitle = {2021 {{International Symposium}} on {{Software Engineering}} for {{Adaptive}} and {{Self-Managing Systems}} ({{SEAMS}})},
  year = 2021,
  month = may,
  pages = {11--17},
  publisher = {IEEE},
  address = {Madrid, Spain},
  isbn = {978-1-6654-0289-7},
  doi = {10.1109/SEAMS51251.2021.00013},
}

@inproceedings{hartsell2021_resonate_risk_assessment,
  author = {Hartsell, Charles and Ramakrishna, Shreyas and Dubey, Abhishek and Stojcsics, Daniel and Mahadevan, Nagabhushan and Karsai, Gabor},
  title = {{{ReSonAte}}: {{A Runtime Risk Assessment Framework}} for {{Autonomous Systems}}},
  booktitle = {2021 {{International Symposium}} on {{Software Engineering}} for {{Adaptive}} and {{Self-Managing Systems}} ({{SEAMS}})},
  year = 2021,
  month = may,
  pages = {118--129},
  publisher = {IEEE},
  address = {Madrid, Spain},
  isbn = {978-1-6654-0289-7},
  doi = {10.1109/SEAMS51251.2021.00025},
}

@article{gerostathopoulos2016_self_adaptation_intensive,
  author = {Gerostathopoulos, Ilias and Bures, Tomas and Hnetynka, Petr and Keznikl, Jaroslav and Kit, Michal and Plasil, Frantisek and Plouzeau, No{\"e}l},
  title = {Self-Adaptation in Software-Intensive Cyber--Physical Systems: {{From}} System Goals to Architecture Configurations},
  journal = {Journal of Systems and Software},
  year = 2016,
  month = dec,
  volume = {122},
  pages = {378--397},
  issn = {01641212},
  doi = {10.1016/j.jss.2016.02.028},
}

@inproceedings{zhou2024_building_nvidia_isaac,
  author = {Zhou, Zhehua and Song, Jiayang and Xie, Xuan and Shu, Zhan and Ma, Lei and Liu, Dikai and Yin, Jianxiong and See, Simon},
  title = {Towards {{Building AI-CPS}} with {{NVIDIA Isaac Sim}}: {{An Industrial Benchmark}} and {{Case Study}} for {{Robotics Manipulation}}},
  booktitle = {Proceedings of the 46th {{International Conference}} on {{Software Engineering}}: {{Software Engineering}} in {{Practice}}},
  year = 2024,
  month = apr,
  pages = {263--274},
  publisher = {ACM},
  address = {Lisbon Portugal},
  isbn = {979-8-4007-0501-4},
  doi = {10.1145/3639477.3639740},
}

@article{cui2023_filtering_out_high,
  author = {Cui, Yangguang and Li, Liying and Tao, Zhe and Chen, Mingsong and Wei, Tongquan},
  title = {Filtering {{Out High Noise Data}} for {{Distributed Deep Neural Networks}}},
  journal = {IEEE Transactions on Automation Science and Engineering},
  year = 2023,
  month = jan,
  volume = {20},
  number = {1},
  pages = {101--111},
  issn = {1545-5955, 1558-3783},
  doi = {10.1109/TASE.2022.3208027},
}

@inproceedings{bao2025_quantification_physics_informed,
  author = {Bao, Tianshu and Liu, Xiaoou and Ma, Meiyi and Johnson, Taylor T. and Wei, Hua},
  title = {Uncertainty {{Quantification}} for {{Physics-Informed Traffic Graph Networks}}},
  booktitle = {Proceedings of the {{ACM}}/{{IEEE}} 16th {{International Conference}} on {{Cyber-Physical Systems}} (with {{CPS-IoT Week}} 2025)},
  year = 2025,
  month = may,
  pages = {1--10},
  publisher = {ACM},
  address = {Irvine CA USA},
  isbn = {979-8-4007-1498-6},
  doi = {10.1145/3716550.3722023},
}

@inproceedings{lin2025_monitoring_learning_enabled,
  author = {Lin, Vivian and Kaur, Ramneet and Yang, Yahan and Dutta, Souradeep and Kantaros, Yiannis and Roy, Anirban and Jha, Susmit and Sokolsky, Oleg and Lee, Insup},
  title = {Safety {{Monitoring}} for {{Learning-Enabled Cyber-Physical Systems}} in {{Out-of-Distribution Scenarios}}},
  booktitle = {Proceedings of the {{ACM}}/{{IEEE}} 16th {{International Conference}} on {{Cyber-Physical Systems}} (with {{CPS-IoT Week}} 2025)},
  year = 2025,
  month = may,
  pages = {1--11},
  publisher = {ACM},
  address = {Irvine CA USA},
  isbn = {979-8-4007-1498-6},
  doi = {10.1145/3716550.3722022},
}

@inproceedings{biglari2022_model_validity_tolerance,
  author = {Biglari, Raheleh and Denil, Joachim},
  title = {Model Validity and Tolerance Quantification for Real-Time Adaptive Approximation},
  booktitle = {Proceedings of the 25th {{International Conference}} on {{Model Driven Engineering Languages}} and {{Systems}}: {{Companion Proceedings}}},
  year = 2022,
  month = oct,
  pages = {668--676},
  publisher = {ACM},
  address = {Montreal Quebec Canada},
  isbn = {978-1-4503-9467-3},
  doi = {10.1145/3550356.3561604},
}

@inproceedings{foldiak2024_probabilistic_graph_queries,
  author = {F{\"o}ldi{\'a}k, M{\'a}t{\'e}},
  title = {Probabilistic {{Graph Queries}} for {{Design Space Exploration Under Uncertainty}}},
  booktitle = {Proceedings of the {{ACM}}/{{IEEE}} 27th {{International Conference}} on {{Model Driven Engineering Languages}} and {{Systems}}},
  year = 2024,
  month = sep,
  pages = {142--148},
  publisher = {ACM},
  address = {Linz Austria},
  isbn = {979-8-4007-0622-6},
  doi = {10.1145/3652620.3688199},
}

@inproceedings{li2017_stochastic_contracts_probabilistic,
  author = {Li, Jiwei and Nuzzo, Pierluigi and {Sangiovanni-Vincentelli}, Alberto and Xi, Yugeng and Li, Dewei},
  title = {Stochastic Contracts for Cyber-Physical System Design under Probabilistic Requirements},
  booktitle = {Proceedings of the 15th {{ACM-IEEE International Conference}} on {{Formal Methods}} and {{Models}} for {{System Design}}},
  year = 2017,
  month = sep,
  pages = {5--14},
  publisher = {ACM},
  address = {Vienna Austria},
  isbn = {978-1-4503-5093-8},
  doi = {10.1145/3127041.3127045},
}

@inproceedings{ali2017_evaluation_mutation_crossover,
  author = {Ali, Shaukat and Li, Yan and Yue, Tao and Zhang, Man},
  title = {An {{Empirical Evaluation}} of {{Mutation}} and {{Crossover Operators}} for {{Multi-Objective Uncertainty-Wise Test Minimization}}},
  booktitle = {2017 {{IEEE}}/{{ACM}} 10th {{International Workshop}} on {{Search-Based Software Testing}} ({{SBST}})},
  year = 2017,
  month = may,
  pages = {21--27},
  publisher = {IEEE},
  address = {Buenos Aires, Argentina},
  isbn = {978-1-5386-2789-1},
  doi = {10.1109/SBST.2017.9},
}

@inproceedings{cai2020_real_out_distribution,
  author = {Cai, Feiyang and Koutsoukos, Xenofon},
  title = {Real-Time {{Out-of-distribution Detection}} in {{Learning-Enabled Cyber-Physical Systems}}},
  booktitle = {2020 {{ACM}}/{{IEEE}} 11th {{International Conference}} on {{Cyber-Physical Systems}} ({{ICCPS}})},
  year = 2020,
  month = apr,
  pages = {174--183},
  publisher = {IEEE},
  address = {Sydney, Australia},
  isbn = {978-1-7281-5501-2},
  doi = {10.1109/ICCPS48487.2020.00024},
}

@inproceedings{ali2015_test_evolving_modelling,
  author = {Ali, Shaukat and Yue, Tao},
  title = {U-{{Test}}: {{Evolving}}, {{Modelling}} and {{Testing Realistic Uncertain Behaviours}} of {{Cyber-Physical Systems}}},
  booktitle = {2015 {{IEEE}} 8th {{International Conference}} on {{Software Testing}}, {{Verification}} and {{Validation}} ({{ICST}})},
  year = 2015,
  month = apr,
  pages = {1--2},
  publisher = {IEEE},
  address = {Graz, Austria},
  isbn = {978-1-4799-7125-1},
  doi = {10.1109/ICST.2015.7102637},
}

@article{feng2021_improving_variational_autoencoder,
  author = {Feng, Yeli and Ng, Daniel Jun Xian and Easwaran, Arvind},
  title = {Improving {{Variational Autoencoder}} Based {{Out-of-Distribution Detection}} for {{Embedded Real-time Applications}}},
  journal = {ACM Transactions on Embedded Computing Systems},
  year = 2021,
  month = oct,
  volume = {20},
  number = {5s},
  pages = {1--26},
  issn = {1539-9087, 1558-3465},
  doi = {10.1145/3477026},
}

@article{camilli2023_enforcing_resilience_equilibrium,
  author = {Camilli, Matteo and Mirandola, Raffaela and Scandurra, Patrizia},
  title = {Enforcing {{Resilience}} in {{Cyber-physical Systems}} via {{Equilibrium Verification}} at {{Runtime}}},
  journal = {ACM Transactions on Autonomous and Adaptive Systems},
  year = 2023,
  month = sep,
  volume = {18},
  number = {3},
  pages = {1--32},
  issn = {1556-4665, 1556-4703},
  doi = {10.1145/3584364},
}

@article{das2024_interpretable_latent_space,
  author = {Das, Suman and Yuhas, Michael and Koh, Rachel and Easwaran, Arvind},
  title = {Interpretable {{Latent Space}} for {{Meteorological Out-of-Distribution Detection}} via {{Weak Supervision}}},
  journal = {ACM Transactions on Cyber-Physical Systems},
  year = 2024,
  month = apr,
  volume = {8},
  number = {2},
  pages = {1--26},
  issn = {2378-962X, 2378-9638},
  doi = {10.1145/3651224},
}

@inproceedings{lu2023_evolutionary_computation_rl,
  author    = {Chengjie Lu},
  title     = {Evolutionary Computation and Reinforcement Learning for Cyber-physical System Design},
  booktitle = {45th {IEEE/ACM} International Conference on Software Engineering: {ICSE} 2023 Companion Proceedings, Melbourne, Australia, May 14--20, 2023},
  year      = {2023},
  pages     = {264--266},
  publisher = {IEEE},
  address   = {Melbourne, Australia},
  doi       = {10.1109/ICSE-Companion58688.2023.00071},
  url       = {https://doi.org/10.1109/ICSE-Companion58688.2023.00071}
}

@article{Haddaway2022Citationchaser,
  author  = {Haddaway, Neal R. and Grainger, Matthew J. and Gray, Charles T.},
  title   = {Citationchaser: A tool for transparent and efficient forward and backward citation chasing in systematic searching},
  journal = {Research Synthesis Methods},
  year    = {2022},
  volume  = {13},
  number  = {4},
  pages   = {533--545},
  doi     = {10.1002/jrsm.1563},
  url     = {https://onlinelibrary.wiley.com/doi/abs/10.1002/jrsm.1563}
}

@article{asmat2023uncertainty,
  title={Uncertainty handling in cyber--physical systems: State-of-the-art approaches, tools, causes, and future directions},
  author={Asmat, Mah Noor and Khan, Saif Ur Rehman and Hussain, Shahid},
  journal={Journal of software: Evolution and Process},
  volume={35},
  number={7},
  pages={e2428},
  year={2023},
  publisher={Wiley Online Library}
}

@article{makelburg2026surveying,
  title={Surveying uncertainty representation: a unified model for cyber-physical systems},
  author={M{\"a}kelburg, Johannes and Perez-Palacin, Diego and Mirandola, Raffaela and Acosta, Maribel},
  journal={Computing},
  volume={108},
  number={5},
  pages={68},
  year={2026},
  publisher={Springer}
}

@incollection{pinto2023analysis,
  title={Analysis and Design of Uncertain Cyber-Physical Systems},
  author={Pinto, Alessandro},
  booktitle={Computation-Aware Algorithmic Design for Cyber-Physical Systems},
  pages={25--53},
  year={2023},
  publisher={Springer}
}

@misc{replication,
  author       = {Khan, Amna  and
                  {Beena} and
                  Di Penta, Massimiliano and
                  Arrieta, Aitor},
  title        = {Replication package of  the paper "Uncertainty in Cyber-Physical Systems: A Literature Survey"},
  month        = jul,
  year         = 2026,
  publisher    = {Zenodo},
  version      = {v1.0.0},
  doi          = {10.5281/zenodo.21604477},
  url          = {https://doi.org/10.5281/zenodo.21604477}
}
